\documentclass[a4paper,11pt]{article}
\pdfoutput=1 

\usepackage{jcappub} 

\usepackage[T1]{fontenc} 
\usepackage{subfigure}
\usepackage{braket}
\title{\boldmath Dynamical equations and primordial GWs in a non-commutative branching cosmology}
\author[a,b]{C\'esar A. Zen Vasconcellos}
\author[c,d,1]{Peter O. Hess \note{Corresponding Author.}}
\author[e]{Jos\'e de Freitas Pacheco}
\author[f,g]{Fridolin Weber}
\author[a]{Benno Bodmann}
\author[a]{Dimiter Hadjimichef}
\author[d]{Marcelo Netz-Marzola}
\author[a]{Geovane Naysinger}
\author[h]{Mois\'es Razeira}
\affiliation[a]{Instituto de F\'isica, Universidade Federal do Rio Grande do Sul (UFRGS), Porto Alegre, Brazil}
\affiliation[b]{International Center for Relativistic Astrophysics Network (ICRANet), Pescara, Roma, Italy}
\affiliation[c]{Universidad Nacional Aut\'onoma de Mexico (UNAM), M\'exico City, M\'exico}
\affiliation[d]{Frankfurt Institute for Advanced Studies (FIAS), Hessen, Germany}
\affiliation[e]{Observatoire de la C\^ote d'Azur, Nice, France}
\affiliation[f]{Department of Physics, San Diego State University (SDSU), San Diego, USA}
\affiliation[g]{Department of Physics, University of California at San Diego (UCSD), La Jolla, USA }
\affiliation[h]{Universidade Federal do Pampa (UNIPAMPA), Ca\c{c}apava do Sul, Brazil}
\emailAdd{cesaraugustozenvasconcellos@gmail.com}
\emailAdd{hess@nucleares.unam.mx}
\emailAdd{fweber@sdsu.edu}
\emailAdd{benno.bodmann@gmx.de}
\emailAdd{dimiter@if.ufrgs.br}
\emailAdd{marzola@fias.uni-frankfurt.de}
\emailAdd{geovane.naysinger@ufrgs.br}
\emailAdd{moisesrazeira@unipampa.edu.br}
\abstract{Branch-cut Gravity (BCG) an extended version of the ontological cosmology domain of General Relativity, which is analytically continued to the complex plane. When combined with the Hawking-Hertog multiverse conception, BCG successfully addresses the issue of the primordial singularity. It consistently portrays the early Universe as a Riemannian foliation in which the singularities of the multiverse merge, giving rise to a smooth branching topological structure that resembles continuously connected Riemann surfaces.
This structure introduces a new cosmic scale factor that is analytically continued into the complex plane. 
In this contribution, we start with the recently developed Wheeler DeWitt-Ho\v{r}ava-Lifshitz non-commutative BCG formulation of quantum gravity. We investigate the impact of a non-commutative mini-superspace of variables obeying Poisson algebra on the accelerated behavior of the branch-cutting cosmic scale factor.
Furthermore, we conduct a preliminary study
on relic gravitational waves based on a linearized BCG approach, which describes the dynamics of a slightly perturbed gravitational field.
Our findings concerning the dynamics of the scale factor reveal that
they are highly sensitive to both the initial conditions and the
different parameterizations of the primordial matter content. These
findings indicate significant acceleration effects when compared to
predictions based on conventional commutative approaches.
These effects drive spacetime acceleration, offering a new perspective on explaining the accelerating expansion rate of our Universe.
As far as primordial relic gravitational waves are concerned, our predictions reveal an intricate transition between the two phases of the branched Universe: a contraction phase preceding the
conventional concept of a primordial singularity and a subsequent expansion phase whose transition region is characterized by a Riemannian topological foliation structure.  Furthermore, this transition is characterized by asymmetric distributions of gravitational wave intensities. These results  
indicate in turn dynamic asymmetries in the distribution of matter and energy, 
arising from the capture of short and long
scales within the non-commutative algebraic structure.
These results are consistent with the expected highly inhomogeneous 
distribution of primordial gravitational waves, in contrast to the 
anticipated future observations of relic gravitational waves, which
are expected to fill the universe as a stochastic homogeneous
background.}


\begin{document}
\maketitle
\flushbottom

\section{Introduction}\label{1}

The chronology of our Universe  indicates that in the primordial times, there was an inflationary-dominated era, followed by dominant phases of radiation, matter, and currently, we are in the dark energy era, which is causing the accelerated expansion of the Universe. 

On a time scale, or conversely, on an inverse temperature scale, the first observed photons (known as the first light), carry remnants of the cosmic microwave background (CMB) radiation, which is the residual light from the Big Bang. These photons originated during the recombination phase, a crucial epoch marked by the decoupling of matter and radiation. This phase occurred approximately 380 million years after cosmic inflation, when the Universe had reached a temperature of approximately T$\sim$ 0.26 eV. During this period, free electrons  began to combine with protons to form hydrogen atoms. Prior to the  recombination era, the photons present in the Universe underwent processes of continuous dispersion by the electric charges that composed the hot, dense primordial plasma. Due to 
Thomson scattering by free electric charges, these continuous dispersion process are supposed to rendering the medium opaque to the propagation of electromagnetic radiation, drastically reducing the mean free path traveled by each photon.

Gravitational waves, ripples in space-time caused by extreme energetic processes as predicted by Albert Einstein~\cite{Einstein1916,Einstein1917} and observed for the first time by the LIGO-Virgo collaboration~\cite{Abbott}, emerge as the most prominent mechanism for accessing information from the evolutionary Universe before the recombination era. In this context, speculations point to the possible  existence of a stochastic gravitational waves background (SGWB)~\cite{Abbott,Abbott2016,Pacheco2023WS}.
This background's signal distribution is expected to be similar to that of the CMB radiation and is originated by the superposition of countless incoherent sources spreading out in all directions. It carries signatures of physical processes in the early Universe, such as quantum fluctuations during the inflation era~\cite{Guth1981,Guth2004}. 
Other mechanisms for generating relic gravitational waves can be mentioned, in particular during the electroweak (EW) phase transition of the primordial Universe, such as, for example, colliding bubbles during their expanding phase, 
the decay of magnetohydrodynamic turbulence produced by bulk motions of the bubbles, and from the propagation of damped sound waves~\citep{Pacheco2023WS}. We will return to these topics later.

In this contribution, using as a starting point a recently developed branch-cut quantum gravity formulation (BCQG) ~\citep{Bodmann2023a,Bodmann2023b}, based on the 
Wheeler DeWitt~\cite{DeWitt1967} and the Ho\v{r}ava-Lifshitz~\cite{Horava} approaches,
we investigate the effects of non-commutative variables in
a mini-superspace that obey Poisson's algebra. This investigation
focuses on the accelerating behavior of the Universe and the branch-cut cosmic scale factor $\ln^{-1}[\beta(t)]$~\cite{Zen2020, Zen2021a, Zen2021b, Bodmann, Pacheco, Zen2022a, Hess, Bodmann2023a,Bodmann2023b} and its impact on relic  gravitational wave signals.

\section{Branch-cut Gravity}

Branch-cut gravity (BCG), an ontological domain extension of General Relativity to the complex plane~\cite{Einstein1916,Einstein1917,Zen2020,Zen2021a,Zen2021b,Zen2022a,Bodmann,Pacheco,Hess,Bodmann2023a,Bodmann2023b}, was developed with the idea of offering a theoretical alternative to inflation models~\cite{Guth1981,Guth2004}, based on the mathematical augmentation technique and notion of closure and existential completeness~\cite{Manders}. Such mathematical procedures proved to be extremely useful in both quantum mechanics~\cite{Dirac1937,Aharonov,Wu}
and in pseudo-complex general relativity (pc-GR)~\cite{Greiner,Hess2017},
with direct physical and cosmological manifestations. 

In 1959, in a seminal work, Y. Aharonov and D. Bohm demonstrated that the electromagnetic potential plays a role in quantum mechanics similar to the (complex) index of refraction in optics~\cite{Aharonov}. This work demonstrated for the first time that expanding the domains of theoretical realization of quantum mechanics to the sector of complex variables was not limited to a mere formal descriptive simplification. 
Numerous investigations were subsequently undertaken in an effort to
comprehend the role of complex variables in quantum
mechanics. Researchers sought to determine whether complex numbers are
indeed indispensable, or if it is feasible to devise an alternative,
coherent formulation that relies solely on real numbers.
In this context, the authors of ref.~\cite{Wu} concluded that the imaginary domain plays a pivotal role in discriminating quantum states.
More precisely, according to these authors, there exist real quantum states which can be perfectly distinguished via local operations and classical communication, but which cannot be distinguished with any nonzero probability if one of the parties has no access to the imaginary sector. The authors confirmed this phenomenon experimentally with linear optics, performing discrimination of different two-photon quantum states by local projective measurements. In conclusion, the results prove that complex numbers are an indispensable part of quantum mechanics. In brief, domain extension of quantum mechanics, as
demonstrated in these examples, reveals predictions related to the influence
of electromagnetic potentials on charged particles within the quantum realm.
This influence persists even in regions where all fields, and consequently forces on particles, vanish~\cite{Aharonov}  or emphasizes the critical role it plays in state discrimination~\cite{Wu}. 
In a recent publication, Marc-Olivier Renou and his collaborators~\cite{Renou} demonstrated that real and complex Hilbert-space formulations of quantum theory yield distinct predictions in network scenarios involving independent states and measurements. This prompted the authors to design a Bell-like experiment, the successful execution of which would disprove real quantum theory, just as conventional Bell experiments disproved the notion of local physics.

Similarly, the incorporation of a pseudo-complex domain into its formalism enabled pc-GR to discern a mechanism for suppressing
 the primordial gravitational singularity and to predict of existence of dark energy both within and outside cosmic mass distributions~\citep{Hess2009,Hess2017,Hess2020,Hess2023WS}.

These findings expanded our understanding that enhancing the
descriptive domain of a theory by incorporating complex or
pseudo-complex variables can lead to a broader perspective on both
infinitesimally small and large scales, with direct and tangible
physical implications.
BGC extends GR to the complex domain, whence one obtains analytically continued Friedmann equations. These new Friedmann equations are then superposed, i.e., the equations are summed over some index $\xi$ which is associated to a set of independent universes, following the concept of multiverse proposed by Hawking-Hertog~\citep{Hawking2018}. 
The Riemann sum is then taken to the integral limit. The resulting set of integro-differential equations - the BCG Friedmann equations - then leads to a new a  scale parameter, $\ln^{-1}[\beta(t)]$, where $\beta(t)$ represents a regularization function. BGC thus provides a possible solution to the issue of a primordial singularity, by depicting the initial phases of the Universe as a set of Riemannian foliations. In other words, in the BCG formulation, the multiverse singularities merge, generating a smooth branching topological structure characterized by continuously connected manifolds.

It should be noted that the introduction of a regularization function  $\beta(t)$ in this formulation must by no means be confused with a change in the integration limits of Friedmann's equations to avoid singularities since essential or real singularities at $t=0$ cannot be removed simply by any coordinate transformation. The technical procedure adopted here results in solutions conformed by branch-cuts that allow to circumvent the singularities, which in turn become branch points. This procedure allows a formal treatment consistent with the Planck scales, which, according to the multiverse concept, mark the region where quantum mechanics and General Relativity converge, scaled by $\beta(t)$.
The line element in the branch-cut gravity is defined as~\cite{Zen2020, Zen2021a, Zen2021b}
\begin{equation}
ds_{[\rm{ac}]}^2 = - \sigma^2 N^2(t) c^2dt^2 + \sigma^2 \bigl( \ln^{-1}[\beta(t) ]\bigr)^2 \Biggl[
\frac{dr^2}{\bigl(1 - kr^2(t) \bigr)}
    + r^2(t) \Bigl(d \theta^2 + \sin^2 \theta d\phi^2 \Bigr) \Biggr] \, . \label{FLRWac2}
\end{equation}
In this expression, $[\rm{ac}]$ denotes analytical continuation to the complex plane, the variables $r$ and $t$ represent real and complex space-time parameters, respectively, and $k$ denotes the spatial curvature of the multiverse, corresponding to negative curvature ($k = -1$), flat ($ k = 0$), or positively curved spatial hypersurfaces ($k = 1$). The quantity $N(t)$ represents the lapse function, 
with $\sigma^2 = 2/3\pi$ denoting a normalization factor. 
The lapse function $N(t)$ is not dynamical but rather a pure gauge variable. Gauge
invariance of the action in General Relativity yields a Hamiltonian constraint that
requires a gauge-fixing condition on the lapse (see~\cite{Feinberg}).
The mapping between the scale parameter of the standard formulation, denoted as $a(t)$, and the corresponding scale factor in BCG, $\ln^{-1}(\beta(t))$, 
 $a(t) \mapsto \ln^{-1}(\beta(t))$  does not correspond to a simple direct scale-factor parametrization based on the real nFriedman-Lema\^{i}tre-Robertson-Walker (FLRW) single-pole metric,  nor does it
 constitute a direct generalization of Friedmann's equations~\cite{Friedman1922}. Due to the non-linearity of Einstein's field equations, such a direct generalization would formally not be consistent. 
  The branch-cut formulation, on the other hand, arises from
 the complexification of the 
 FLRW metric~\cite{Friedman1922,Lemaitre1927,Robertson1935,Walker1937}, resulting in a sum of field equations associated to continuously distributed  single-poles  arranged along a line in the complex plane, each with infinitesimal residues (for details, see refs.~\cite{Zen2020, Zen2021a, Zen2021b, Bodmann, Pacheco, Zen2022a, Hess, Bodmann2023a,Bodmann2023b}). Through
 a Riemann integration process, this complexification gives rise to the new
 scale factor, denoted as $\ln^{-1}(\beta(t))$, and a topological
foliated spacetime structure. 
 Branch-cut gravitation provides a gateway to the evolutionary phase 
 preceding the primordial singularity, often referred to as a {\it mirror world}.
 In both its classical and quantum versions, this concept  replaces the singularity
 with a topological transition between the contraction and expansion phases of the 
 evolutionary universe. During
this transition, spacetime takes on a helix-like topological shape, as
proposed by branch-cut cosmology around a branch-point. Importantly,
this process preserves the fundamental conservation laws of
thermodynamics. 

In the classical scenario of the branch-cut cosmology, the universe evolves continuously from the negative complex cosmological time sector, prior to a primordial singularity, to the positive one, circumventing continuously a branch cut, and no primordial singularity occurs in the imaginary sector, only branch points. The branch-cut universe involves a continuous sum of an infinite number of infinitesimally (originally) separated poles, surrounding a primordial branch point, arranged along a line in the complex plane with infinitesimal residues.  And similarly to the primordial branch point singularity, the argument of the resulting analytic function, can be mapped from a single point in the domain to multiple points in the range. 

In short, in addition to branch cuts, there are `singularities', --- the branch points ---, but at the same time there are multiple points that configure continuous paths in the Riemann sheets. This enables continuous solutions of the primordial singularity, which, in general relativity, is inescapable. For this to happen, the presumption at the level of a local continuity prevails, i.e.,  that there is some neighborhood of the branch point, let's call it $z_0$, close enough although not equal to $z_0$, where one can find a small region around (local patches) where $\ln^{-1}[\beta(t)]$ is single valued and continuous. The cuts in the branch cut are shaped by the $\beta(t)$ function which defines as stressed before the range of  $\ln^{-1}[\beta(t)]$. What is the largest region possible for this range? This is an interesting question. At the moment we do not have a definitive answer. However, it is crucial to model the generation of the structures observed today, as for instance, heavy black hole seeds in the early Universe, via primordial fluctuations. 

In the primordial phase, the scale factor $\ln^{-1}[\beta(t)]$ shrinks to a finite critical size, which is shaped by the $\beta(t)$ function which characterizes its range, foliation regularization, and domain extension. 
Its range in particular extends beyond
the Planck length as per the Bekenstein  criterion~\citep{Pacheco}. In view of this criterion, the impossibility of  confining energy and entropy within a finite size makes the transition between the contraction and expansion phases exceptionally peculiar, in which spacetime shapes itself topologically around a branch-point. In the contraction phase, as the patch size reduces linearly with  $\ln[\beta(t)]$, light travels along geodesics within each leaf of the Riemann foliation, continually  circumventing the branch-cut. Although the horizon size increases with $\ln^ {\epsilon}[\beta(t)]/\ln[\beta(t)]$, where $\epsilon$ denotes the dimensionless thermodynamic connection, the length of the path that  light must traverse compensates for the difference in scale between the patch and the horizon sizes. Under these conditions, causality between the size of the horizon and the size of the patch can be achieved through the accumulation of branches in the transition region between the current state of the universe and 
past events~\citep{Pacheco}. In addition to causality, cosmological  dilemmas such
as the flatness problem and the horizon problem come into focus.
The flatness problem pertains to the value of the ratio between the total density of the universe and the critical density, resulting in a very small Planck value for the dimensionless and time-dependent cosmic spatial factor $\Omega_c$~\citep{Ijjas2014,Ijjas2018,Ijjas2019}. This factor scales as $\ln^{2\epsilon}[\beta(t)]/\ln^2[\beta(t)]$. The horizon problem, 
on the other hand,
arises precisely because the patch corresponding to the observable universe has never been causally connected in the past~\citep{Ijjas2014,Ijjas2018,Ijjas2019}. The restoration of causality in BCG provides 
an additional reliable perspective on the resolution of these 
cosmological puzzles~\cite{Zen2022a}.

\section{Non-commutative Branch-cut Quantum Gravity}

The starting point of this study is a recently developed formulation based on the Wheeler-DeWitt-Ho\v{r}ava-Lifshitz quantum gravity~\cite{Bodmann2023a,Bodmann2023b}. Unlike General Relativity, Ho\v{r}ava-Lifshitz theory
is a renormalizable theory of gravity that 
retains Lorentz invariance at low energy while breaking 
this symmetry at high energy, which is a consequence of the implicit presence of a minimal 
length~\cite{Horava,Bertolami2011}. 
The Ho\v{r}ava-Lifshitz approach incorporates
high-order curvature contributions into the Lagrangian density,
preserving the diffeomorphism property of General Relativity~\cite{Kiefer}, an isomorphism of smooth manifolds, as well as the usual foliation of the Arnowitt-Deser-Misner (ADM) formalism at the Infrared Limit~\cite{Compean}. 
The Ho\v{r}ava-Lifshitz formulation, in combination with the Wheeler-DeWitt equation (WdW)~\citep{DeWitt1967}, is ghost-free and thus suitable for describing quantum effects of the gravitational field~\cite{Compean}. The solutions of the WdW equation, represented by a geometric functional of compact manifolds and matter fields, describe the evolution of the quantum wave function of the Universe~\citep{HH,Hawking1982}.
An intriguing aspect 
of the WdW equation is the absence of a time
variable,
although it is linked to a classical image of space-time~\citep{Kiefer}.
This feature is characteristic of the classical Hamilton-Jacobi formulation of General Relativity, where the observable Universe does not exhibit time-reversal symmetry, giving rise to a quantum description of events involving cosmic constituents~\citep{Lukasz,Rovelli2004,Rovelli2011}. The dependence of the quantum wave function of the Universe on a ``3-geometry'', corresponds to the equivalence class of metrics under a diffeomorphism, 
independent of the specific coordinates of the metric tensor~\citep{Rovelli2004,Rovelli2011,Rovelli2019}.
As for the quantum cosmological interpretation of the wave function of the Universe, the prevailing view involves a functional restricted to a superspace configuration, encompassing three surfaces and matter fields denoted generically by $\Phi$. In this structure, the metric is represented by $h_{ij}$. The associated WdW wave function, denoted as $\Psi(h_{ij}, \Phi)$, can be understood as a description of the evolution of $\Psi(\Phi)$ with respect to the physical variable $\Phi$.

Using this formulation, we investigate the effects of non-commutativity in a mini-superspace of variables obeying Poisson's algebra on the accelerating behavior of the wave function of the Universe~\cite{Bodmann2023a}, on the dynamical equations involving the branch-cut scale factor and on relic gravitational waves. 
The  Ho\v{r}ava-Lifshitz action, ${\cal S}_{HL}$, is given by~\cite{Horava,Bertolami2011,Compean,Cordero2019,Hess,Bodmann2023a,Bodmann2023b,Bezerra,Abreu}:
\begin{eqnarray}
 {\cal S}_{HL} & = & \frac{M_P^2}{2} \int d^3x \, dt \, N  \sqrt{g}  \Biggl(K_{ij}K^{ij}  -  \lambda K^2  - g_0 M^2_P  
  -  g_1 {\cal R} \nonumber \\ && -   g_ 2 M^{-2}_P {\cal R}^2  -  g_3 M_P^{-2}  {\cal R}_{ij} {\cal R}^{ij}   
   -    g_4 M^{-4}_P {\cal R}^3   -  g_5 M_P^{-4}  {\cal R}({\cal R}^i_j {\cal R}^j_i) \nonumber \\ &&  -  g_6 M_P^{-4} {\cal R}^i_j {\cal R}^j_k {\cal R}^k_i
  -    g_7 M^{-4}_P {\cal R} \nabla^2  {\cal R}  \! -   g_8 M^{-4}_P \nabla_i {\cal R}_{jk} 
  \nabla^i R^{jk}  \Biggr)\, .
 \label{HL} 
\end{eqnarray}
In the BCG formulation, ${\cal S}_{HL}$ depends on the branching scalar curvature of the Universe, ${\cal R}$ and on its derivatives, in different orders~\cite{Horava,Bertolami2011,Compean,Cordero2019,Hess,Bodmann2023a,Bodmann2023b,Bezerra,Abreu}. In expression (\ref{HL}),
$g_i$ denote running coupling constants, $M_P$ is the Planck mass,  $\nabla_i$ represents covariant derivatives, and
the branching Ricci components of the three dimensional metrics 
may be determined 
by imposing a maximum symmetric surface foliation~\citep{Hess} which gives:
\begin{equation}
    {\cal R}_{ij} = \frac{2}{\sigma^2 u^2(t)} g_{ij}\, , \quad  \mbox{and} \quad {\cal R} = \frac{6}{\sigma^2 u^2(t)} \, ,
\end{equation}
where the variable change $u(t) \equiv \ln^{-1}[\beta(t)]$, with $du \equiv d\ln^{-1}[\beta(t)]$, was introduced. 
$K = K^{ij} g_{ij}$ represents in expression (\ref{HL}) the trace of the extrinsic curvature tensor $K_{ij}$~\citep{Hess,Bodmann2023a,Bodmann2023b}:
  \begin{equation}
K = K^{ij} g_{ij} =  - \frac{3}{2\sigma N u(t)}  \frac{du(t)}{dt}.
\end{equation}

 Applying standard canonical quantization procedures and thus promoting the 
 canonical conjugate momentum into an operator, 
 i.e. $p_u \mapsto -i \frac{\partial}{\partial u}$, 
 the Hamiltonian is also elevated to an operator. 
 The canonical quantization Dirac procedure applied to the Einstein-Hilbert action results in a second-order functional differential equation 
 defined in general terms in a configuration superspace, whose solutions depend on a three-dimensional metric and on matter fields~\cite{DeWitt1967,HH,Hawking1982,Lukasz}. Among the different quantization methods,  the 
 canonical quantization procedure allows the preservation of the original formal structure of a classical theory, as well as its symmetries and corresponding underlying conservation laws.
 
 The Hamiltonian, the new complex dynamical variable $u(t)$, representing the helix-like scale factor analytically continued to the complex plane, along with the 
 corresponding conjugate momentum $p_{\rm{u}}$, are then treated as operators, denoted respectively as $\hat{\cal H}(t)$, $\hat{u}(t)$, and $\hat{p}_{\rm{u}}(t)$.  To simplify notation, the operators $\hat{\cal H}(t)$, $\hat{u}(t)$, and $\hat{p}_{\rm{u}}(t)$, as well as most equations involving the time-dependent variable $u(t)$, do not typically use the hat symbol and neither explicit time-dependence indication, only in very special cases. This leads to the formulation of the branching Hamiltonian given by~\cite{Bodmann2023a,Bodmann2023b} (see also \cite{Bertolami2011})
\begin{equation}
 {\cal H}  =  \frac{1}{2} \frac{N}{u} \Bigl[- p^2_{u} + g_r  - g_k u^2 +     
 g_{\Lambda} u^4 
 + \frac{g_s}{u^2}  \Bigr],  \label{HH}
\end{equation} 
with $p_{\rm{u}} =  - \frac{u(t)}{N}\frac{d u(t)}{dt}$.
In this expression, $p_{\rm{u}}$ represents the conjugate momentum of the original branching gravitation dynamical variable $\ln^{-1}[\beta(t)]$, 
$g_k$, $ g_{\Lambda}$, $g_r$, and $g_s$  represent respectively the curvature, cosmological constant, radiation, and stiff matter running coupling constants~\citep{Bertolami2011,Maeda}
\begin{eqnarray}  g_k  & \equiv &  \frac{2}{3 \lambda -1}; 
  g_{\Lambda}  \equiv  \frac{\Lambda M^{-2}_{PI}}{18 \pi^2 \bigl(3\lambda - 1 \bigr)^2};   g_r  \equiv 24\pi^2\bigl(3g_2 + g_3\bigr); \nonumber \\ 
 g_s  & \equiv  & 288 \pi^4\bigl(3\lambda -1\bigr)(9g_4 + 3g_5 + g_6).
\end{eqnarray}
The $g_r$, and $g_s$ running coupling constants can be positive or negative, without affecting the stability of the solutions. 
Stiff matter contribution in turn is determined by the $p~=~\omega \rho$ condition in the corresponding equation of state.  In ref.~\cite{Bodmann2023a,Bodmann2023b}, we supplemented the Hamiltonian with two additional terms, $g_m u$, that describes the contribution of baryon matter combined with dark matter,  and $g_q u^3$, a quintessence contribution, a time-varying, spatially-inhomogeneous and negative pressure component of the cosmic fluid~\cite{Caldwell,Zlatev}, which allows approaching the ``coincidence problem'':
\begin{equation}
 {\cal H} =  \frac{1}{2} \frac{N}{u} \Bigl[- p^2_{u} + g_r - g_m u - g_k u^2 - g_q u^3 +     
 g_{\Lambda} u^4
 + \frac{g_s}{u^2}  \Bigr] \, .   \label{H*}
\end{equation} 
The coincidence problem refers to the initial conditions necessary to produce the quasi-coincidence of the densities of matter and quintessence in the current stage of the Universe~\cite{Caldwell,Zlatev}.
The parametrizations of 
curvature, cosmological constant, radiation, stiff matter, baryon matter combined with dark matter, and   quintessence running coupling constants
are in tune with the Wilkinson Microwave Anisotropy Probe (WMAP) observations~\cite{Hinshaw,Rogerio}.

The next steps for building a formalism based on a non-commutative algebra, follows the paths shown in~\citep{Abreu}: (a) The insertion, in the Ho\v{r}ava-Lifshitz formalism, of the action of a perfect fluid, characterized by a dimensionless number $\omega$, associated with the variable $v(t)$, a quantum variable that span, with $u(t)$, dual reciprocal spaces, and
whose canonically conjugated momentum is represented by $p_v$. (b) The former commutative variables 
$\{{u},{p}_u,{v},{p}_v\}$ satisfy now a non-commutative algebra, defined as:
\begin{eqnarray}
\{{u},{v}\} & = & \sigma \, ; \quad \{{p}_u,{p}_v\} = \alpha \, ; \quad \{{u},{p}_v\} = \gamma \, ; \nonumber \\ 
\quad \{{v},{p}_u\} & = & \chi \, ; \quad \{{u},{p}_u\} = \{{v},{p}_v\} = 1 \, , \label{NCA}
\end{eqnarray}
where ${p}_u$ and ${p}_v$ represent the canonically conjugated momenta associated to ${u}$ and ${v}$.
The final step in building the formalism is (c) to carry out a linear transformation of the original non-commutative phase space configuration
into a commutative representation; this transformation allows the incorporation of the non-commutative algebra, through the insertion of the new set of variables 
$\{\tilde{u},\tilde{p}_u,\tilde{v}, \tilde { p}_v\}$
and the consequent modification of the quantum gravity phase space, into the intrinsic structure of the cosmic quantum dynamics. In this new set,
the momentum variables can be substituted by derivatives in terms of the coordinates, as done in the standard
quantization procedure.

We then introduce 
a
mapping which relates the commutative $\{\tilde{u},\tilde{p}_u,\tilde{v},\tilde{p}_v\}$ and the non-commutative phase space set of variables $\{{u},{p}_u,{v},{p}_v\}$~\cite{Bodmann2023b} (for the details, see Appendix: 
$\tilde{p}_u = (p_u -\chi p_v);  
   ~\tilde{p}_v =(-\gamma p_u + \alpha u -  \alpha v + p_v)$).  
 By inverting these equation, using (\ref{H*}) and
 the contribution of the perfect fluid, mentioned above
 (for details, please consult \citep{Bodmann2023b}),
 we arrive at
the following expression for the super-Hamiltonian results~\cite{Bodmann2023b}
\begin{eqnarray}
 {\cal H} 
      & = & \frac{1}{2}\frac{N}{u} \Biggl[- \Bigl(\tilde{p}_u  - \chi \tilde{p}_v \Bigr)^2
-  \frac{1}{u^{3 \alpha-1}} \Bigl(\gamma \tilde{p}_u - \alpha u + \alpha v - \tilde{p}_v \Bigr)  \nonumber \\
&& +  \Biggl( g_r - g_m u -  g_k u^2 - g_q u^3 +     
 g_{\Lambda} u^4 
    + \frac{g_s}{u^2} \Biggr) \Biggr] \, . 
    \label{HTinverted} 
\end{eqnarray}

Canonical quantization procedures applied to the Hamiltonian (\ref{HTinverted}), allow
 the variables $u(t)$ and $v(t)$ along with their corresponding conjugate momenta $p_{\rm{u}}$ and $p_{\rm{v}}$, to be treated as operators.
 Making the replacements $\tilde{p}_u \to -i \frac{\tilde{\partial}}{\partial u}$ and $\tilde{p}_u \to -i \frac{\tilde{\partial}}{\partial v}$, we obtain
the following differential equation to describe the evolution of the wave function of the Universe in the non-commutative approach~\cite{Bodmann2023b}, giving the  condition
${\cal H}\Psi(u,v)  =  0$:
\begin{eqnarray}
{\cal H}\Psi(u,v) & = & \frac{1}{2}\frac{N}{u}\Biggl[ \Biggl(\frac{\tilde{\partial}^2}{\partial u^2}  -   2 \chi \frac{\tilde{\partial}}{\partial u} \frac{\tilde{\partial}}{\partial v}  
+ \chi^2  \frac{\tilde{\partial}^2}{\partial v^2} \Biggr)
+  \frac{1}{u^{3 \alpha-1}} \Biggl(i \gamma \frac{\tilde{\partial}}{\partial u} - i \frac{\tilde{\partial}}{\partial v} + \alpha u - \alpha v  \Biggr) \nonumber \\
&&
+   \Biggl( g_r - g_m u -  g_k u^2 - g_q u^3 +     
 g_{\Lambda} u^4 
    + \frac{g_s}{u^2} \Biggr) \Biggr] \Psi(u,v)  = 0 \, . \label{main}
\end{eqnarray}
In ref.~\cite{Bodmann2023b}, the parameters $\chi$ and $\gamma$ are treated as complex numbers. Moreover,  to make contact with conventional formulations, in special regarding the insertion of a set of factor-ordering parametrization to overcome ambiguities in the time-ordering of quantum operators~\cite{Steigl,Page,Bezerra,Bodmann2023a,Bodmann2023b},
and maintain the complex nature of the variables $u$ and $v$,  only the real component of the parameter $\alpha$ was taken into account. In addition, the following simplified notation was adopted: $\gamma~=~i|\gamma|  \mapsto i\gamma$, so $i\gamma = i^2 |\gamma| 
= i \gamma = - |\gamma|$, so the previous equation becomes:
\begin{eqnarray}
{\cal H}\Psi(u,v) & = & \Biggl[ \Biggl(\frac{\tilde{\partial}^2}{\partial u^2}   -   2 \chi \frac{\tilde{\partial}}{\partial u} \frac{\tilde{\partial}}{\partial v}  
+ \chi^2  \frac{\tilde{\partial}^2}{\partial v^2} \Biggr)
-  \frac{1}{u^{3 \alpha-1}} \Biggl( |\gamma| \frac{\tilde{\partial}}{\partial u} + i \frac{\tilde{\partial}}{\partial v} - \alpha u + \alpha v  \Biggr) \nonumber \\ &&
   +    \Biggl( g_r - g_m u -  g_k u^2 - g_q u^3 +     
 g_{\Lambda} u^4 
    + \frac{g_s}{u^2} \Biggr) \Biggr] \Psi(u,v)  = 0 \, . \label{mainNew}
\end{eqnarray}
For simplicity, in the following we do not use the symbol {\it tilde} in the partial derivatives as identification of the new variables in the scope of non-commutative algebra.
Equation~(\ref{mainNew}) can be rewritten in the general form:
\begin{equation}
\Biggl\{\Biggl[\! a(u,v) \frac{\partial^2 }{\partial u^2} - 2b(u,v)  \frac{\partial^2 }{\partial u \partial v} 
+ c(u,v) \frac{\partial^2 }{\partial v^2} \! \Biggr]
-  \Biggl[\! d(u,v) \frac{\partial}{\partial u} + e(u,v) \frac{\partial}{\partial v} + F\bigl(u,v \bigr) \! \Biggr]\! \Biggr\} \Psi(u,v) = 0 \label{parabolic}
\end{equation}
with 
\begin{eqnarray} 
a(u,v) & = & 1 \, ; \quad  
b(u,v)  =  \chi \, ; \quad 
c(u,v) = \chi^2 \, ; \quad 
d(u,v) = \frac{|\gamma|}{u^{3 \alpha-1}} \, ; \quad 
e(u,v)  =  \frac{i}{u^{3 \alpha-1}} \, ; \nonumber \\
 F\bigl(u,v \bigr) & = &   -\Bigl(g_r - g_m u - g_k u^2 - g_q u^3  + g_{\Lambda} u^4  + \frac{g_ s}{u^2} - \frac{\alpha}{u^{3\alpha-2}} + \frac{\alpha v}{u^{3\alpha-1}}\Bigr) \, ,
\end{eqnarray}
where $a(u,v)$, $b(u,v)$, and $c(u,v)$
represent functions of the independent variables $u$ and $v$, and have continuous derivatives up to the second-order.
Since the 
$b^2(u,v) - a(u,v)c(u,v) = 0$ expression
 (\ref{parabolic}) belongs to the mathematical group of parabolic equations. In order to
  reduce this equation to a canonical form, one should first write out the characteristic
equation~\cite{Polyanin}
 \begin{equation}
 a~du^2  - 2 b~du~dv + c~dv^2 = 0 \, , 
 \end{equation}
 which splits into two equations
 \begin{equation}
 a~d\eta - \bigl(b \pm \sqrt{b^2 - ac} \bigr) d\xi = 0 \, .
 \end{equation}
And then, one should find their general integrals. In the case of a parabolic equation, the two previous solutions coincide, resulting in a common general integral $\varphi(u,v) = {\cal I}_G.$ This allows the variables $u$ and $v$ to be changed to new independent variables $\xi$ and $\eta$, in accordance with
 \begin{equation}
 \xi = \varphi(u,v) , \quad \mbox{and} \quad \eta = \eta(u,v) \, , 
 \end{equation} 
 where $\eta(u,v)$ is a differentiable function that satisfies the non-degeneracy condition of the Jacobian $D(\xi,\eta)/D(u,v)$ in the given domain. As a result, Equation~(\ref{parabolic}) is reduced to the canonical form
\begin{eqnarray}
\frac{\partial^2\Psi(\xi,\eta)}{\partial \eta^2}  =  G \Biggl(\xi, \eta,\Psi, \frac{\partial}{\partial \xi}, \frac{\partial}{\partial \eta}  \Biggr) \, .
\end{eqnarray}
For $\eta$ one can take $u$ or $v$. We take, for convenience, $u$.

 In the Faddeev--Jackiw formalism, the variables $u$ and $v$ are non-commutative, and after the variable transformation,
$\xi$ and $\eta$ are also non-commutative. In this sense, $\xi$ and $\eta$ are canonically conjugate dual variables, which span reciprocal spaces, so the following relation between these variables holds:
\begin{equation}
\xi = \frac{1}{\sqrt{2\pi}} \int_{-\infty}^{\infty} A(\eta) e^{i \xi \eta} d \eta \, .
\end{equation}

In order to reduce this equation to a canonical form, on basis of its characteristic
equation~\cite{Bodmann2023b,Polyanin}, a change of the variables $u$ and $v$ in terms of new independent variables $\xi$ and $\eta$ was introduced, in accordance with
 \begin{equation}
 \xi = \varphi(u,v) , \quad \mbox{and} \quad \eta = \eta(u,v) \, , 
 \end{equation} 
 where $\eta(u,v)$ is a differentiable function that satisfies the non-degeneracy condition of the Jacobian $D(\xi,\eta)/D(x,y)$ in the given domain and $\varphi(u,v)$ represents the general integral. 
 As a result, equation (\ref{mainNew}) is reduced to the canonical form
\begin{eqnarray}
{\cal H}(\xi,\eta)\Psi(\xi,\eta) 
 & = &  \frac{1}{2} \frac{N}{\eta} \Biggl[\frac{\partial^2}{\partial \eta^2} 
  + \frac{\gamma}{\eta^{3 \alpha-1}} \frac{\partial}{\partial \eta}  +
 g_r - g_m \eta  -  g_k \eta^2 - g_q \eta^3 + g_{\Lambda} \eta^4 \nonumber \\ && + \frac{g_s}{\eta^2}  +    \frac{\alpha}{\eta^{3\alpha-2}}   - \frac{\alpha \xi}{\eta^{3\alpha-1}} - \frac{i}{\eta^{3\alpha-1}}\frac{\partial}{\partial \xi} \Biggr]\Psi(\xi,\eta) = 0 \, ,\label{Hsupersuper*}
\end{eqnarray}
which may be cast in the form
\begin{eqnarray}
{\cal H}(\xi,\eta)\Psi(\xi,\eta)  
 & = & \frac{1}{2}\frac{N}{\eta} \Biggl[-p^2_{\eta,\gamma,\alpha} +
 g_r - g_m \eta   -   g_k \eta^2 - g_q \eta^3  + g_{\Lambda} \eta^4 \nonumber \\ && + \frac{g_s}{\eta^2}    +    \frac{\alpha}{\eta^{3\alpha-2}}   - \frac{\alpha \xi}{\eta^{3\alpha-1}} + \frac{1}{\eta^{3\alpha-1}} p_{\xi} \Biggr]\Psi(\xi,\eta) = 0 \, .\label{Hsupersuper}
\end{eqnarray}
In this expression, $-p^2_{\eta,\gamma,\alpha}$ is defined as
\begin{eqnarray}
-p^2_{\eta,\gamma,\alpha} & \equiv & \frac{\partial^2}{\partial \eta^2} 
  + \frac{\gamma}{\eta^{3 \alpha-1}} \frac{\partial}{\partial \eta} \\ 
  & = & - \Bigl(-i \frac{\partial}{\partial \eta}\Bigr) \Bigl(-i \frac{\partial}{\partial \eta}\Bigr) + \frac{i |\gamma|}{\eta^{3 \alpha-1}} \frac{\partial}{\partial \eta}
  \equiv - p^2_{\eta} - p_{\eta,\gamma,\alpha} \, , \label{p2} \nonumber 
\end{eqnarray}
with
\begin{equation}
    p_{\eta,\gamma,\alpha} \equiv - \frac{i |\gamma|}{\eta^{3 \alpha-1}} \frac{\partial}{\partial \eta}  \mapsto \frac{\gamma}{\eta^{3 \alpha-1}}
   \Bigl(-i\frac{\partial}{\partial \eta}\Big) =  \frac{\gamma}{\eta^{3 \alpha-1}} p_{\eta} \, . 
\end{equation}

With the particular choice $\alpha = 1/3$, that allows a separation of variables, with $\Psi(\xi,\eta) = \Psi(\xi)\Psi(\eta)$, equation (\ref{Hsupersuper}),  reduces to the following equations~\cite{Bodmann2023b}
\begin{equation}
   \Biggl( \frac{\partial^2}{\partial \eta^2}
+ \gamma \frac{\partial}{\partial \eta} + V(\eta)\Biggr) \Psi(\eta) = 0 \, , \label{psieta}
\end{equation}
 where
 \begin{equation}
     V(\eta) =  \tilde{g}_r - \tilde{g}_m \eta - g_k \eta^2 - g_q \eta^3 + g_{\Lambda} \eta^4  + \frac{g_s}{\eta^2} \, , \label{Veta}
 \end{equation}
 with $\tilde{g}_r \equiv g_r - {\cal C}$ and $\tilde{g}_m \equiv g_m - 1/3$,
and
\begin{equation}
 \Biggl(i\frac{\partial}{\partial \xi}  -  \frac{1}{3} \xi + {\cal C}\Biggr)  \Psi(\xi) = 0 \, ; \label{WFU}
 \end{equation}
in these equation, 
${\cal C}$ is a separation constant. 
The solution to the second equation above up to an additional constant is
\begin{equation}
    \Psi(\xi) =  i\Biggl( {\cal C} \xi - \frac{1}{3} \xi^2 \Biggr) \, . 
\end{equation}

\subsection{Probability interpretation of the wave function of the Universe}

Given the Klein-Gordon equation for a free particle with mass $m$
\begin{equation}
\Bigl( \partial_{\mu} \partial^{\mu} + m^2 \Bigr) \Phi(x) = 0 \, , \label{DE}
\end{equation}
the four-current \begin{equation}
j = \frac{i}{2} \Bigl(\Phi^{*} \, \nabla \cdot \Phi -    \Phi \,  \nabla \cdot \Phi^{*} \Bigr),   
\end{equation}
in covariant notation may be written as
\begin{equation}
j^{\mu}(x) = i \Bigl[\Phi^*(x) \partial^{\mu} \Phi(x) - \Bigl( \partial^{\mu} \Phi^*(x )\Bigr) \Phi(x) \Bigr] . \label{Jc}
\end{equation}
Combining (\ref{DE}) and (\ref{Jc}), we obtain $\partial_ {\mu} j^{\mu}(x) =0$, so
the four-current $j^{\mu}(x)$ is a conserved quantity. However, the density
\begin{equation}
\rho(x) = i \Bigl[\Phi^*(x) \frac{\partial \Phi(x)}{\partial t} - \frac{\partial 
\Phi^*(x)} {\partial t} \Phi(x) \Bigr] , \label{rhoc}
\end{equation}
and the corresponding inner product
\begin{equation}
\langle \Phi_1(x) | \Phi_2(x)\rangle = \frac{i}{2} \int_{{\cal R}^3} d^3x \rho(x)_{\Phi_1\Phi_2}  \,, \label{inner}
\end{equation}
with
\begin{equation}
 \rho(x)_{\Phi_1\Phi_2} = \Bigl[\Phi_2(x) \frac{\partial \Phi^*_1(x)}{ \partial t} - \frac{\partial \Phi^*_2(x)}{\partial t} \Phi_1(x) \Bigr]\, , 
\end{equation}
are not positive defined. 
As a result, the Klein-Gordon wave function cannot be interpreted as a probability amplitude.

The Wheeler-DeWitt formulation for quantum gravity consists in constraining a wave-function which applies to the universe as a whole, the so called wave function of the Universe. The significance of the wave-function of the Universe can be ascribed only to the intrinsic dynamics of the universe~\citep{WDW}, or to the probability amplitude for the universe to have some space geometry, or to be found in some point of the Wheeler super-space~\citep{Shestakova}. In accordance with the Dirac recipe: 
\begin{equation}
\hat{\cal H} \Psi = 0 \, ,
\end{equation}
i.e., the Wheeler DeWitt equation corresponds to a stationary, timeless equation, instead of a time-dependent quantum mechanics
wave equation as for instance
\begin{equation}
i \frac{\partial}{\partial t} \Psi = \hat{H} \Psi \, . 
\end{equation}
Here, $\hat{H}$ denotes the Hamiltonian operator of a quantum subsystem,  
while $\hat{\cal H}$ in the previous equation represents a quantum operator which describes a general relativity constraint, resulting in a second-order hyperbolic  equation of gravity variables\footnote{More precisely, the scale factor $a(t)$, the density $\rho(t)$, the pressure $p(t)$, and the gravitation constant $\Lambda$.}. The WdW equations thus constitutes a Klein-Gordon-type equation, having therefore a `natural' conserved  associated current (${\cal J}$),
\begin{equation}
{\cal J} = \frac{i}{2} \Bigl(\Psi^{\dagger} \, \nabla \cdot \Psi -    \Psi \,  \nabla \cdot \Psi^{\dagger} \Bigr); \, \, \, \mbox{with} \, \, \nabla \cdot {\cal J} = 0 \, . \label{J}  
\end{equation}
A similar problem arises in handling the Wheeler-DeWitt equation due to its Klein-Gordon structure, generating divergent interpretations about a probabilistic interpretation of the wave function of the universe.

For Rovelli~\citep{Rovelli2004,Rovelli2011,Rovelli2015,Rovelli2019} the absence of time is a feature of the classical Hamilton-Jacobi formulation of general relativity, and the wave function is only a function of the “3-geometry”, namely the equivalence class of metrics under a diffeomorphism, and not of the specific coordinate dependent form of the metric tensor.
According to the second law of thermodynamics, forward in time represents the direction in which entropy increases and in which we obtain information, so the flow of time would represent a subjective feature of the universe, not an objective part of physical reality~\citep{Rovelli2004,Rovelli2011,Rovelli2015,Rovelli2019}. In this realm, in which the observable universe does not exhibit time-reversal symmetry, events, rather than particles or fields, are the basic constituents of the universe, implying that the evolution of physical quantities is related to the description of the relationship between events~\citep {Rovelli2004,Rovelli2011,Rovelli2015,Rovelli2019}. For instance, given the wave function of the universe as a functional constrained to a region configuration of a super-space that contains a three-surface $\Sigma$ and matter fields configuration, represented by $\phi$, where the metric is described by $h_{ij}$ the corresponding WdW wave function $\Psi(h_{ij},\phi)$ may be interpreted according to~\citep{Rovelli2004,Rovelli2011,Rovelli2015,Rovelli2019}, as stressed before, as describing the evolution of $\Psi(\phi)$ in the physical variable $\phi$. In the following we assume the proposal made by Hawking~\citep{Page} for which $|\Psi[h _{ij},\phi,\Sigma]|^2$ is proportional to the 
probability $P({\cal A})$ of finding a 3-surface $\Sigma$ with metric $h _{ij}$ and matter field configuration $\phi$:
\begin{equation}
P({\cal A}) \propto \int_{\cal A} |\Psi[h _{ij},\phi,\Sigma]|^2 \, ^{*}1  \label{HawkingPrescription}
\end{equation}
where $^*$ is the Hoge dual in the metric $\Gamma(N)$, i.e. $^*1$ is the volume element~\citep{Page}.  $\Gamma(N)$ in turn represents the 
superspace metric which does not depend linearly on $N$. In what follows, we adopt similar interpretations to Hawking's prescription for $|\Psi(\eta)|^2$ and $|\Psi(\xi)|^2$, and to establish a connection with standard quantum mechanics we use the denomination ``probability density".

Briefly, as highlighted by Hartle~\cite{Hartle2021}, in standard quantum mechanics, the probabilities associated with wave functions are represented by squares of amplitudes, and additionally, a criterion is needed to specify which sets of histories can have probabilities consistently assigned to them. And yet, as highlighted by Hartle, in standard quantum mechanics that criterion is measurement – probabilities can be consistently assigned to histories of measured alternatives and usually not otherwise. Hartle then recalled that a criterion based on measurements or observers cannot be fundamental in a quantum theory that seeks to explain the early universe where neither existed. A more general criterion for closed systems assigns probabilities to just those sets of histories for which there is vanishing interference between its individual members as a consequence of the system’s initial quantum state. Such sets of histories are said to decohere. Decoherent sets of histories are what may be used for prediction and retrodiction in quantum cosmology for they may be assigned probabilities~\citep{Hartle2021}. And this type of identification, in our view, with decoherent sets of histories, is in perfect harmony with Hawking's prescription and with the modes of interpretation adopted by us for the dynamic evolutionary processes of the Universe's wave function and the scale factor in branch-cut gravity.

\subsection{The wave function of the Universe in the non-commutative approach}
\begin{figure*}[htpb]
\centering
\includegraphics[scale=0.35]{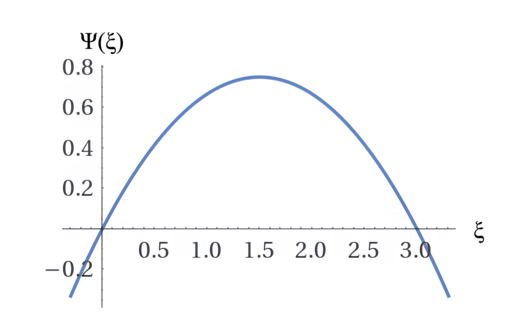} 
\includegraphics[scale=0.35]{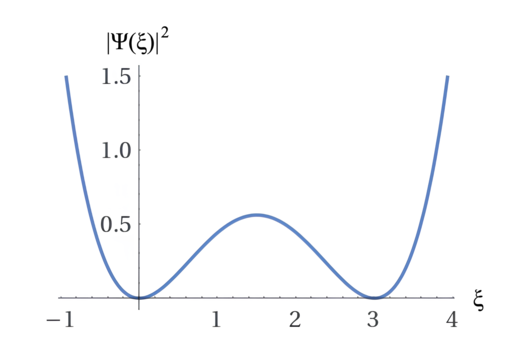}
\caption{On the left side we show the complex plane solution of the wave function $\Psi(\xi)$, which is the dual complementary to the wave function of the Universe $\Psi(\eta)$ when utilizing the commutative approach given by
equation~(\ref{WFU}).
On the right side, we show the probability density of the wave function $\Psi(\xi)$ using Hawking's prescription given by equation (\ref{HawkingPrescription}).} \label{NCWFUxi}
\end{figure*}

In Figure \ref{NCWFUxi}, the left side displays the behavior of the dual solution, $\Psi(\xi)$.
The most astonishing outcome in this formulation pertains to the curve on the right of figure \ref{NCWFUxi}, which illustrates the probability density $|\Psi(\xi)|^2$, as per Hawking's prescription~\citep{Page}. The behavior of this curve has exactly the same shape as the potential conceived to describe the inflationary scenario~\cite{Guth1981,Guth2004}. In this theory inflation is driven by a scalar field perched on a plateau of the potential energy diagram, whose curve is completely similar to the behavior of $|\Psi(\xi)|^2$ shown in Figure \ref{NCWFUxi}. A scalar field of this kind is commonly referred to as the inflaton. If the plateau is flat
enough, such a physical state can be stable for successful inflation~\cite{Guth1981,Guth2004}. The curves on the left side of figure \ref{NCWFUxi}, in turn, bear a striking resemblance to the inverse of the potential required to induce chaotic inflation, which could be represented in the form $V(\phi) = - \frac{1}{2} m \phi^2$, where $\phi$ represents the inflaton field~\cite{Guth1981,Guth2004}. Due to the dual characteristics involving the mutually correlated variables $\eta$ and $\xi$, which spam complementary spaces, intrinsically correlated by the algebraic Poisson's structure, we can infer from these results that the behavior of $\Psi(\xi)$ reveals an underlying dynamics as a realization of the non-commutative algebraic spacetime structure configuration, which leads to a kind of inflation stage of the Universe. 
Hence, it appears to be crucial, in order to capture small scales within a non-commutative framework, that there exists a duality between the variables linked to the Poisson algebra or another similar algebra.

In the following calculations, we seek to expand the epistemological basis, as well as the methodologies adopted and the logical reasoning used systematically with a view to both the formal development of the theory and the expansion of our knowledge on the topics under study. 
 The most relevant global phenomena in quantum mechanics are confined to shorter length scales, and in practice, it is therefore not necessary to explicitly include dynamics on longer length scales, where low energy and curvature phenomena predominate. However, when describing the short-scale properties of the Universe's evolution, in the quantum gravity domain, where high curvatures and high energies predominate, the effects of long-range scales are in general still implicitly contained in the coupling parameters of a field theory and in the expansion constants of the wave function of the Universe. 

In the study of the evolution of the Universe, we come across an epistemic limitation of realism: the under-determination of theory by evidence.
Data under-determination poses a substantial problem for the high-energy frontier of fundamental physics, most notably
within the fields of particle physics and quantum gravity. 
Therefore, an organizing and guiding principle is required to assess the viability of models in a non-empirical manner, enabling consistent and accurate calculations.
As first proposed by Weinberg~\citep{Weinberg1972} the principle of naturalness serves as a conventional method for classifying and organizing the terms within highly intricate approaches. It also provides guidance for understanding the various interaction couplings associated with the dynamic composition of matter, energy, and the primordial sources of gravitational waves, and this principle will be adopted. 
Autonomy of scales (AoS) provides the most cogent definition of naturalness~\citep{Dijkstra}, in which the underlying parameters are all the same size in appropriate units, or more precisely, the idea that a given quantum field theory can only describe nature at energies below a certain scale, or cutoff~\cite{Weinberg1972}. To date, there are no compelling reasons for the generic decoupling of the laws of nature into quasi-autonomous physical domains. 
Decoupling of scales in the quantum realm is often claimed to be entailed by the Decoupling Theorem~\citep{Cao}, despite its claimed weakness to underwrite quasi-autonomous physical domains in quantum field theories due to the demanding imposition of naturalness~\citep{Dijkstra}. 
Consequently, violations of naturalness would carry ontological implications. Unnatural parameters would not be adequately described by effective field theories but would instead be addressed by field theories that exhibit, for instance, a UV/IR interplay, as exemplified in the work by Dijkstra~\citep{Dijkstra}.
The same analysis applies, more generally, involving other separations of scales, which remain valid in commutative quantum field theory, being the paradigmatic basis of renormalization group theory, making it possible in this way to organize physical phenomena according to the scale energy or distance scale. In the particular case of UV and IR radiation, short-distance ultraviolet physics does not directly affect the qualitative characteristics of long-distance infrared physics and vice versa. However, as we will see later, there is evidence that in non-commutative field theory, the mixture of scales appears as an element that inhibits these conclusions in principle. 
Nonetheless, as our starting point, we uphold the application of the naturalness principle to the theory's constants. We do so under the belief that these conclusions do not alter the very nature of the phenomena, which is the fundamental foundation of the concept of naturalness. This principle remains constrained to the dual nature of the fields introduced within a non-commutative algebraic formulation.
In this context, we favor calculations based on appropriate choices for the arbitrary constants that arise from solving the relevant differential equations. When setting values for these constants, we adhere to the principles of naturalness, normalizing them to unity. We have adopted this practice, taking into account the considerable uncertainties associated with the amplitude of these signals, which are generally quite feeble. Our current priority lies in studying their structural characteristics.
 
In what follows, we use the powerful iterative Range-Kutta-Fehlberg numerical analysis for solving differential equations to find solutions for equation (\ref{psieta}), without introducing numerical approximations, in order to describes the evolution of the wave function of the Universe, $\Psi(\eta)$. The boundary conditions of the solutions are based on the Bekenstein criterion, which provides an upper limit for the Universe's entropy, following the proposition presented in~\cite{Bodmann2023a,Bodmann2023b}. 
The total entropy of a black hole, according to the Bekenstein limit, is proportional to the number of Planck areas needed to cover the event horizon, where each area corresponds to one unit of entropy. In non-commutative branched gravitation, we assume that the primordial singularity is equally covered by a certain number of Planck areas, whose numerical value in turn corresponds to the total primordial entropy of the Universe. 
We assume that the dimensions of this boundary region correspond to the farthest points observable while still respecting causality. For this, we consider an
appropriate distance, denoted as $d(t)$, between
a pair of objects at any given moment $t$, and the corresponding
distance, denoted as $d(t_0)$, at a reference time $t_0$. We
establish this relationship as $d(t) = |\eta(t)|d(t_0)$. 
This means
that the relationship between the two distances is modulated by the
scale factor of the branch-cut Universe. This implies that for $t = t_0$
we have $|\eta(t_0)| = 1$.
From a quantum probabilistic point of view, this condition implies a maximum probability of observation, $|\Psi(1)|^2 =1$, assuming a normalized wave function. Thus, the boundary conditions considered in this contribution are, in the contraction sector $\Psi(-1) = -1$, while in the expansion region, $\Psi(1) =1$.

Complex equations similar to Friedmann's equations underlie the branched gravitation scenarios, in which the primordial singularity is replaced by a foliated transition region, described by helix-shaped cosmological factor $\eta(t)$, analytically continued to the complex plane, interposing two distinct evolutionary stages of the Universe, a contraction phase and an expansion phase. The consequences of these scenarios on the behavior of the universe's wave function are notable in that they imply the evolutionary description of $\Psi(\eta)$ in both regions.

\begin{figure*}[htpb]
\centering
\includegraphics[scale=0.125]{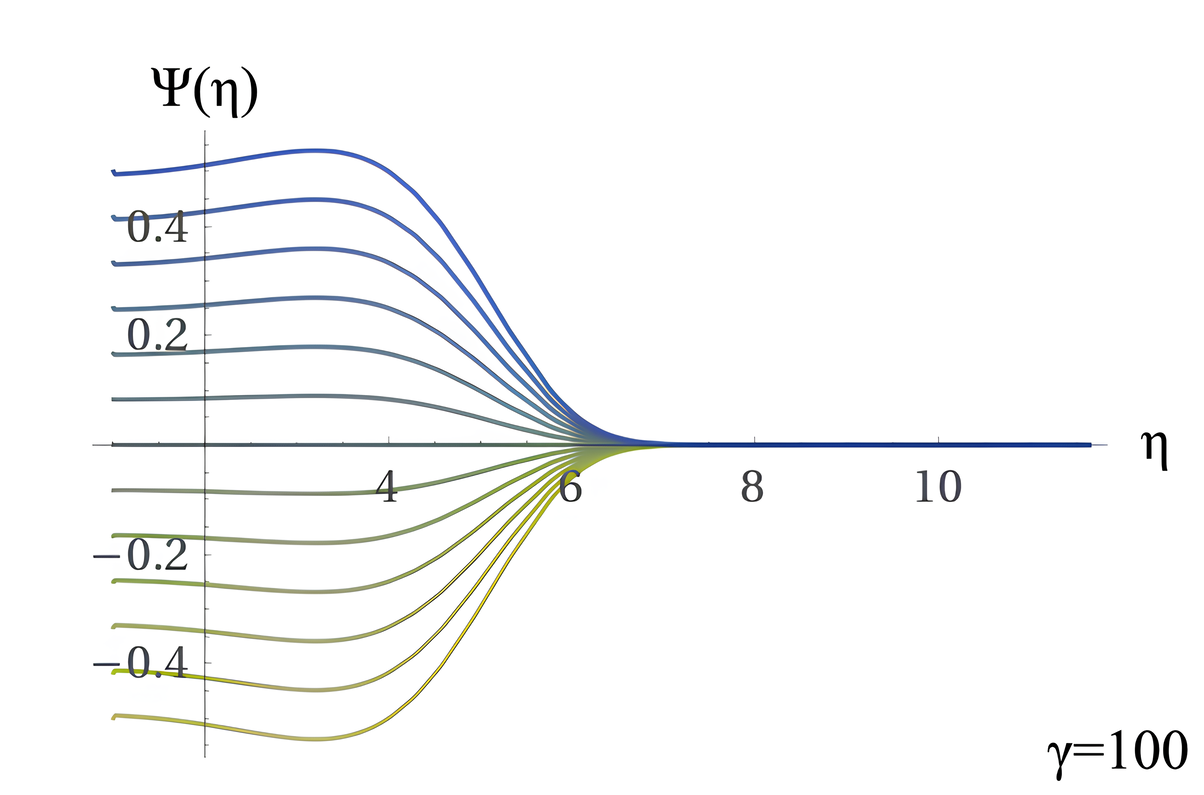} \hspace{-0.15cm}
\includegraphics[scale=0.248]{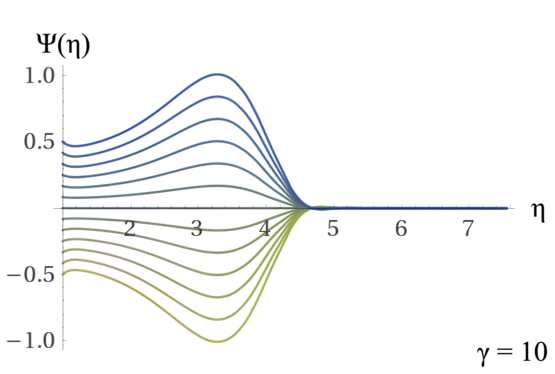} \hspace{-0.25cm}
\includegraphics[scale=0.248]{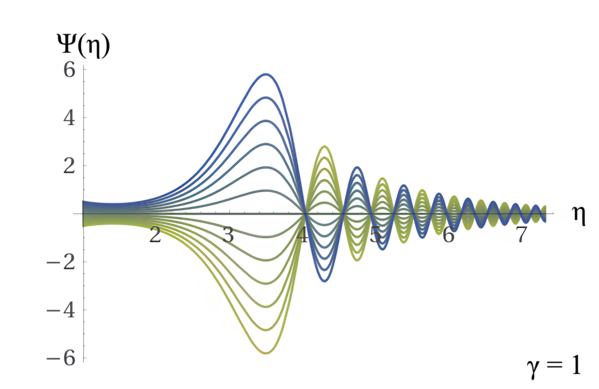} 
\includegraphics[scale=0.248]{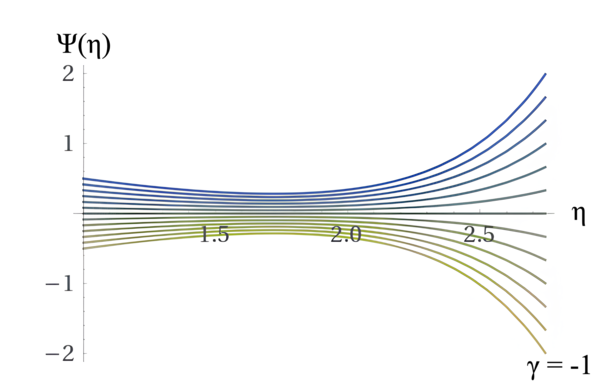} \hspace{-0.25cm}
\includegraphics[scale=0.248]{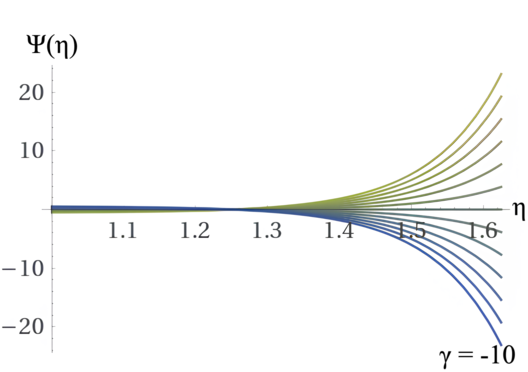} \hspace{-0.25cm}
\includegraphics[scale=0.248]{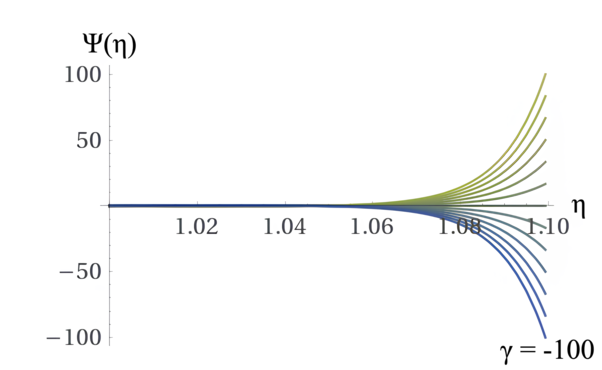}
\caption{Typical sample solutions of the wave function of the Universe $\Psi(\eta)$ utilizing the non-commutative approach given by equation (\ref{psieta}). The solutions are for initial conditions $\Psi(1)$ and $\Psi'(1)$ and different values of the parameter $\gamma$. The parameter $\alpha$ was fixed at $1/3$. The parameter $\chi$ is implicitly embedded in the intrinsic structure of the scale factor $\eta$ as well as in the dual complementary scale factor $\xi$.  The values of the running coupling constants and parameters are: $\tilde{g}_r = 0.4$; $\tilde{g}_m = 0.6185$; $g_k = 1$; $g_q = 0.7$; $g_{\Lambda} = 0.333$; $g_s = 0.03$; $\alpha = 1/3$; ${\cal C} = 1$.} \label{WaveFunction1}
\end{figure*}

\begin{figure*}
    \centering
   \subfigure[]{\includegraphics[width=0.35\textwidth]{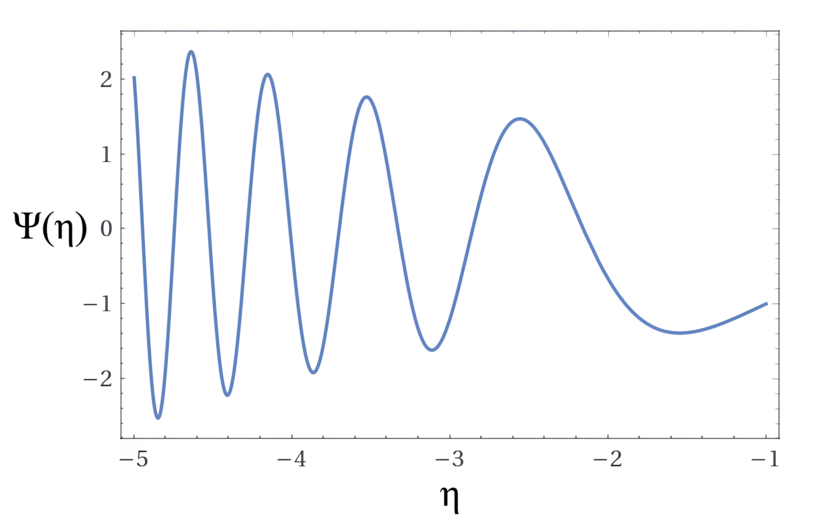}}
   \subfigure[]{\includegraphics[width=0.347\textwidth]{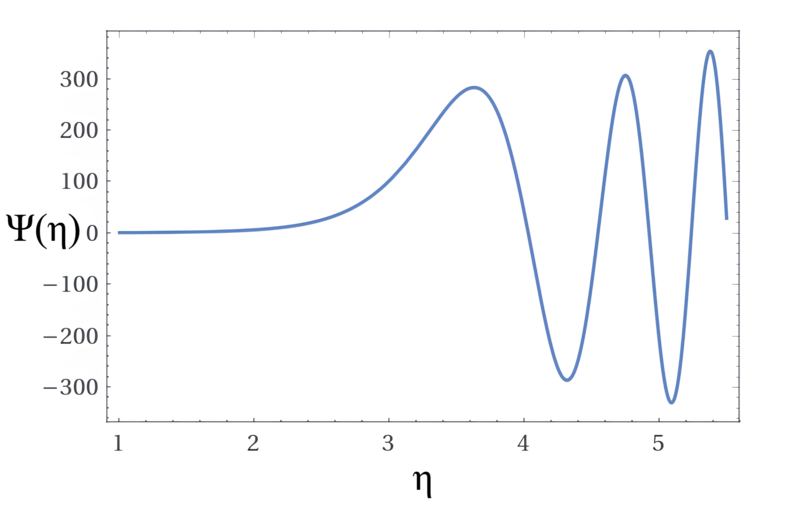}}
  \subfigure[]{\includegraphics[width=0.36\textwidth]{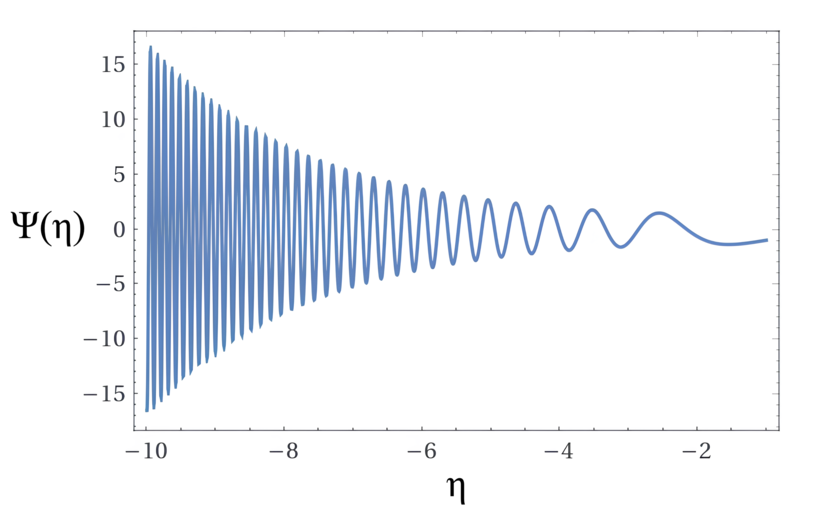}}
    \subfigure[]{\includegraphics[width=0.353\textwidth]{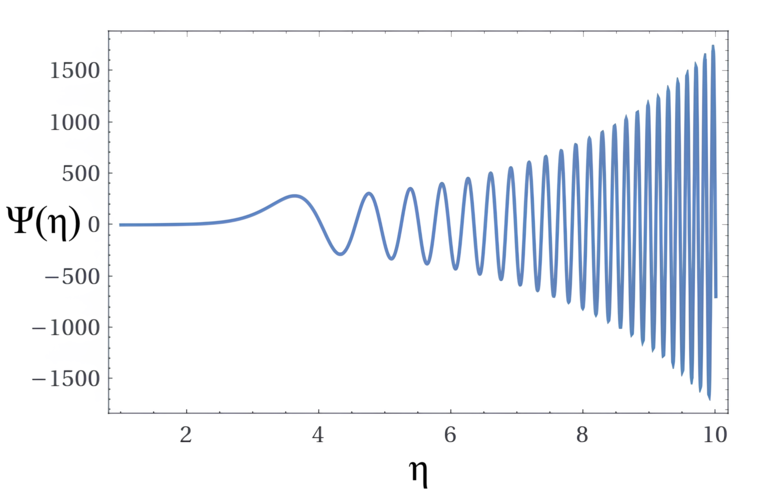}}
     \subfigure[]{\includegraphics[width=0.365\textwidth]{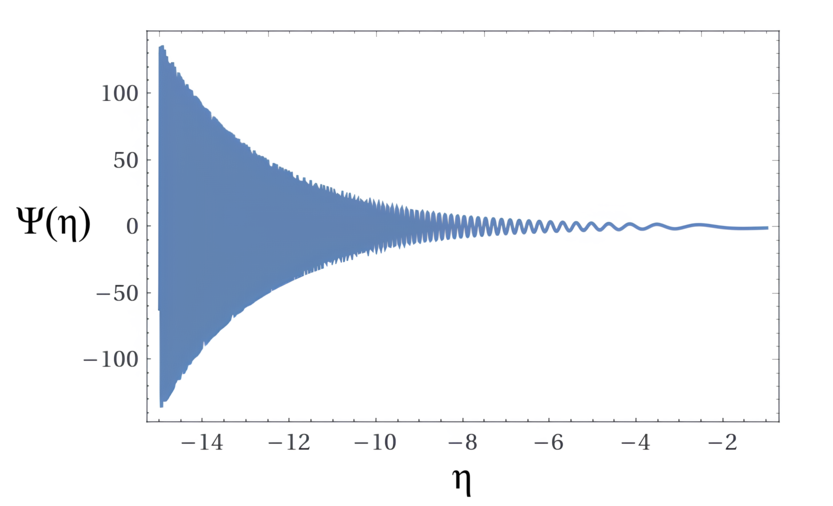}}
    \subfigure[]{\includegraphics[width=0.3499\textwidth]{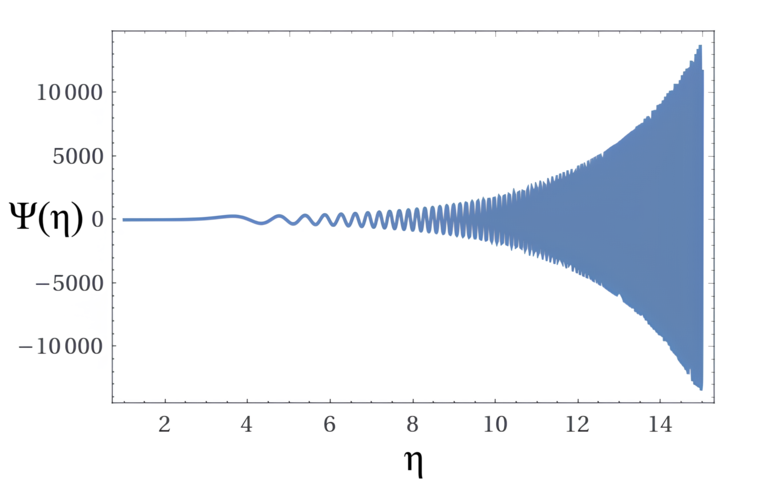}}
    \caption{Typical solutions of the wave function of the Universe $\Psi(\eta)$ using the non-commutative approach given by equation (\ref{psieta}).  The parameter $\alpha$ was fixed at 1/3. The parameter $\chi$ is implicitly embedded in the intrinsic structure of the scale factor $\eta$ as well as in the dual complementary scale factor $\xi$. On the left side, in figures ((a), (c), and (e)) the values of the running coupling constants and parameters are
    as follows: $\tilde{g}_r = 0.4$; $\tilde{g}_m = 0.6185$; $g_k = 1$; $g_q = 0.7$; $g_{\Lambda} = 0.333$; $g_s = -0.03$; $\alpha = 1/3$; $\gamma=1$; ${\cal C} = 1$. The boundary condition is set as $\Psi(-1)=-1$. On the right
    side, in figures ((b), (d), and (f)) , the values of the running coupling constants and parameters are: $\tilde{g}_r = 0.4$; $\tilde{g}_m = 0.6185$; $g_k = 1$; $g_q = 0.7$; $g_{\Lambda} = 0.333$; $g_s = -0.03$; $\alpha = 1/3$; $\gamma=-1$; ${\cal C} = 1$. The boundary condition is set as $\Psi(1)=1$.}
  \label{WaveFunction2}
\end{figure*}

Figures \ref{WaveFunction1} and \ref{WaveFunction2} depict the behavior of the solutions to equation~(\ref{psieta}), which pertains to the wave function of the Universe $\Psi(\eta)$ within a non-commutative approach.
These results align with the predictions of branch-cut gravity ,
suggesting that the present Universe did not originate from nothing (see refs.~\cite{Guth1981,Guth2004,Vilenkin}) or from a quantum loop~\cite{Ellis}. Instead, it appears to have emerges from a prior phase before the current expansion phase. The implications of the non-commutative algebra on the acceleration of the Universe are most pronounced in figure \ref{WaveFunction2}, indicating the presence of a dynamic acceleration of the Universe driven by a force, whose work can be expressed as $W = -p_f dV$, where $p_f $ represents the strength of the pressure in the expanding region.

Several possible explanations for the acceleration of the Universe can be suggested. One explanation, as proposed by Albert Einstein, would be the presence of dark energy as an intrinsic property of spacetime, which would not be diluted as space expands. On the contrary, as more space comes into existence, more of this intrinsic energy of space-time would appear, causing the Universe to expand at an accelerated rate. From the point of view of a non-commutative formulation, the configuration of this intrinsic energy would probably be affected to the extent that the algebraic structure of non-commutative geometry captures the short-scales properties of spacetime.

Another possible explanation for spacetime acquiring energy arises from the quantum fluctuation of matter fields, where empty space (vacuum) is permeated by an indefinite amount of pairs of virtual particles that may exist for a short moment, in accordance with the uncertainty principle, and would subsequently annihilate each other. However, similar to what occurs near a black hole's event horizon, where one particle may fall into the black hole where the other goes into space, so the escaping particle causes the black hole to evaporate (Hawking Radiation), here the annihilation between virtual particles could be avoided  
by the escaping of particles from each pair towards the mirror Universe, while the others would remain in their current cosmic counterpart, so the escaping particles evaporation would cause spacetime acquiring energy.
However, the proposition that virtual particles contributed to the creation of the observable universe as well as to its acceleration demands to search for signatures, possibly similar to Hawking radiation, an open problem yet. 
Another explanation could be based on the contribution of a new kind of dynamical energy fluid or field, that fills all space, whose effect on the expansion of the universe would be the opposite of that of ``normal'' matter and  energy, as for instance the ``quintessence" and ``stiff'' contributions, parameterized by the terms corresponding to the running coupling constants $g_q$ and $g_s$. Our results indicate that the term corresponding to stiff matter has expressive implications on the conformation of the potentials adopted in the present formulation, particularly with regard to the transition regions between the contraction and expansion phases of the branch-cut Universe. 

Finally, when we compare the Universe expansion results corresponding to the commutative and non-commutative formulations, the configuration of matter, represented by the potentials (\ref{Vu}), (\ref{Vut}), (\ref{VetatNew}), and (\ref{vetat}), show expressive differences demonstrating that the non-commutative formulation imprints on the structure of spacetime a non-symmetrical redistribution of matter which captures, as stressed before, the short-distance scales. 

The transition region between the two universes could serve as a source of matter/particles and energy, which drives the acceleration of the Universe.
Speculations regarding the composition of particles in the two Universes, consider in general theories based on the product $G \times G'$ of two identical gauge factors, based on the minimal symmetry $G_{SM} \times G'_{SM}$. Here, $G_{SM} = SU(3) \times SU(2) \times U(1)$ stands for the standard model of observable particles that corresponds to three families of quarks and leptons and the Higgs. $G'_{SM}$ in turn corresponds to the mirror gauge counterpart $G'_{SM} = [SU(3)\times SU(2)\times U(1)]'$ with analogous particle content, more precisely, three families of mirror quarks and leptons and the mirror Higgs. Furthermore, besides the gravity, the two sectors could communicate by means of ordinary photons which could have a kinetic mixing with mirror photons, ordinary active neutrinos could mix with mirror sterile neutrinos, or even the two sectors could have a common flavor gauge symmetry. 

Briefly, the results involving the contraction phase as well as the expansion phase of the branch-cut Universe in the non-commutative domain, indicate an acceleration of the wave function $\Psi(\eta)$ in the expansion phase, in tune with the predictions of the inflation model, as well as a deceleration in the contraction phase, both predictions in tune with the BCG prognosis. 

In a similar line of research, although adopting a commutative formulation, Naoki Sato~\citep{Sato} investigated how the distribution of a statistical set of matter is modified if the particles feel the curvature of space-time resulting from the imposition of the principles of general relativity. Sato demonstrated that in this case ambiguities arise in the definition of the thermodynamic arrow of time in relation to which statistical processes evolve, thus directly affecting the notions of temperature and thermodynamic equilibrium. By assuming that the statistical set is made up of particles that obey the geodesic equations, which define the phase space of the system, Sato discovered that the curvature of space-time gives a heterogeneous structure to the distribution of particles~\cite{Sato}. This heterogeneous distribution of matter, in the case of non-commutative BCG, is particularly expressive, resulting in structural modifications of spacetime that are a relevant indication, although they require further in-depth studies, of the short-scales capture by the non-commutative algebraic structure. These results further indicate that the consequent reconfiguration of matter has dynamic implications for the structure of spacetime in small dimensions that materialize in the acceleration of the primordial Universe.

This heterogeneous distribution of matter and energy and its implications in terms of the expansion dynamics of the Universe can be visualized, as we will see later, particularly in the configuration of the color palette of the effective potentials of the non-commutative formulation of the branch-cut gravity. The intensities of these potentials, which are reflected in the accelerated evolution of the wave function and the scale factor of the Universe, are intrinsically linked to the spectrum of these color palettes. In the 3D colored graphical representations of these potentials, lighter colors identify regions of greater intensity, while darker colors represent regions of lower intensity. The palettes also allow us to visualize the asymmetry of these color combinations, which reflect the heterogeneity of the distribution of matter and energy in the primordial Universe.

\subsection{Dual spaces in quantum mechanics}

The discussion of this topic, quantum dual spaces, despite not being a central point of our study, requires to consider the representation of a physical state in quantum mechanics. Such a state may be represented by an abstract vector, defined in the realm of the complex Hilbert vector space with an inner product. As for its dimensions, we can consider both an infinite space, complete or closed, as well as a finite-dimensional space. Using Dirac notation for the {\it ket} abstract representation of a vector $u$ in quantum mechanics, $\ket{u}$, the inner product of two vectors, $\bra{u}$ and $\ket{v}$, is written as 
$\braket{v|u}$. An inner product, as an operation between two vectors, produces a scalar number. 
Given two vectors $|u>$ and $|v>$, elements of a Hamiltonian ${\cal H}$, then the internal product  $\braket{v|u}$ corresponds to a scalar assuming its defining properties: (a) $\braket{u|v}^* = \braket{v|u}$; (b) $\braket{v|u}$ is linear in $\bra{u}$ and anti-linear in $\bra{v}$; (c) $\braket{u|u} \geq 0$.
Thus, inner products allow to define the length or norm of a vector,  as $||\Psi|| = \sqrt{\braket{\Psi|\Psi}}$. 

In mathematical (and physical) terms, in comparison to {\it kets}, {\it bras} are different objects. In the case where a given vector $u$ belongs to the Hilbert space $U_H$, its {\it bra} counterpart belong to the $U_H^*$ space, dual to $U_H$. Elements of $U^*$ thus correspond to a linear mapping from $U_H$ to $\mathbb{C}$. Using mathematics notation, given that $u \in U_H$, and a linear function $\phi \in U_H^*$, such that $\phi(v)$ denotes the action of the function $\phi$ on the vector $v$, this function is a
number. In the bracket notation the following replacements can be identified:
$u \to \ket{u}$; $\phi \to \bra{v}$; $\phi_v(u) \to \braket{v|u}$.

A {\it bra} is the conjugate transpose of the {\it ket}, representing its dual vector and is used to represent the properties of a quantum state, such as its probability or expectation value. In this sense, the probability density $|\Psi(\xi)|^2$, has its physical meaning intrinsically correlated with $|\Psi(\eta)|^2$, 
since the inner product between these dual vectors $\braket{\xi | \eta}$ is different from zero, measuring this way the degree of overlap between these quantities. Accordingly, the behavior of $|\Psi(\xi)|^2$ is an indication of an underlying mechanism similar to inflation, originated by the non-commutative structure of the BCG, identified with the only element that may alter the distribution of matter in the early Universe, 
- if we limit ourselves to the formulation adopted in this work -,
the $V(\eta)$ potential.  Due to the change of variables introduced to reduce the original Hamiltonian to a canonical form, based on its characteristic equation~\cite{Bodmann2023b,Polyanin}, the principle of superposition in quantum mechanics is an indication of the degree of superposition of $\ket{\eta}$ (with the essence, or more precisely) with the the intrinsic {\it inflationary} nature of $\bra{\xi}$. The corresponding quantum implications are revealed in the behavior of $|\Psi(\xi)|^2$,  which is shaped and captured in an underlying manner by the behavior at short scales of $\Psi(\eta)$.
This intrinsic duality can be expressed and shaped in a representation of Fourier transformation in the form
\begin{equation}
    \Psi(\eta) = \frac{1}{\sqrt{2\pi}} \int_{-\infty}^{\infty} \Psi(\xi) e^{i\xi\eta} d\xi \, . 
\end{equation}

\section{Dynamical Equations}

In the following we investigate the behavior of the scale factors in the commutative ($u(t)$) and non-commutative ($\eta(t)$) approaches.  

\subsection{Dynamical equations: commutative approach}

Going a few steps back, the Ho\v{r}ava-Lifshitz commutative BCG action (\ref{HL}) is related to the Lagrangian density ${\cal L}_{HL}$, expressed in terms of the standard variable $u(t)$ in the form~\cite{Bodmann2023a} (see also refs.~\cite{Bertolami2011,Abreu,Bezerra})
\begin{equation}
  S_{HL} =  \int dt {\cal L}_{HL} \, ,
  \end{equation}
  resulting in the following expression
 \begin{equation} 
S_{HL} =   \frac{1}{2} \! \int \! dt  \frac{N}{a} \! \Biggl[ -  p^2_u - 
 g_r +  g_m a  +  g_k a^2 + g_q a^3 
 - g_{\Lambda} a^4    -   \frac{g_s}{a^2} 
    \Biggr] \, , 
\end{equation}
with
\begin{equation}
    p_u = \frac{\partial {\cal L}_{HL}}{\partial \dot{u}} = - \frac{u \dot{u}}{N} \, .
\end{equation}
The Ho\v{r}ava-Lifshitz mini-superspace commutative BCG Hamiltonian density ${\cal H}_{HL}$ 
\begin{equation}
    {\cal H}_{HL}  =  p_u \dot{u} - {\cal L}_{HL} \, , 
\end{equation}    
may be cast as 
\begin{equation}
 {\cal H}_{HL}    =  \frac{1}{2} \frac{N}{u} \Biggl[-p^2_u + 
 g_r -  g_m u  -  g_k u^2 - g_q u^3 
 + g_{\Lambda} u^4 + \frac{g_s}{u^2} \Biggr] \, . \tag{\ref{H*} revisited}
\end{equation}
\begin{figure}[htpb]
\centering
\includegraphics[scale=0.3]{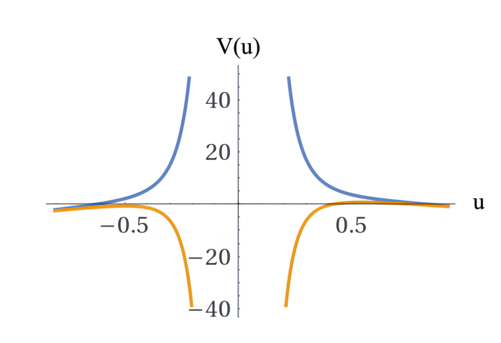} 
\includegraphics[scale=0.3]{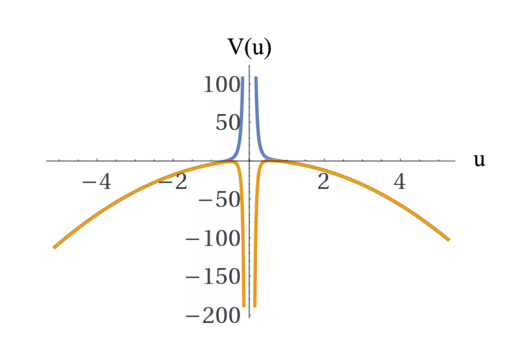}
\caption{Graphical illustrations of the potential $V(u)$ of equation (\ref{Vu}), emphasizing the contour lines. The values of the running coupling constants and parameters are, in the left figures, $g_k = 1$; $g_q = 0.7$; $g_{\Lambda} = 0.333$; $g_r = 0.4$; $g_s = 0.03$; $\omega = -1$; ${\cal C} = 1$.  In the right figures, the values  are: $g_k = 1$; $g_q = 0.7$; $g_{\Lambda} = 0.333$; $g_r = 0.4$; $g_s = - 0.03$; $\omega = -1$ ${\cal C} = 1$. } \label{V12}
\end{figure} 
In order to introduce a non-commutative formalism, based on the perfect fluid conception of Hermann Weyl~\cite{Weyl}, using the formalism of Bernard F. Schutz~\cite{Bernard}, an additional term was inserted in the Lagrangian formalism, whose Hamiltonian may be expressed as~\cite{Bodmann2023b} (see also \cite{Abreu}):
\begin{equation}
{\cal H}_{fp} = \frac{1}{2}N \frac{p_v}{u^{3 \omega}} \, , \quad \mbox{with} \quad p_v =  - \frac{v \dot{v}}{N} \, , 
\end{equation}
resulting, from equation (\ref{H*}), 
assuming the gauge $N=1$, and simplifying the notation, eliminating the HL label,
in the super-Hamiltonian:
\begin{equation}
  \label{SuperHamiltonian2} 
 {\cal H}  \! = \! \frac{1}{2} \Biggl[- \frac{1}{u} p^2_{u}  +    \frac{g_r}{u}  -  g_m   -  g_k u  -  g_q u^2  
 +   
 g_{\Lambda} u^3 
    + \frac{g_s}{u^3} + \frac{1}{u^{3\omega}} p_v \Biggr] . 
\end{equation} 
From equation (\ref{SuperHamiltonian2}) the following Hamilton equations result:
\begin{eqnarray}
    \dot{u} & = & \frac{\partial {\cal H}}{\partial p_u} = -  \frac{p_u}{u} \, ; \quad 
     \dot{v} =  \frac{\partial {\cal H}}{\partial p_v} = - \frac{1}{2u^{3 \omega}} \, ; \nonumber \\ 
     \dot{p}_v & = &  - \frac{\partial {\cal H}}{\partial v} = 0 \rightarrow p_v = {\cal C} \, , 
\end{eqnarray}
and
\begin{eqnarray}
\label{Heqs}  
    \dot{p}_u     =   - \frac{\partial {\cal H}}{\partial u} = - \frac{1}{2u} \Biggl[\frac{p^2_u}{u}  - \frac{g_r}{u}  -  g_ku  -  2 g_q u^2  
    -   3 g_{\Lambda} u^3 
      -  3 \frac{g_ s}{u^3} - 3 \omega \frac{{\cal C}}{u^{3 \omega}}\Biggr]
    \, , 
\end{eqnarray}
where ${\cal C}$ is a constant in the time sector.  
It is important to point out that the relative sign between the term labeled by $g_{\Lambda}$ has changed, in the derivative 
procedure, in comparison with the  $g_r$ and $g_s$ contributions (see equation (\ref{SuperHamiltonian2})).


\subsubsection{Second-order dynamical equations: commutative formulation}

Since 
\begin{equation}
p_u = - u \dot{u} \, , \quad  \mbox{and} \quad  \dot{p}_u  = - \dot{u}^2 - u \ddot{u} \, ,  
\end{equation}
we obtain from (\ref{Heqs}) the following dynamical equation for the scale factor $u$:
\begin{equation}
         2u\ddot{u} + \dot{u}^2  +  V(u) = 0 \, .   
      \label{deu}
\end{equation}
In this expression, the potential $V(u)$ is defined as
\begin{equation}
    V(u) =  g_k + 2 g_q u - 3 g_{\Lambda} u^2 + \frac{g_r}{u^2} + 3 \frac{g_ s}{u^4} + 3 \omega \frac{{\cal C}}{u^{3 \omega + 1}} \, . \label{Vu}
\end{equation}
Figure \ref{V12} show typical graphical illustrations of the potential $V(u)$, eq. (\ref{Vu}). The main difference between the curves is related to the sign of the stiff matter running coupling constant, for positive signal it gives rise to a repulsive potential (upper curves) and for negative values an attractive potential around a singularity in the origin.

\begin{figure*}[htpb]
\centering
\includegraphics[scale=0.24]{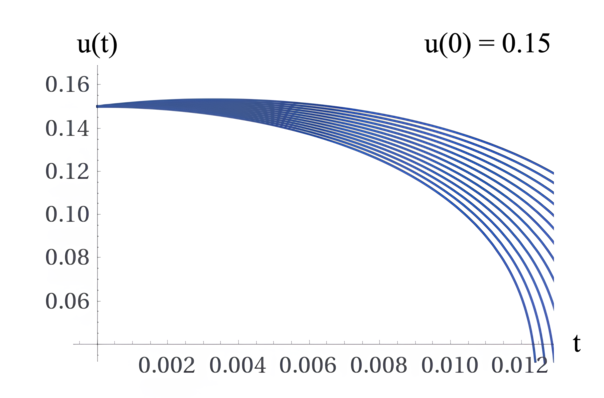} \hspace{-0.25cm}
\includegraphics[scale=0.24]{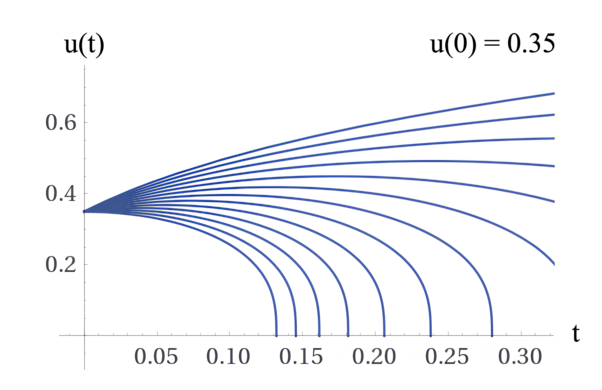}\hspace{-0.25cm}
\includegraphics[scale=0.24]{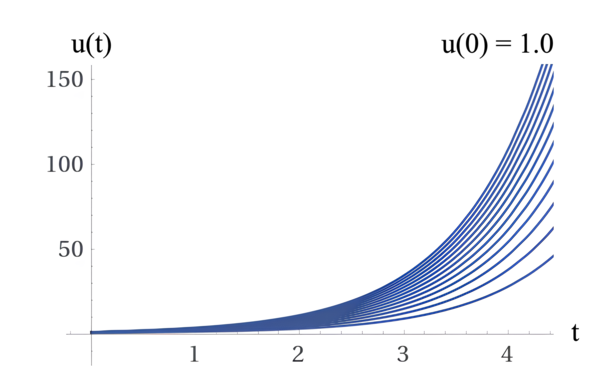}
\caption{Sample solutions for the scale factor $u(t)$ given by equation~(\ref{deu}) for initial conditions $u(0)=0.15$, 0.35, and 1.0. The values of the running coupling constants and parameters are: $g_k = 1$; $g_q = 0.7$; $g_{\Lambda} = 0.333$; $g_r = 0.4$; $g_s = 0.03$; $\omega = -1$; ${\cal C} = 1$.} \label{C++-}
\end{figure*} 
Figures \ref{C++-} show typical sample solutions of the second-order time derivative of the scale factor $u(t)$ given in equation (\ref{deu}). For all values of the running coupling constants, the results indicate, differently from generally assumed, that the Universe does not originate from a singularity or out of {\it nothing} as originally proposed by Alan Guth~\cite{Guth1981,Guth2004} in the inflation model (see also ref. \cite{Vilenkin}). Our results are more in tune with a transition phase between the contraction and expansion topological boundaries intermediated by a Riemannian foliated transition region as proposed by the branch-cut gravity. 

As a corollary, it is important to highlight that this conclusion is not a simple ad hoc consequence of the BCG theoretical conception regarding the boundary conditions which imply in  a quantum leap in the transition region according to the Bekenstein criterion~\cite{Pacheco}. In fact, this conclusion is more related to the intrinsic nature of the mathematical equations and solutions themselves, whose convergence only occurs in the case where the boundary conditions $\Psi(-1) = -1$ and $\Psi(1)=1$ are established, in good accordance with the theorem of existence and uniqueness for solving differential equations. 

\begin{figure*}[htpb]
\centering
\includegraphics[scale=0.165]{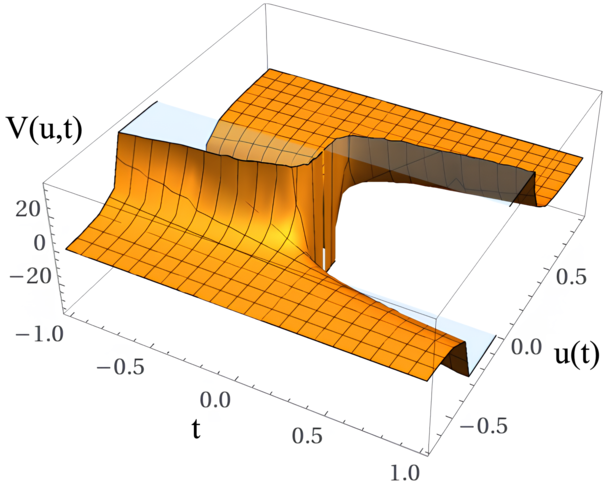} \hspace{-0.25cm}
\includegraphics[scale=0.165]{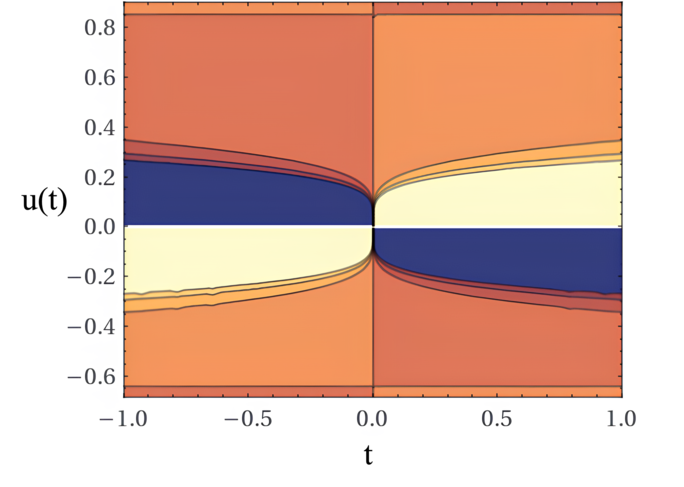} \hspace{-0.25cm}
\includegraphics[scale=0.165]{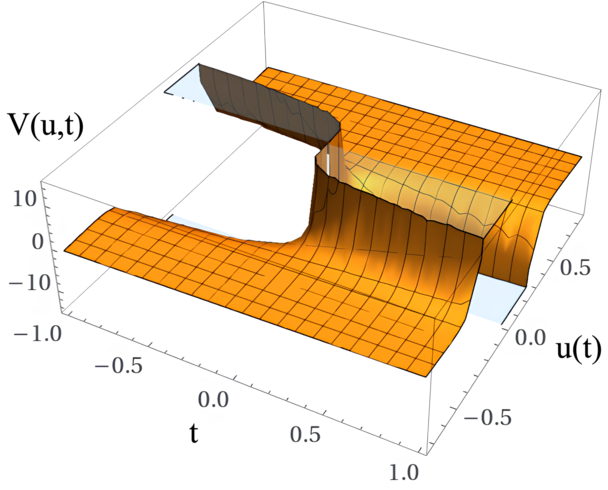} \hspace{-0.15cm}
\includegraphics[scale=0.165]{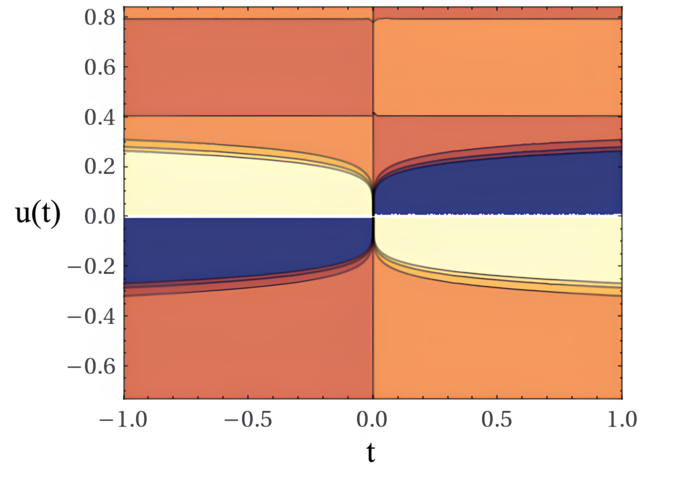}
\caption{Graphical illustrations of the potential $V(u,t)$ and the scale factor of equations (\ref{1stu}) and (\ref{Vut}) for the commutative approach. 
In the left-pair of figures, the values of the running coupling constants and parameters are: $g_k = 1$; $g_q = 0.7$; $g_{\Lambda} = 0.333$; $g_r = 0.4$; $g_s = 0.03$; $\omega = -1$; ${\cal C} = 1$.  In the right-pair of  figures, the values are: $g_k = 1$; $g_q = 0.7$; $g_{\Lambda} = 0.333$; $g_r = 0.4$; $g_s = - 0.03$; $\omega = -1$ ${\cal C} = 1$. } \label{Vutime}
\end{figure*} 
\subsubsection{First-order dynamical equations: commutative formulation}

In what follows, we use Cauchy's implicit function theorem, assuming that the variable $u$ has an explicit time dependence, so the variables $u$ and $t$ can be separated, thus ensuring that these variables are differentiable. We also assume, for simplicity, in view of the high difficulty to solve the above equations, that $\dot{u}$ and $u$ act as independent variables, so from equation (\ref{deu}) we have
\begin{equation}
2 u \frac{d\dot{u}}{dt} + \dot{u} \frac{du}{dt} = - V(u)  \rightarrow 2 u d\dot{u} + \dot{u} du = - V(u) dt \, .
\end{equation}
This assumption allows the integration on time of the equation above giving
\begin{equation}
2u \int d \dot{u} +  \dot{u} \int du 
= 3 u(t) \dot{u} = - V(u) \int dt 
\end{equation}
which may be cast as
\begin{equation}
\dot{u}(t)   +  V(u,t)  = 0  \, , 
 \label{1stu}
\end{equation}
where we define
\begin{equation}
    V(u,t) \equiv \frac{1}{3} \Biggl( \frac{g_k}{u}  + 2 g_q  - 3 g_{\Lambda} u + \frac{g_r}{u^3} + 3 \frac{g_ s}{u^5} + 3 \omega \frac{{\cal C}}{u^{3 \omega + 2}} \Biggr) t \, . \label{Vut}
\end{equation}

Figure \ref{Vutime} presents graphical representations of equation (\ref{Vut}). In these contour plots, varying colors indicate the behavior of the potential $V(u, t)$ in regions of dominance, either repulsion or attraction. Lighter colors signify higher intensity or amplitude, while darker colors indicate the opposite.
The blue color signifies the region with lower values, typically surrounding
 the singularity or the repulsive part of the potential, while the cream color designates regions with higher values,  where either the singularity or the repulsion of the potential predominates. The white regions, albeit not always clearly visible in some of the figures, corresponds to the exact region of the singularity or the maximum of repulsion. The results regarding the potential $V(u,t)$ are characterized by symmetric distributions of color compositions, indicating that the commutative algebraic structure contributes in a moderate way to the configuration and distribution of matter in the early Universe. 

\begin{figure*}[htpb]
\centering
\includegraphics[scale=0.24]{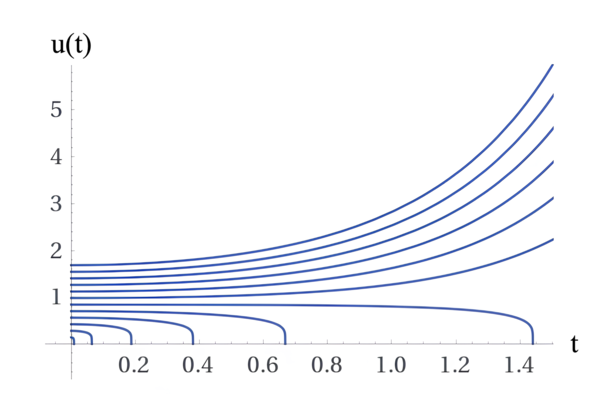} 
\includegraphics[scale=0.24]{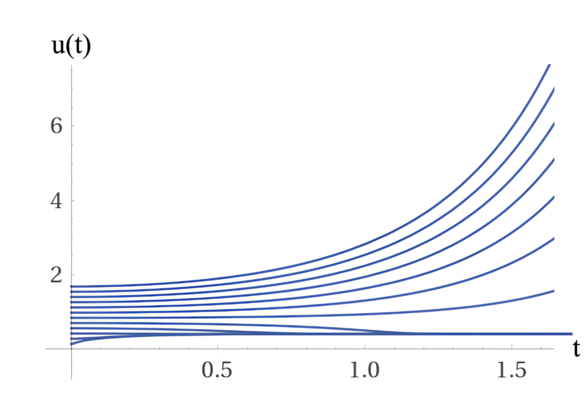} 
\includegraphics[scale=0.24]{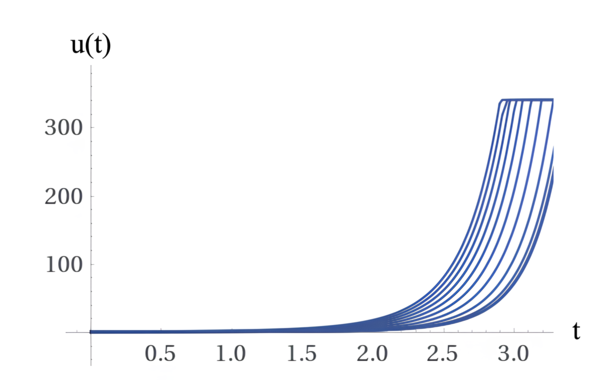}
\caption{Sample solutions of the first-order time derivative of $u(t)$ as given in equation (\ref{1stu}) for the commutative formulation. 
The initial condition is $u(0)$. The values of the running coupling constants and parameters are as follows. Left figure: $g_k = 1$; $g_q = 0.7$; $g_{\Lambda} = 0.333$; $g_r = 0.4$; $g_s = 0.03$; $\omega = -1$; ${\cal C} = 1$; 
Middle figure: $g_k = 1$; $g_q = 0.7$; $g_{\Lambda} = 0.333$; $g_r = 0.4$; $g_s = -0.03$; $\omega = -1$;  ${\cal C} = 1$; Right figure: $g_k = 1$; $g_q = 0.7$; $g_{\Lambda} = 0.333$; $g_r = -0.4$; $g_s = -0.03$; $\omega = -1$; ${\cal C} = 1$.} \label{SS+-}
\end{figure*}

Figure \ref{SS+-} displays typical sample solutions of equation (\ref{1stu}) for various sets of initial conditions. The outcomes obtained from the dynamic equations suggest that, contingent upon the initial conditions, phases that lead to the big crunch or moderately accelerated phases of the early Universe tend to prevail.

The consistency of the assumptions that led to equation (\ref{1stu}) and to the potential $V(u,t)$ given by equation (\ref{Vut})
will be tested in the following by comparing the results that describe the behavior of the solutions of the dynamic equations (\ref{deu}) and (\ref{1stu}). 
It is important to emphasize that articles addressing this topic typically do not provide solutions to the equations describing the behavior of the Universe's wave function due to the formal challenges associated with these descriptive equations.
These approaches are constrained to performing computational or algebraic calculations for dynamic equations, typically relying on approximations that significantly restrict their scope and predictive capabilities.
In our case, from a computational perspective, we utilized the Runge-Kutta-Fehlberg method for solving differential equations, known for its high computational efficiency. This approach, extending to other subsequent calculations, enabled us to numerically solve equations (\ref{deu}) and (\ref{1stu}) with a high degree of convergence, without the need to resort to approximations that might otherwise restrict the descriptive potential of the present formulation.
Moreover, it is crucial to highlight that when deriving equation (\ref{deu}), which describes the behavior of dynamic solutions corresponding to the second-order derivatives of the scale factor $\eta$, we made no approximations or descriptive impositions. However, when obtaining equation (\ref{1stu}), which characterizes the behavior of dynamic solutions corresponding to the first-order derivatives of the scale factor $\eta$, we introduced two impositions: we assumed an explicit time dependence for the variable $u$, allowing the separation of variables $u$ and $t$, and we treated $\dot{u}$ and $u$ as independent variables.
 Typical solutions of equation (\ref{1stu}) are shown in figure \ref{SS+-}. 
 When comparing the results of the solutions of equations (\ref{deu}) and (\ref{1stu}), it becomes evident that there is a similar behavior in both solutions, at least on a qualitative level. This suggests that the impositions introduced during the development of equation (\ref{1stu}) are consistent enough to be applied to the subsequent equations. 
  The first and second-order differential equations for u(t) will prove essential for the determination of the corresponding equations that describe primordial gravitational waves in the realm of a non-commutative formulation.
Our results also indicate  a strong dependence on the boundary initial conditions applied to the scale factor $u(t)$. For positive $g_s$,  depending on the boundary initial conditions, our results indicate  the presence of a big crunch on the evolution of the Universe, which instead of expanding, violently contracts on itself. For other boundary initial conditions, the Universe experience a deceleration period followed by a moderate acceleration period. For negative $g_s$ the results indicate only the deceleration period followed by an moderate acceleration period. Our study reveals the presence of a plethora of solutions in the commutative expansion sector.
 The solutions are significantly dependent on the choices of parameters, on the running coupling constants, and on boundary conditions.


\subsection{Dynamical equations: Non-commutative approach}
In what follows, we analyze the role of the non-commutative algebraic structure in the expansion of the Universe.  In the analysis that we will carry out in the following, we seek to examine the role of the non-commutative structure in the early time accelerated expansion of the Universe, more precisely in the inflation period.
\begin{figure*}[htpb]
\centering
\includegraphics[scale=0.1875]{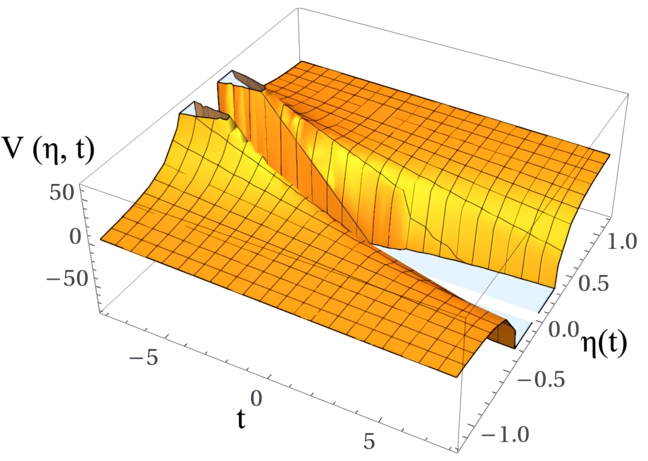} \hspace{0.025cm}
\includegraphics[scale=0.14]{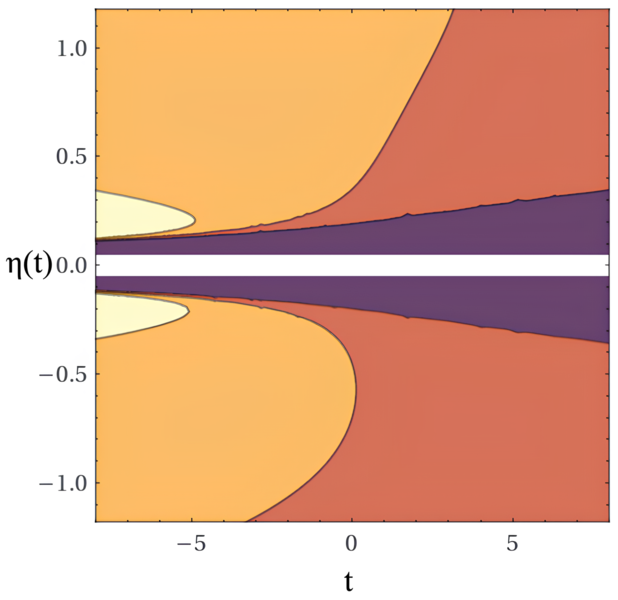} \hspace{0.025cm}
\includegraphics[scale=0.1875]{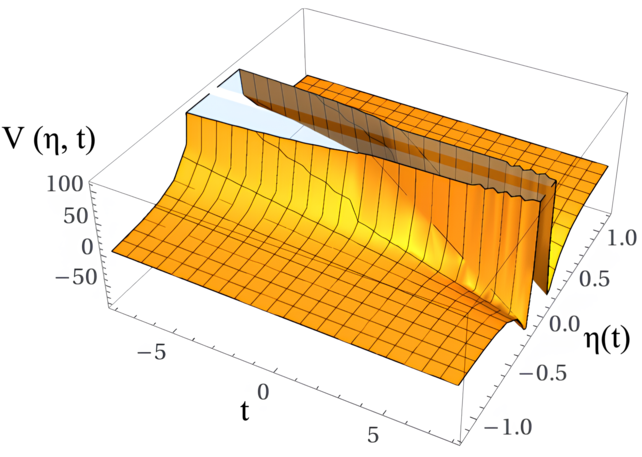}  \hspace{0.025cm}
\includegraphics[scale=0.145]{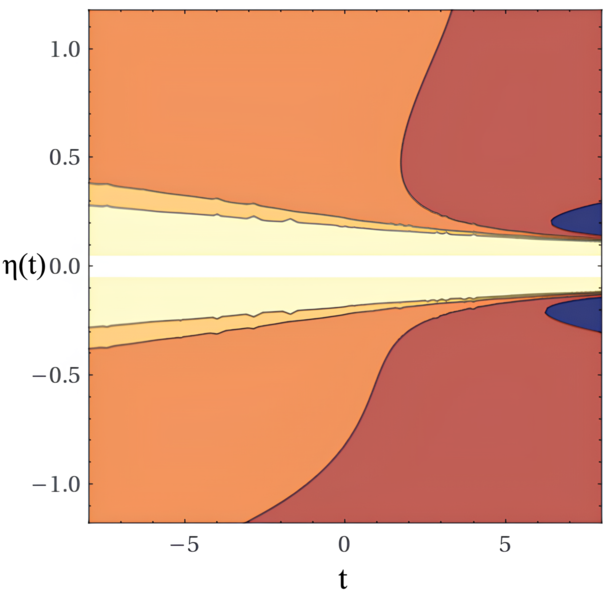}
\caption{Graphical illustrations of the potential $V(\eta,t)$ from equation (\ref{VetatNew}) and corresponding contour plots.  
In the left-pair of figures, the values of the running coupling constants and the parameters of the non-commutative algebra are as follows: $g_k = 1$; $g_q = 0.7$; $g_{\Lambda} = 0.333$; $g_r = 0.4$; $g_s = -0.03$; $\alpha = 1/2$; ${\cal C} = 1$. In the right-pair of figures,
the values of the running coupling constants and the parameters of the 
non-commutative algebra are as follows: $g_k = 1$; $g_q = 0.7$; $g_{\Lambda} = 0.333$; $g_r = 0.4$; $g_s =  0.03$; $\alpha = 1/2$; ${\cal C} = 1$.
} \label{VNC2ndVetatNew}
\end{figure*} 

From equations (\ref{Hsupersuper*}) and (\ref{Hsupersuper}), the starting point of our approach is based on the super-Hamiltonian which reads, in the non-commutative BCG, as:
\begin{equation}
{\cal H}^{NC}  =      \frac{1}{2}N \Biggl[-\frac{1}{\eta}p^2_{\eta} -  \frac{\gamma}{\eta^{3\alpha}} p_{\eta}   + 
 \frac{g_r}{\eta} - g_m    -   g_k \eta - g_q \eta^2  + g_{\Lambda} \eta^3 + \frac{g_s}{\eta^3}  +   \frac{\alpha}{\eta^{3\alpha-1}}   -  \frac{\alpha \xi}{\eta^{3\alpha}} + \frac{1}{\eta^{3\alpha}} p_{\xi} \Biggr]\, . \label{sH}
\end{equation}
The Hamilton equation (\ref{sH}) characterizes the time dynamical evolution of the branch-cut Universe parameterized by a dimensionless cosmic scale factor represented, in the present case in which we consider a non-commutative algebraic formalism, by the variable $\eta(t)$.
As a peculiarity, the results referring to commutative algebra indicate a late time moderate expansion of the wave function of the Universe and the scale factor.

\subsubsection{Second-order dynamical equations: non-commutative formulation}

In the following, we consider second-order dynamical equations for the scale factor $\eta(\tau)$ and its counterpart $\xi(\eta)$. 

From expression (\ref{sH}), assuming the gauge fixing condition $N=1$, the following dynamical equations for  $\eta(\tau)$ and $\xi(\eta)$ result:
\begin{eqnarray}
    \dot{\eta} & = &  \frac{\partial {\cal H}^{NC}}{\partial p_{\eta}}  =  -  \frac{p_{\eta}}{\eta} - \frac{\gamma}{2 \eta^{3 \alpha}} \, ; \\
     \dot{\xi} & = & \frac{\partial {\cal H}^{NC}}{\partial p_{\xi}}  =  \frac{1}{2\eta^{3 \alpha}} \, ;  \\
      \dot{p}_{\xi} &  =   & - \frac{\partial {\cal H}^{NC}}{\partial {\xi}} = -\frac{\alpha}{2\eta^{3 \alpha}}\, ;
\end{eqnarray}
     and
     \begin{eqnarray}
     \label{Heqs2}
  \!\!\!\! \!\!\!\!\! \dot{p}_{\eta}   & \! = \! &  - \frac{\partial {\cal H}^{NC}}{\partial \eta} \nonumber \\
    \!\!\!\!\!\!\!\!\! &=& \! - \frac{1}{2 \eta} \Biggl[\frac{1}{\eta}p^2_{\eta} + \frac{3 \alpha\gamma}{\eta^{3 \alpha}}p_{\eta}  - \frac{g_r}{\eta} - g_k \eta  
    -    2 g_q \eta^2 \! - \! 
 3 g_{\Lambda} \eta^3 - 3 \frac{g_ s}{\eta^3} 
 - \frac{\alpha \bigl(3 \alpha - 1 \bigr)}{\eta^{3 \alpha -1}} + \frac{3 \alpha^2 t}{\eta^{6 \alpha -1 }} \Biggr] \, 
    , 
\end{eqnarray}
where we have assumed an explicit time-dependence on the variables $\eta$ and $\xi$; thus, all known solutions are of the separation of variables type, where time and space dependence are treated separately. 
Thus, from the previous equations, we obtain the following expressions for $\eta$ and $\xi$:
\begin{equation}
 \xi  =  \int dt \dot{\xi} = \frac{t}{2 \eta^{3 \alpha}} \, , \quad
 {p}_{\xi}  =  \int dt \dot{p}_{\xi} =   -  \frac{\alpha t}{2\eta^{3 \alpha}} \, . 
\end{equation}
The time derivative of the conjugate canonical momentum $p_{\eta} = - \eta \dot{\eta} /N $ in the $N=1$ gauge may be written as
\begin{equation}
\dot{p}_{\eta} =  \frac{\partial p_{\eta}}{\partial t} = -\frac{\partial  (\eta \dot{\eta})}{\partial t} = -\dot{\eta}^2 - \eta \ddot{\eta}, \label{eqeta}
\end{equation}
Equation (\ref{Heqs2}) combined with (\ref{eqeta})
may be recast in the form
\begin{equation}
    2 \eta(t) \ddot{\eta}(t)+ \dot{\eta}^2(t) + \frac{3 \alpha \gamma \dot{\eta}}{\eta^{3 \alpha}} + V(\eta,t) = 0 \, , \label{etaNC}
\end{equation}
where the potential $V(\eta,t)$ has an explicit time dependence: 
\begin{equation}
    V(\eta,t) \! = \! g_k + 2 g_q \eta -  3 g_{\Lambda} \eta^2 + \frac{g_r}{\eta^2} + 3 \frac{g_s}{\eta^4} + \frac{\alpha \bigl(3 \alpha - 1 \bigr)}{\eta^{3 \alpha}} 
    - \frac{3 \alpha^2 t}{\eta^{6 \alpha  -1}} \, . 
\label{VetatNew}
\end{equation}
Figures  \ref{VNC2ndVetatNew} 
show 3D graphical illustrations of the potential $V(\eta,t)$ of equation (\ref{VetatNew}) and the corresponding contour plots characterizing the dependence of the scale factor $\eta(t)$ on time. 
The contour plots in the commutative formulation, as seen in figure \ref{Vutime}, exhibit a symmetrical distribution of intensities or amplitudes of the potential $V(u, t)$ intensities or amplitudes around the lines $t = 0$ and $u(t) = 0$.
In the non-commutative formulation, this symmetry is broken, indicating a mixture of intensities or amplitudes of the potential $V(\eta,t)$. The potential $V(\eta,t)$ simulates the presence of different compositions of matter in the primordial Universe that imply structural modifications of the spacetime structure, shaping this way its curvature that depends locally on the amount and distribution of matter or, equivalently, energy. This symmetry breaking reveals the potentiality of a non-commutative formulation in terms of its implications in affecting not only the curvature of space-time,
but furthermore, the capture of short and long scales, boosting the evolution dynamics of the wave function of the Universe and the cosmic scale factor. Insofar, as the presence of the potential is associated with a force, of a gravitational character, which may constitute the propelling element of the acceleration of the primordial Universe. These results provide conceptual elements that indicate a reconfiguration of matter on small scales of dimensions. This reconfiguration of matter, as previously observed, represents a typical characteristic of a non-commutative formulation, which drives, through a non-symmetric gravitational force, the acceleration of the Universe. Furthermore, due to the dual role of gravity, spacetime is both warped by matter and matter in turn experiences a warped spacetime due to the presence of other distributions of matter, which intensifies the effects of higher curvature terms.

The color palette used in the contour plots is indicative of these dynamic processes. Lighter colors suggests the materialization of greater amplitudes or intensities of the effects associated with the composition of matter and energy, while darker colors signify the opposite effects. It is precisely within the
interplay of these effects, where symmetry breaks along the axis $\eta(t)=t=0$,
that the accelerated evolutionary dynamics of the branch-cut Universe finds 
the necessary components for its manifestation.

\begin{figure*}
    \centering
       \subfigure[]{\includegraphics[width=0.325\textwidth]{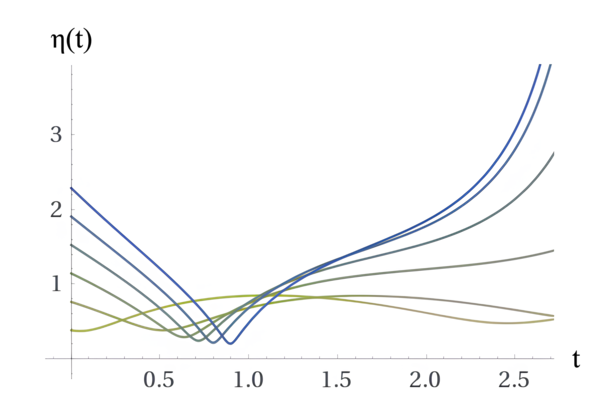}} 
    \subfigure[]{\includegraphics[width=0.325\textwidth]{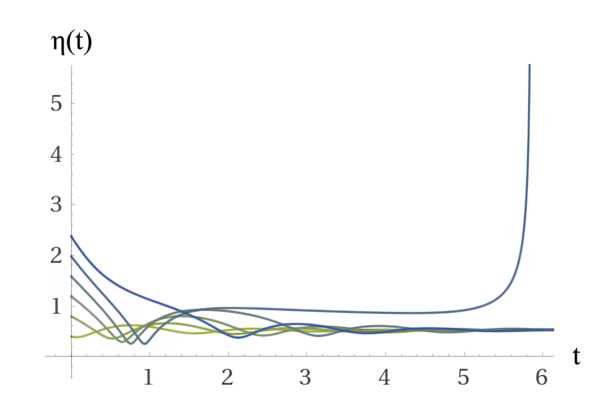}}
    \subfigure[]{\includegraphics[width=0.325\textwidth]{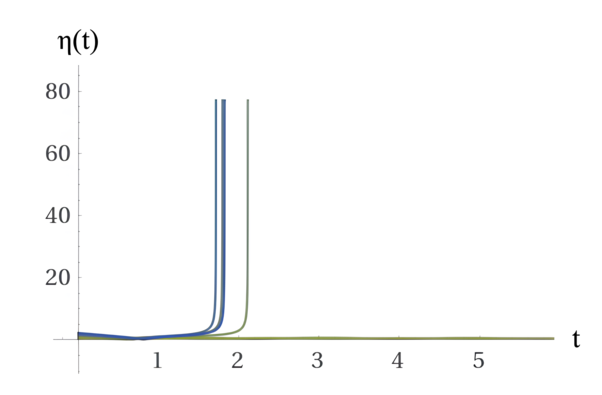}}
    \caption{Typical sample family solutions for the scale factor of the branch-cut Universe, $\eta(t)$, using the second-order time derivative 
    non-commutative approach equation (\ref{etaNC}).
The solutions are for varying initial conditions $\eta(0)$ and $\eta'(0)$.  In the left figure (a), the values of the running coupling constants and parameters are as follows:  $\tilde{g}_r = -0.4$; $g_k = 1$; $g_q = 0.7$; $g_{\Lambda} = -0.333$; $g_s = -0.03$; $\alpha = -1/2$; $\gamma=1$; ${\cal C} = 1$.
   In figure (b), the values of the running coupling constants and parameters remain the same, except for  $\alpha = -3/4$; $\gamma=-1$. In figure (c), the values for $\alpha$ and $\gamma$ are $\alpha = -3/4$ and $\gamma=-1/3$.}
  \label{NCSF2nd}
\end{figure*}

\subsubsection{First-order dynamical equations: non-commutative formulation}

As mentioned earlier, the dynamic equations governing the evolution of the scale factor $\eta(t)$, both in terms of first-order and second-order time derivatives, are crucial for describing primordial gravitational waves. Thus, in the following, we will elaborate on the corresponding first-order dynamic equations.

Upon integrating equation (\ref{eqeta}), while assuming that $\eta$ possesses explicit time dependence allowing for separation from time, and further considering that $\dot{\eta}$ and $\eta$ function as independent variables, we obtain
\begin{equation}
    \dot{\eta} + v(\eta,t) = 0  \, , \label{NC1steq} 
\end{equation}
where the potential $v(\eta,t)$ is defined as
\begin{equation}
  v(\eta,t) =  \frac{1}{3\eta} \Biggl\{- \frac{3 \alpha \gamma}{\bigl(3\alpha-1\bigr)\eta^{3\alpha - 1}} + \Biggl( g_k + 2 g_q \eta - 3 g_{\Lambda} \eta^2 + \frac{g_r}{\eta^2} + 3 \frac{g_s}{\eta^4} + \frac{\alpha \bigl(3 \alpha - 1 \bigr)}{\eta^{3 \alpha}} \Biggr) t
-  \frac{3}{2}\frac{ \alpha^2 t^2}{\eta^{6 \alpha  -1}} \Biggr\} \, . \label{vetat}
\end{equation}
Figure \ref{VNC1stTorsion} presents graphical representations of the potential $v(\eta, t)$ from equation (\ref{vetat}), indicating the presence of a torsion deformation.
More specifically, it indicates a deformation that resembles a kind of torsional conformation of an object due to the application of an ``external'' torque.

\subsubsection{Torsion deformation}

\begin{figure}[htpb]
\centering
\includegraphics[scale=0.2]{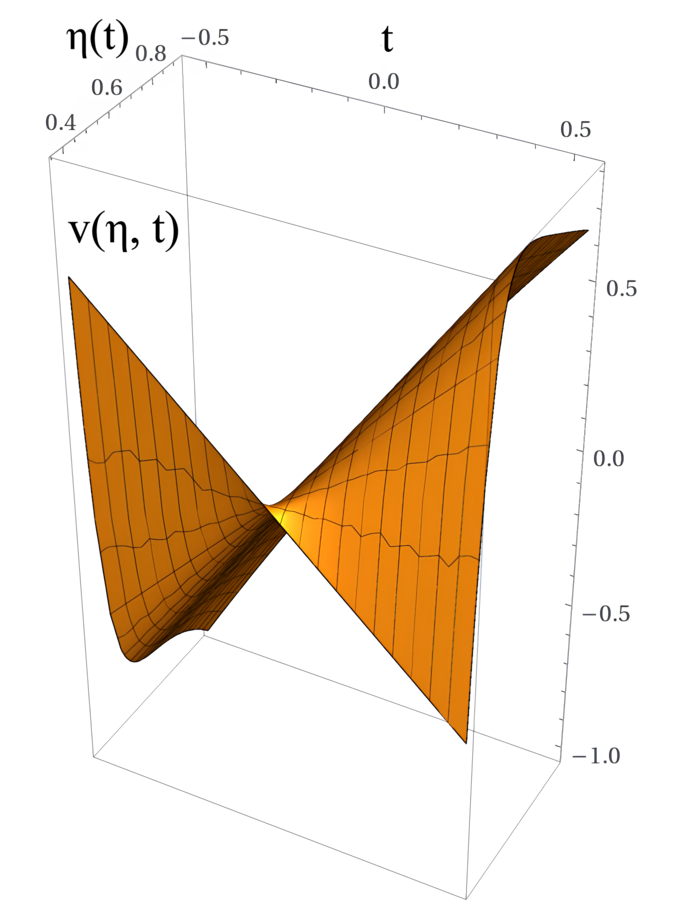} 
\includegraphics[scale=0.3]{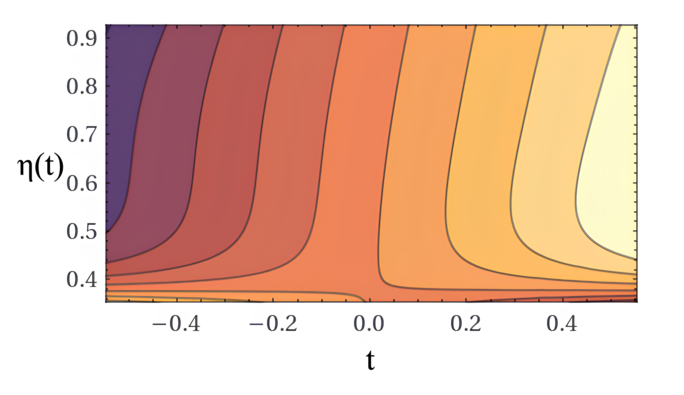}
\caption{Graphical illustrations of the potential $v(\eta,t)$ of equation (\ref{vetat}) and contour plots. 
The values of the running coupling constants and the parameters of the non-commutative algebra for the figures are: $g_k = 1$; $g_q = 0.7$; $g_{\Lambda} = 0.333$; $g_r = 0.4$; $g_s =  -0.03$; $\alpha = -5/6$; $\gamma = 3/5$ ; ${\cal C} = 1$.} \label{VNC1stTorsion}
\end{figure}

A widely used method for constructing non-commutative spacetime algebraic structures and non-commutative field theories involves the concept of torsion (or twist) deformation.
The graphical representations of the potential $v(\eta, t)$ from equation (\ref{vetat}) and the accompanying contour plots in figure \ref{VNC1stTorsion} in turn resemble, in particular, a characteristic torsion (or twist) deformation of the spacetime geometry. In the following, we focus on the perspectives and implications of this type of space-time distortion within the context of a background branch cut. 
\begin{figure}
    \centering
    \subfigure[]{\includegraphics[width=0.4\textwidth]{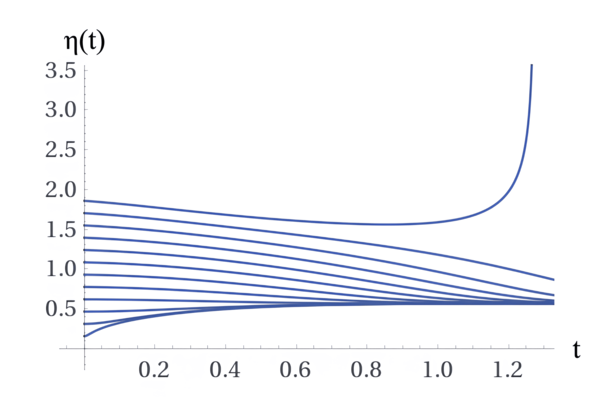}} \hspace{0.15cm}
      \subfigure[]{\includegraphics[width=0.4\textwidth]{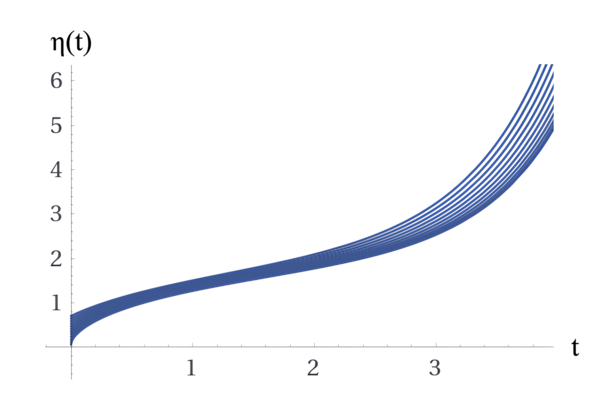}} 
    \caption{Typical sample family solutions for the scale factor of the branch-cut Universe, $\eta(t)$, 
    with variations in the initial condition $\eta(0)$ and different sets of 
    values for the running coupling constants and parameters.
In the left figure (a), the values are as follows: $\tilde{g}_r = -0.4$; $g_k = 1$; $g_q = 0.7$; $g_{\Lambda} = -0.333$; $g_s = -0.03$; $\alpha = -3/4$; $\gamma=-1$; ${\cal C} = 1$. In the middle figure (b), the values are:
   $\alpha = -3/4$; $\gamma=-1$; $\tilde{g}_r = -0.4$; $g_k = 1$; $g_q = 0.7$; $g_{\Lambda} = -0.333$; $g_s = -0.03$; $\alpha = 1/4$; $\gamma=-$; ${\cal C} = 1$. 
}
  \label{NCSF1st}
\end{figure}
In the context of a twist deformation, the geometry of spacetime, as
described by the Ricci tensor $R_{\mu\nu}(\Gamma^{\alpha}_{\mu\nu})$
and the affine connection $\Gamma^{\alpha }_{\mu\nu}$, undergoes a
change. In the presence of a twist, the affine connection
$\Gamma^{\alpha}_{\mu\nu}$ becomes an antisymmetric
object~\cite{Lima},
\begin{equation}
\Gamma^{\alpha}_{\mu\nu} = \tilde{\Gamma}^{\alpha}_{\mu\nu}+\Omega^{\alpha}_{\mu\nu}
\end{equation}
where $\tilde{\Gamma}^{\alpha}_{\mu\nu}$ represents the symmetric connection part and $\Omega^{\alpha}_{\mu\nu}$ defines the torsion tensor.
$\Omega^{\alpha}_{\mu\nu}$ may be decomposed as 
\begin{equation}
\Omega^{\alpha}_{\mu\nu} = S^{\alpha}_{\mu\nu} + S_{\mu\nu}^{\alpha} + S_{\nu\mu}^{\alpha} 
\end{equation}

In the realm of the algebra of diffeomorphisms on non-commutative spacetime, a torsion  generator ${\cal F}$ and its inverse ${\cal F}^{-1}$ may be written as~\cite{Paolo}
\begin{equation}
{\cal F} = e^{-\frac{i}{2} \theta^{\mu\nu}\partial_{\mu} \otimes \partial_{\nu}} \, ; \quad 
{\cal F}^{-1} = e^{\frac{i}{2} \theta^{\mu\nu}\partial_{\mu} \otimes \partial_{\nu}} \, .
\end{equation}
As a general principle, non-commutative products may be constructed by composing commutative products with the torsion generator ${\cal F}$.
The star-product  between two functions ($f$ and $g$) can be obtained from the usual pointwise product via the action of the torsion operator, namely,
\begin{equation}
f \star g := \mu \circ {\cal F}^{-1}(f\otimes g) \,
\end{equation}
where $\mu$ is the usual pointwise product between the functions,
$\mu(g \otimes g) =fg$.
The star product, also called Moyal product or Weyl–Groenewold product is an associative, non-commutative 
 phase-space star product.
The twist generator may be decomposed into the form:
\begin{equation}
{\cal F} = f^{\alpha} \otimes f_{\alpha} \, ; \quad \mbox{with} \quad {\cal F}^{-1} = \bar{f}^{\alpha} \otimes \bar{f}_{\alpha} \, ,
\end{equation}
only if
\begin{equation}
f \star g := \bar{f}^{\alpha}(f) \bar{f}_{\alpha}(g) \, .
\end{equation}
Accordingly, a deformed version  a bilinear
map $\mu$, denoted by $\mu_{\star}$
corresponds to
\begin{equation}
\mu_{\star}:= \mu \circ {\cal F}^{-1} \, .
\end{equation}
As an example, the pair of spacetime coordinates, $x$ and $y$, under a torsion deformation, suffer the following
change
\begin{equation}
(\mathbf{x},\mathbf{y}) \mapsto \mu_{\star}(\mathbf{x},\mathbf{y})(x,y)=\mu(\bar{f}^{ \alpha}(\mathbf{x}, \bar{f}^{\alpha}(\mathbf{x}).
\end{equation}
In view of these results, we may speculate that the potential $V(\eta,t)$ acts as a kind of spacetime deformation generator, more precisely, as a spacetime torsion deformation generator, although evidently this assumption requires further investigation. The key point that guides this assumption, seeking to identify the mechanism(s) that drive the acceleration of the Universe, is the breaking of fundamental symmetries of spacetime in the commutative scope.  In particular, a torsion (or twist) deformation of the commutative spacetime geometry into its non-commutative counterpart and where functions and tensors are star-multiplied, spacetime diffeomorphisms may be twisted into non-commutative diffeomorphisms, indicating, even still speculatively, an spontaneous symmetry breaking at small scales.


\subsubsection{Results: non-commutative dynamical equations}

Figure \ref{NCSF2nd} presents typical sample family solutions of the scale factor of the branch-cut Universe, $\eta(t)$, obtained through the combination of second-order in time $\ddot{\eta}(t)$ and first-order in $\dot{\eta}(t)$ non-commutative approach given by equation (\ref{etaNC}). These solutions are for variations in the initial conditions $\eta(0)$ and $\eta'(0)$.
Figure \ref{NCSF1st}  shows 
typical sample family solutions of $\eta(t)$ using the first-order in time $\dot{\eta}$ non-commutative approach given by equation (\ref{NC1steq}), with
variations in the value of $\eta(0)$.
These results demonstrate the significant impact of the
non-commutative formulation on the dynamics of the early Universe's
evolution. The findings reveal that the behavior of the scale factor
$\eta(t)$ is notably different from what is predicted in a commutative
formulation, particularly concerning the acceleration components of
the early Universe.
While the Universe may experience a deceleration phase in its initial
moments of expansion for specific parameter sets, it abruptly
undergoes a drastic acceleration. This acceleration is characterized
by an evolutionary curve that approaches a ninety-degree angle with
the x-axis, nearly parallel to the y-axis. These results align with
one of the key propositions of this work, which aims to understand the
mechanisms driving the acceleration of the Universe's expansion.

 The formal structure of the super-Hamiltonian obtained enables us to associate the new variable $\eta(t)$ with the scale factor of branch-cut gravitation, $\ln^{-1}[\beta(t)]$. However, it is evident that, as a result of the imposed variables transformations, this incorporates elements characterizing a non-commutative algebra at a fundamental level. And given its non-commutative nature, this algebraic structure allows us to identify the $\xi(t)$ variable as the dual quantum counterpart of $\eta(t)$, with both variables scanning reciprocal quantum complex spaces. It is important to note that, even though
 the new variables $\eta$ and $\xi$ are treated as linearly independent dual variables, they carry new effective identities when compared to the original bare variables. This is due to the nature of the coordinate transformations, which imbue these variables with underlying mutual and complementary properties and identities.

In classical  formulations, the statistical distribution of matter is typically assumed to be homogeneous and is determined by the dynamics of the Hamiltonian. However, systems with higher complexity can exhibit topological constraints that  may be independent of the Hamiltonian and can affect the shape of statistical matter distribution functions~\cite{Sato}. In general relativity, the curvature of spacetime arises as a source of heterogeneity in the statistical distribution of matter~\cite{Sato}. In our formulation we can identify an additional source of statistical distribution heterogeneity, specifically, the non-commutative algebraic representation of branch-cut gravity combined with the Ho\v{r}ava-Lifshitz formulation.  This formulation leads to the realization of a potential that incorporates different matter composition contributions. These
contributions are represented by algebraic terms dependent on $\eta(t)$
and appear in different orders. Their formal structure, apart from the derivative terms, is similar to  the dependence of the Ho\v{r}ava-Lifshitz formulation on the scalar curvature of the Universe. The implications of this formulation suggest a significant impact not only on the curvature of spacetime and the statistical distribution of matter but also on the evolutionary dynamics of the early Universe, which drives its acceleration.

In revisiting the cosmological implications of the results presented in 
figure~\ref{VNC1stTorsion}, which  indicate the presence of a torsion or twist deformation of spacetime in the context of the branch-cut non-commutative algebra, we encounter an intriguing outcome. The color palette associated with the contour plots unveils a remarkable pattern. There appears to be a progressive transition in color intensity, ranging from darker colors (representing lower potential intensity values) to lighter colors (indicating higher potential intensity values) when examining the contour plots of the potential $v(\eta,\tau)$.
When we relate these colors to the regions corresponding to the 3D graph of the potential, a significant revelation emerges. The point where $t=0$, the boundary region between the present universe and its mirrored counterpart, aligns with the point of maximum torsion in the potential $v(\eta, \tau)$. This suggests that the accelerated expansion of our universe may be the result of a folded memory
shared by both universes. This concept implies that spacetime possesses a fold-memory (or twist-memory or torsion-memory), which, when subjected to a twist in the mirrored counterpart and subsequently regulated and shaped by this fold-memory, spontaneously unfolds in response to ``external'' stimuli. This unfolding process propels the acceleration of our universe's expansion.
\section{Gravitational Waves from the Primordial Universe}
Similarly to the CMB, the gravitational wave background is expected to propagate as a homogeneous noise, although it is supposedly too weak to be measured by current detection systems.
When investigating relic gravitational waves and their production in the first evolutionary stages of the branch-cut universe, we adopt a proposition that follows the conceptual approach of general relativity, the application of perturbation theory to the metric tensor that describes the geometry of space-time in the weak field domain, assuming a linearized gravity BCG approach. 

In the following, we develop a preliminary study involving the non-commutative branch-cut cosmology and relic gravitational waves from the early Universe.

\subsection{Linearized BCG equations}

Linearized branching cosmology, similarly to linearized general relativity, describes the dynamics of a slightly perturbed gravitational field in such a way as to describe the dynamics of gravitational waves as small ripples in flat spacetime. Therefore, we consider a metric tensor,  analytically continued to the complex plane, decomposed into the Minkowski metric and a small perturbation, in the form
\begin{equation}
    g^{[\rm ac]} = {\zeta}^{[\rm ac]}_{\mu \nu} + {h}^{[\rm ac]}_{\mu \nu} \, , \quad \mbox{with} \quad |{h}^{[\rm ac]}_{\mu \nu}| << 1\, .
    \label{acmt}
\end{equation}
In this expression, the condition $|{h}^{[\rm ac]}_{\mu \nu}| << 1$ implies that higher orders of ${h}^{[\rm ac]}_{\mu \nu}$ are omitted.

In branch-cut gravity (BCG), an ontological domain extended version of General Relativity analytically continued to the complex plane~\cite{Bodmann2023a,Manders}, the conditions imposed for its formulation imply that the equations that describe the branching Universe may be cast in the form
\begin{equation}
    G^{[\rm ac]}_{\mu \nu} = R^{[\rm ac]}_{\mu \nu} - \frac{1}{2} g^{[\rm ac]}_{\mu \nu}R^{[\rm ac]} = \frac{8 \pi G}{c^4} T^{[\rm ac]}_{\mu \nu} \, .
\end{equation}
The conceptual procedure for such analytical continuation, as previously mentioned, is based on the mathematical augmentation technique and the notions of closure and existential completeness ~\cite{Manders}.

Similarly to General Relativity, the branch-cut gravity equations relate the spacetime geometry, encoded in the metric $g^{[\rm ac]}$, to matter described by the energy-momentum tensor $T^{[\rm ac]}_{\mu \nu}$ analytically continued to the complex plane. The Ricci tensor $R^{[\rm ac]}_{\mu \nu}$ and Ricci scalar $R^{[\rm ac]}$ for the linearized theory are computed following the usual scheme, starting from the analytically continued Christoffel symbol which will lead to an analytically continued Riemann curvature tensor.
From equation (\ref{acmt}), following similar mathematical procedures of the standard theory, we obtain for the
analytically continued linearized Christoffel symbol 
\begin{eqnarray}
    \Gamma^{{[\rm ac]}\rho}_{\mu \nu} & = &  \frac{1}{2} g^{{[\rm ac]}\rho \sigma} \Bigl[\partial_{\mu} g^{[\rm ac]}_{\nu \sigma} + \partial_{\nu} g^{[\rm ac]}_{\mu \sigma} - \partial_{\sigma} g^{[\rm ac]}_{\mu \nu}\Bigr] \nonumber \\
    & = &  \frac{1}{2} \zeta^{{[\rm ac]}\rho \sigma} \Bigl[\partial_{\mu} h^{[\rm ac]}_{\nu \sigma} + \partial_{\nu} h^{[\rm ac]}_{\mu \sigma} - \partial_{\sigma} h^{[\rm ac]}_{\mu \nu}\Bigr] \nonumber \\
   & + & {\cal O}({h^{[\rm ac]}}^2) \, .
\end{eqnarray}

This expression allows to define the analytically continued 
Riemann curvature tensor in the form
\begin{eqnarray}
      R^{{[\rm ac]}\mu}_{\nu \sigma \rho} & = & \partial_{\sigma} \Gamma^{{[\rm ac]} \mu}_{\nu \rho} - \partial_{\rho} \Gamma^{{[\rm ac]} \mu}_{\nu \sigma} + \Gamma^{{[\rm ac]} \mu}_{\sigma \lambda}\Gamma^{{[\rm ac]} \lambda}_{\nu \rho}  
      + \Gamma^{{[\rm ac]} \mu}_{\rho \lambda}\Gamma^{{[\rm ac]} \lambda}_{\nu \sigma}  \, , \nonumber \\
      & = & \frac{1}{2} \zeta^{{[\rm ac]}\mu \lambda} \Bigl[\partial_{\sigma} \partial_{\nu}  h^{[\rm ac]}_{\rho \lambda} - \partial_{\sigma}\partial_{\rho} h^{[\rm ac]}_{\nu \lambda} - \partial_{\sigma} \partial_{\lambda} h^{[\rm ac]}_{\nu \rho}
      - (\sigma \leftrightarrow \rho)
      \Bigr] + {\cal O}({h^{[\rm ac]}}^2) \, .
\end{eqnarray}
It is crucial to note that in the upper expression, the second-order terms involving $\Gamma$ do not contribute to the first-order branch-cut equations.

Based on these expressions, the Ricci tensor, analytically continued to the complex plane, may be cast as:
\begin{eqnarray}
    R^{[\rm ac]}_{\mu \nu} & = &  R^{{[\rm ac]}\rho}_{\mu \rho \nu}  
    \nonumber \\
  &  = & \frac{1}{2}\Bigl[\partial_{\rho} \partial_{\mu}  h^{{[\rm ac]}\rho}_{\nu} + \partial_{\rho}\partial_{\nu} h^{{[\rm ac]}\rho}_{\mu} - \partial_{\mu} \partial_{\nu} h^{[\rm ac]}
      - \Box h_{\mu \nu}
      \Bigr]
       +  {\cal O}({h^{[\rm ac]}}^2) \, .
\end{eqnarray}

Similarly, the Ricci scalar is
\begin{equation}
    R^{[\rm ac]} = g^{{[\rm ac]}\mu \nu} R^{[\rm ac]}_{\mu \nu } = \partial_{\mu} \partial_{\nu} h^{{[\rm ac]}\mu \nu} - \Box h^{[\rm ac]}  + {\cal O}({h^{[\rm ac]}}^2) \, .
\end{equation}
In this expression, 
 $h=h^{\rho}_{\rho}$ is the trace of the metric perturbation, $\partial^{\alpha} \equiv \eta^{\alpha \beta} \delta_{\beta}$ and $\Box \equiv \partial_{\mu} \partial^{\mu}$. 

From these expressions, we can construct the Einstein tensor, again to first order in the metric perturbation, as
\begin{equation}
    G^{[\rm ac]}_{\mu \nu } = - \frac{1}{2} \Bigl[\Box h^{[\rm ac]}_{\mu \nu} +  \zeta^{[\rm ac]}_{\mu\nu} \partial^{\rho} \partial^{\sigma} h^{[\rm ac]}_{\rho \sigma} - \zeta^{[\rm ac]}_{\mu\nu}\Box h^{[\rm ac]} 
    - \partial^{\rho} \partial_{\nu} h^{[\rm ac]}_{\mu \rho} - \partial_{\rho} \partial_{\mu} h^{{[\rm ac]}\rho}_{\nu}
    + \partial_{\nu} \partial_{\mu} h \Bigr] + {\cal O}({h^{[\rm ac]}}^2) \, .
\end{equation}
As usually adopted  in standard cosmology, the above equation may be simplified by inserting the trace reversed quantity $\bar{h}^{[\rm ac]}_{\mu \nu}  \equiv h^{[\rm ac]}_{\mu \nu} - \frac{1}{2} \zeta^{[\rm ac]}_{\mu\nu} h$: 
\begin{equation}
    G^{[\rm ac]}_{\mu \nu } = - \frac{1}{2} \Bigl[\Box \bar{h}^{[\rm ac]}_{\mu \nu} +  \zeta^{[\rm ac]}_{\mu\nu} \partial^{\rho} \partial^{\sigma} \bar{h}^{[\rm ac]}_{\rho \sigma} 
    - \partial^{\rho} \partial_{\nu} \bar{h}^{[\rm ac]}_{\mu \rho} - \partial^{\rho} \partial_{\mu} \bar{h}^{{[\rm ac]}}_{\nu \rho}
 \Bigr] + {\cal O}({h^{[\rm ac]}}^2) \, .
\end{equation}

\subsection{Stochastic gravitational wave background}

Speculations concerning stochastic gravitational wave backgrounds (SGWBs)
involve their definition as the superposition of relic gravitational waves with different wave numbers $k$, encompassing variations in both magnitude and direction. These SGWBs are envisaged to exhibit characteristics such as
isotropy, lack of polarization, and Gaussian distribution. They could originate from diverse sources, encompassing astrophysical and cosmic phenomena like inflation, the presence of primordial black holes, various primordial cosmic seeds, cosmic strings, and phase transitions. 
These speculations draw parallels between SGWBs and the cosmic microwave background (CMB) radiation originating from the primordial electromagnetic spectrum. The key distinction between these two types of primordial emissions lies in the fact that SGWBs have the potential to provide insight into earlier evolutionary stages of the Universe that predate the recombination phase characterized by the decoupling of matter and radiation. This is because gravitational waves can travel freely through a primitive hot plasma, which is not transparent to photons. These considerations are particularly important for BCG, as one of its scenarios involves a violent transition between two phases of the Universe: a contracting phase preceding the conventional concept of a primordial singularity, and a subsequent expanding phase. The region of transition is mediated by a Riemannian foliation structure. 

We define the non-commutative gravitation branch-cut metric in the form
\begin{equation}
    ds^{{[\rm ac]}2} \equiv - dt^2 + \eta^2(t) dx^i dx_i = - \eta^2(\tau) \bigl(d\tau^2 - g^{{[\rm ac]}}_{ij} dx^i dx^j \bigr) \, ,
\end{equation}
where $d \tau = dt/\eta(t)$ defines the conformal time. 
Expanding this metric around a flat homogeneous cosmological background $ g^{{[\rm ac]}}_{ij} = \delta^{{[\rm ac]}}_{ij} + {h}^{{[\rm ac]}}_{ij}$, from the previous equations, the linearized field equations for the implicit dependence of the scale factor $\eta(\tau)$ on the conformal $\tau$,  may be expressed as \begin{equation}
\Box \bar{h}^{[\rm ac]}_{ij}(\mathbf{x},\tau) - 2\frac{ \eta^{'}(\tau)}{\eta(\tau)}\bar{h}^{[\rm ac]'}_{ij}(\mathbf{x},\tau) = 16 \pi GT^{[\rm ac]}_{ij} \, , \label{1stGWeq}
\end{equation} 
with the symbol $'$ denoting derivatives with respect to the conformal time $\tau$.

Following standard procedures, we introduce a Fourier transformation of this expression and we define $\tilde{h}^{[\rm ac]}_{\lambda} ~\equiv~\eta(\tau)~h_{\lambda}$,
so
the field equation (\ref{1stGWeq}) may be recast in the form
\begin{equation}
 {h}''_{\lambda}(\mathbf{k},\tau)+ \Bigl(k^{2}- \frac{\eta''(\tau)}{\eta(\tau)}\Bigr){h}_{\lambda}(\mathbf{k},\tau)   =  16 \pi G\eta(\tau)T_{\lambda}(\mathbf{k},\tau) \, , \label{1stGWeq2}
\end{equation}
where $k$ represents the co-moving wave number,  $\lambda = +, \times$ denotes the two polarization modes of gravitational waves, and we have set $G=1$ to simplify the notation.

This equation can be simplified by considering two main cases: 1. 
The sub-horizon case, characterized by the condition $k^2 >> (\eta H)^2$.
2. The super-horizon case, defined by 
the condition $k^2 << (\eta H)^2$, where $H \equiv {\eta'} /\eta$.
Below we briefly discuss the implications of theses approximations to establish
a connection with the standard formulation. 

\subsubsection{Sub-horizon condition with no source}
\begin{figure*}[htpb]
\centering
\includegraphics[scale=0.25]{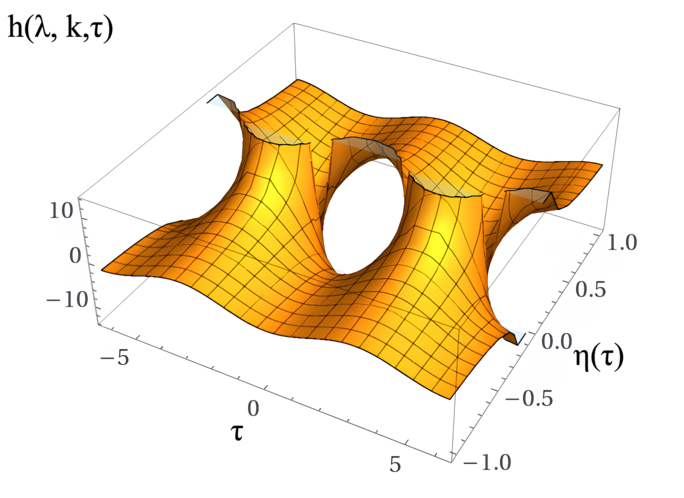} \hspace{0.5cm}
\includegraphics[scale=0.2]{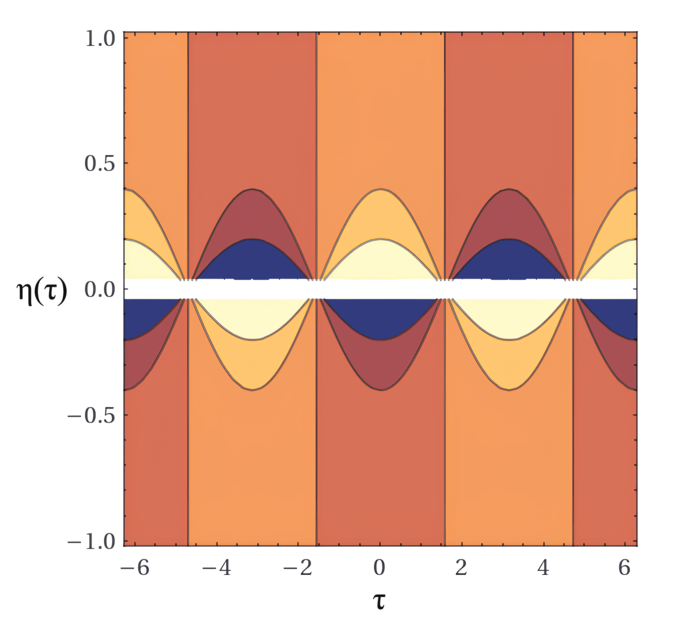} 
\includegraphics[scale=0.25]{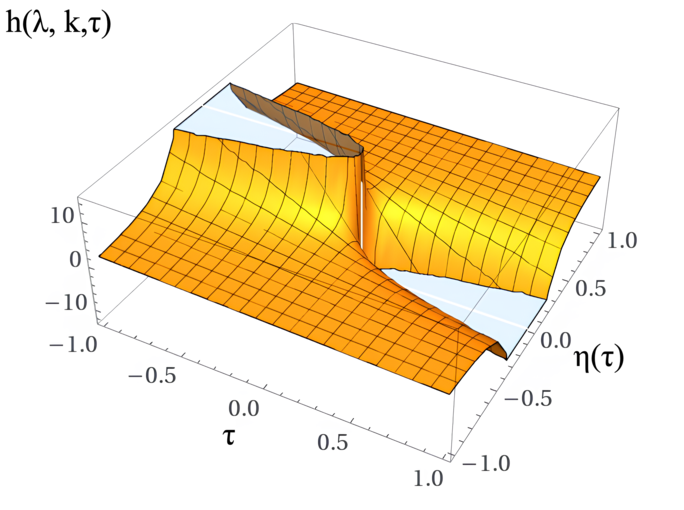} \hspace{0.5cm}
\includegraphics[scale=0.2]{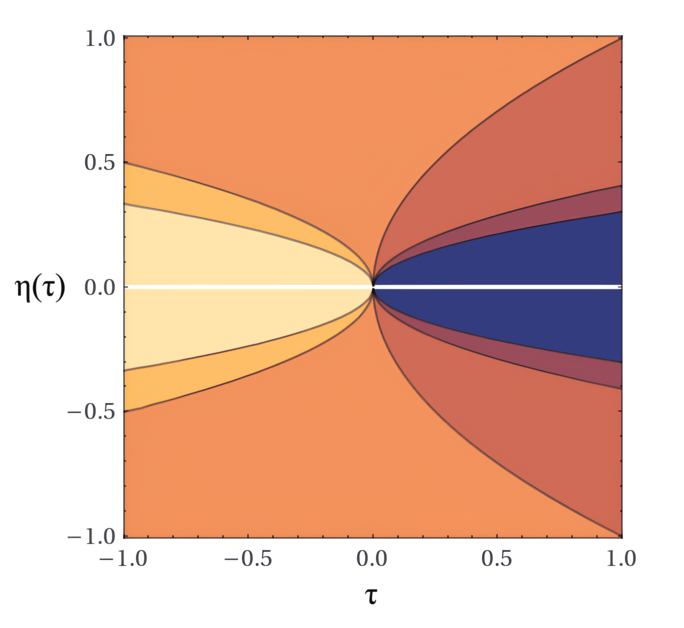}
\caption{Typical 3D solutions and contour plots of 
${h}(\lambda,\mathbf{k},\tau)$ as a function of $\eta(\tau)$ and $\tau$ for a fixed values of $k$ and the dimensional constants.  The
results are presented in a pair of figures above, in accordance
with equation (\ref{Sub}), and in another pair of figures below, in line with equation (\ref{h**}).}
\label{sub}
\end{figure*} 
With respect to the term $T_{\lambda}$ we adopt the following representation, in tune with the conventional standard  Einstein equations:
\begin{equation}
T^{[\rm ac]}_{\lambda}(\mathbf{k},\tau) \quad \rightarrow \quad p_{\lambda}(\eta(\tau)) =  -\eta(\tau) \eta'(\tau) \, , \label{T}
\end{equation}

The sub-horizon condition (sub-Hubble scales) involves neglecting the term $\eta''/ \eta$ in comparison to $k^2$, effectively assuming  $k^2 >> (\eta H)^2$. This condition leads to the following equation:
\begin{eqnarray}
{h''}_{\lambda}(\mathbf{k},\tau) + \Bigl( k^2  - \frac{\eta''(\tau)}{\eta} \Bigr) {h}_{\lambda}(\mathbf{k},\tau) = -16 \pi \eta^3 \frac{{\eta'}}{\eta}  \nonumber \\
 \rightarrow 
 {h''}_{\lambda}(\mathbf{k},\tau) + k^2 {h}_{\lambda}(\mathbf{k},\tau) = -16 \pi \eta^3 \frac{{\eta'}}{\eta}  \, . \label{shc}
\end{eqnarray}
In the standard formulation,
when considering a generic scale factor obeying a power law such as
$\eta(\tau) = \eta_n \tau ^n$~\cite{Caprini}, which covers the cases of radiation domination ($n = 1$), matter domination ($n = 2$), as well as De Sitter inflation ($n = -1$), 
the general solution of equation (\ref{1stGWeq}), with the source term set to zero, is given by~\cite{Polyanin,Caprini}
\begin{equation}
     {h}(\lambda,\mathbf{k},\tau) = \frac{A_{\lambda}(\mathbf{k})}{\eta(\tau)} e^{i k \tau} + \frac{B_{\lambda}(\mathbf{k})}{\eta(\tau)} e^{-i k \tau}, \quad \mbox{for} \quad  k >> {\cal H} \, , \label{Sub}
\end{equation}
with $A_{\lambda}(\mathbf{k})$ and $B_{\lambda}(\mathbf{k})$ denoting dimensional constants to be determined from the initial conditions. 

Figure \ref{sub} shows typical 3D solutions and contour plots of 
${h}(\lambda,\mathbf{k},\tau)$ as a function of $\eta(\tau)$ and $\tau$ for fixed values of $k$ and the dimensional constants, based on equations (\ref{Sub}) for the left pair of figures and equation (\ref{h**}) for the right pair of figures.
The results presented in these figures depict systematic variations in the spacetime configuration of primordial matter due to the non-commutative formulation. These variations suggest a unique dynamics in the production of relic gravitational waves. 
Additionally, the color palette associated with the distribution of matter in these figures indicates greater intensity in the region prior to the expansion phase, which has predictable consequences for the acceleration of the Universe and the production of gravitational waves.
It is worth noting that the solutions presented in the right figures are expected to remain constant in time for modes outside the Hubble radius~\cite{Caprini}. However, these expectations may not align with the format of the present equation as well as with the evolutionary behavior of the dynamical equations.


\subsubsection{Super-horizon condition with no source}
\begin{figure*}[htpb]
\centering
\includegraphics[scale=0.2]{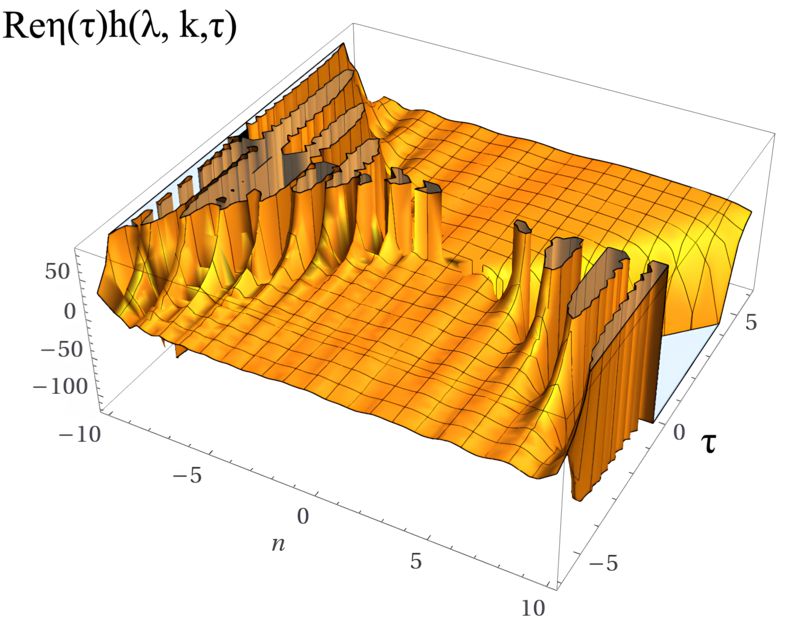}\hspace{0.5cm}
\includegraphics[scale=0.2]{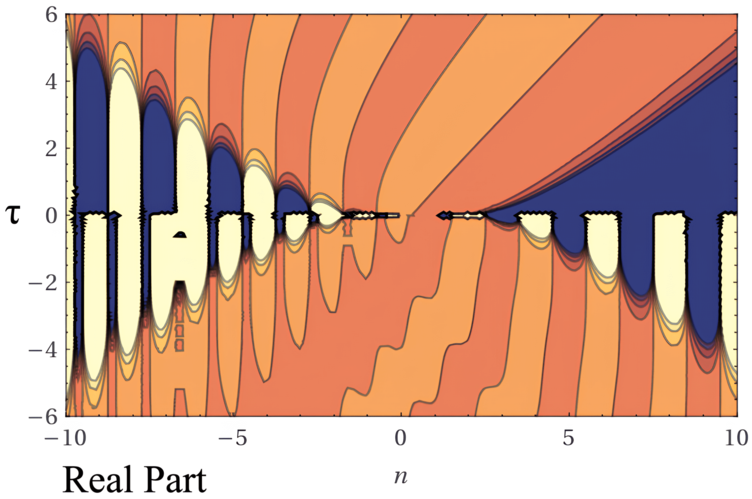} 
\includegraphics[scale=0.2]{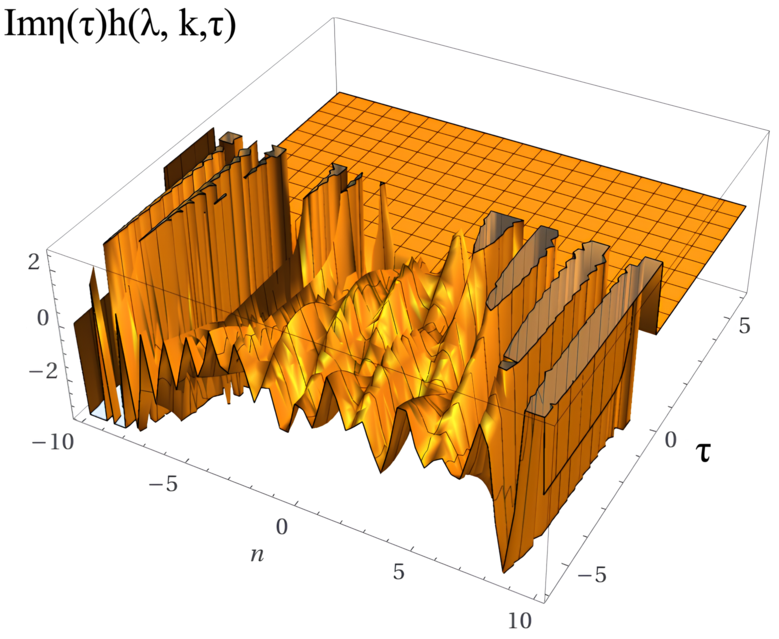} \hspace{0.5cm}
\includegraphics[scale=0.2]{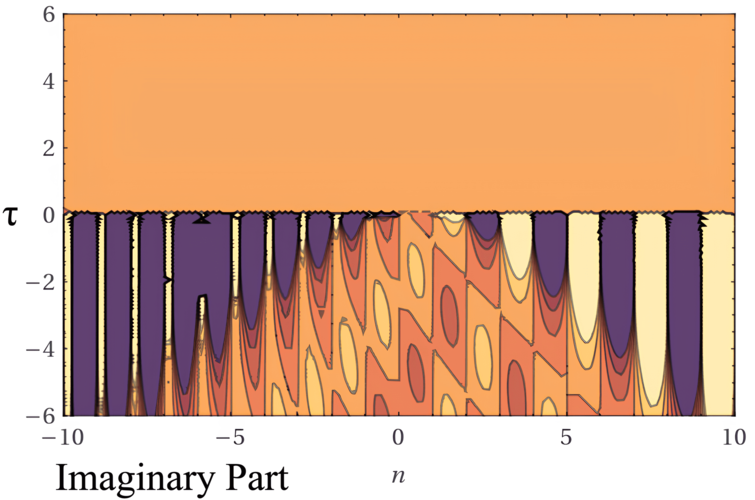} 
\caption{Real and imaginary parts of 3D solutions and contour plots of the product  $\eta(\tau){h}(\lambda,\mathbf{k},\tau)$ as a function of $\tau$ for different values of $n$, according to equation (\ref{finalGWs}).} \label{etay}
\end{figure*} 
\begin{figure*}[htpb]
\centering
\includegraphics[scale=0.2]{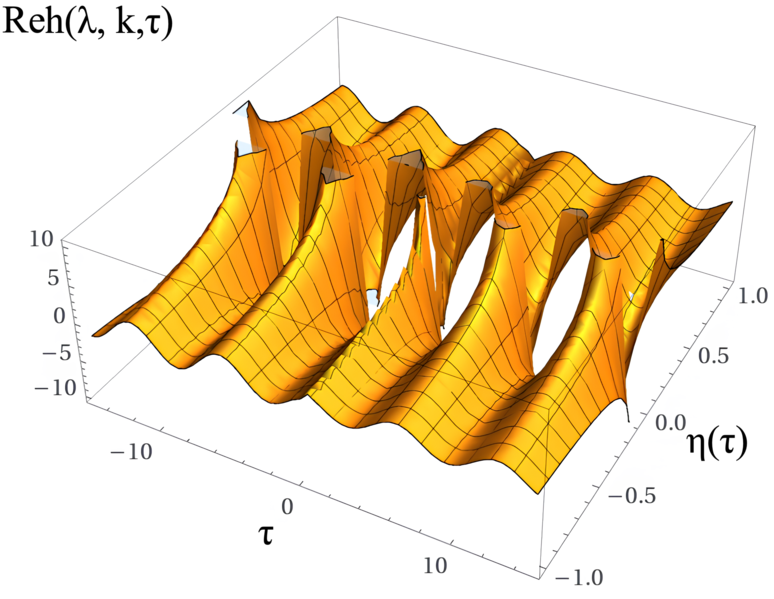}  \hspace{0.5cm}
\includegraphics[scale=0.04]{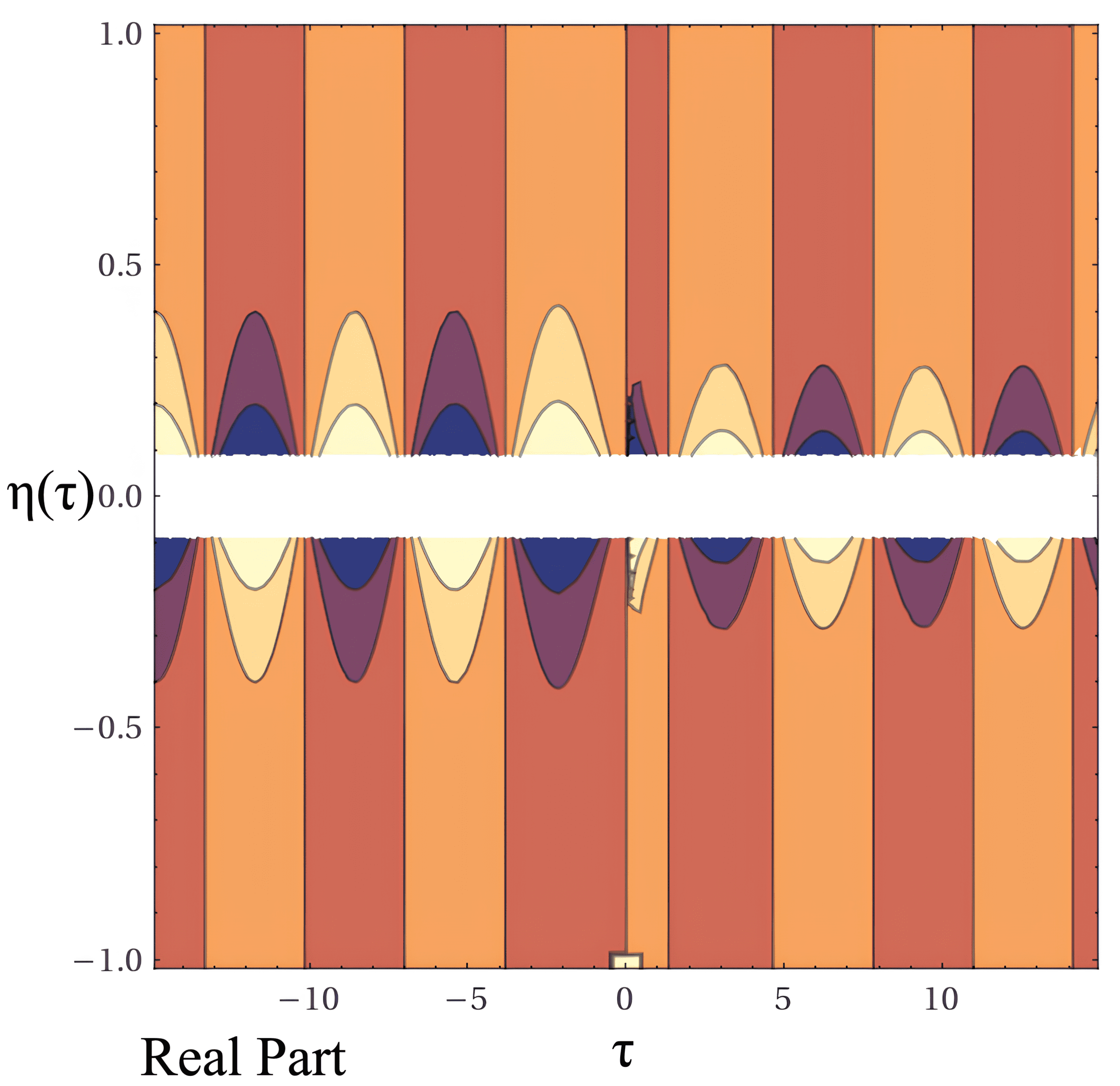}  
\includegraphics[scale=0.2]{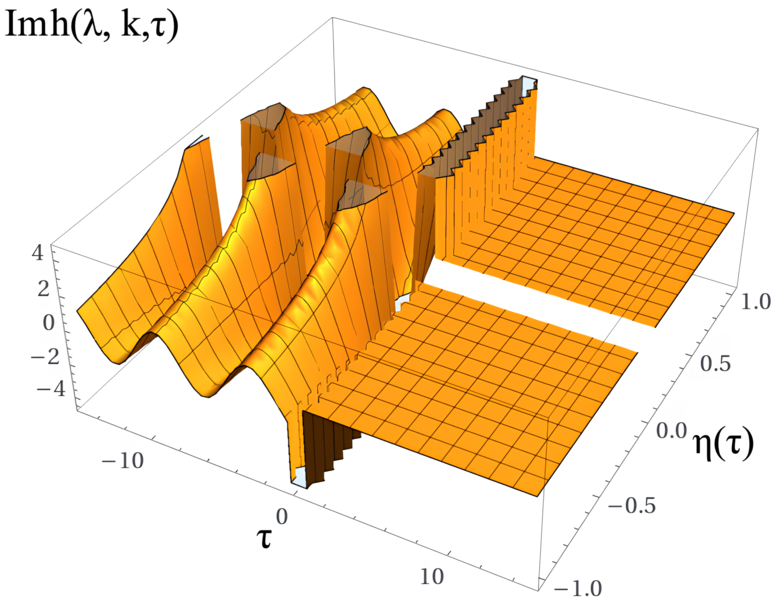}  \hspace{0.5cm}
\includegraphics[scale=0.15]{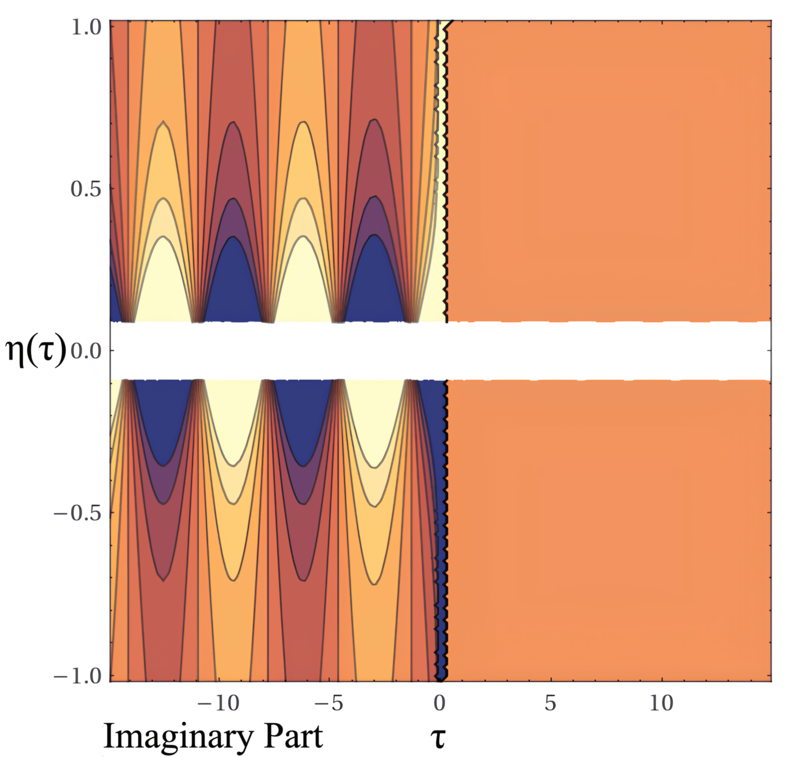} 
\caption{Real and imaginary parts of 3D solutions and contour plots of equation (\ref{finalGWs}). } \label{y*}
\end{figure*}
The super-horizon condition, which corresponds to assuming $k^2 << (\eta H)^2$,  results in the following equation:
\begin{eqnarray}
{h''}_{\lambda}(\mathbf{k},\tau) + \Bigl( k^2  - \frac{\eta''(\tau)}{\eta} \Bigr) {h}_{\lambda}(\mathbf{k},\tau) = -16 \pi \eta^3 \frac{{\eta'}}{\eta}  \nonumber \\
 \rightarrow 
 {h''}_{\lambda}(\mathbf{k},\tau) - \frac{\eta''(\tau)}{\eta}\tilde{h}_{\lambda}(\mathbf{k},\tau) = -16 \pi \eta^3 \frac{{\eta'}}{\eta} \, . \label{Main}
\end{eqnarray}

Assuming, as in the standard formulation, the power-law time dependence $\eta(\tau) = \eta_n \tau^n$~\cite{Caprini} and setting the source to zero, the typical solutions of equation (\ref{1stGWeq}) are as follows:
\begin{equation}
     {h}(\lambda,\mathbf{k},\tau) = A_{\lambda}(\mathbf{k}) + B_{\lambda}(\mathbf{k}) \int_0^{\tau }  \frac{d \tau}{\eta^2(\tau)}  \quad \mbox{for} \quad  k << {\cal H} \, . \label{h**}
\end{equation}


\subsubsection{General solution with no source}


Indeed, due to the systematic contribution at the core of the differential equation from various terms related to the potential $V(\eta, \tau)$, we can make a good approximation by setting the source term to zero. This allows us to express the general solution of equation (\ref{1stGWeq}) as
\begin{equation}
 {h}(\lambda,\mathbf{k},\tau) = \frac{A_{\lambda}(\mathbf{k})}{\eta(\tau)} \tau J_{n-1}(k\tau) + \frac{B_{\lambda}(\mathbf{k})}{\eta(\tau)} \tau Y_{n-1}(k\tau) \, , \label{finalGWs}
\end{equation}
where $J_{n-1}(k\tau)$ and $Y_{n-1}(k\tau)$ represent spherical Bessel functions. 

Figures \ref{etay} and \ref{y*} display typical sample solutions of equation (\ref{finalGWs}). These figures provide insight into the behavior of
$\eta(\tau){h}(\lambda,\mathbf{k},\tau)$ and
${h}(\lambda,\mathbf{k},\tau)$ under arbitrary parameter choices,
focusing on their structural behavior.
In figure \ref{etay}, we
observe the behavior of $\eta(\tau){h}(\lambda,\mathbf{k},\tau)$ as a
function of $\tau$ for different values of $n$. In figure \ref{y*}, we
examine the behavior of ${h}(\lambda,\mathbf{k},\tau)$ as a function
of $\eta(\tau)$ for a fixed value of $n$. These results suggest a
turbulent and violent behavior of relic gravitational waves. It is
important to note that the stochastic gravitational wave background,
characterized by a homogeneous distribution, refers not to their
primordial distribution but to their collective behavior at
present. According to the authors of ref.~\cite{Caprini}, the integral
in equation (\ref{h**}) decays as the Universe expands, indicating that
gravitational waves are ``frozen'' outside the Hubble horizon, a
mechanism similar to what occurs during inflation. Once they re-enter
the horizon, tensor perturbations become sub-horizon modes again.


\subsection{General solution in the presence of a source: super-horizon condition}
\begin{figure*}[htpb]
\centering
\includegraphics[scale=0.18]{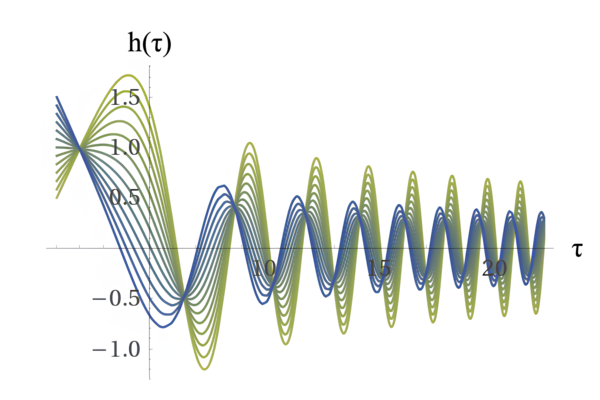} \hspace{-0.35cm}
\includegraphics[scale=0.18]{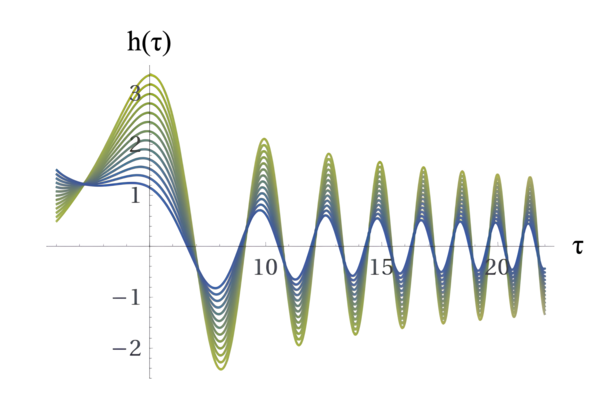}\hspace{-0.25cm}
\includegraphics[scale=0.18]{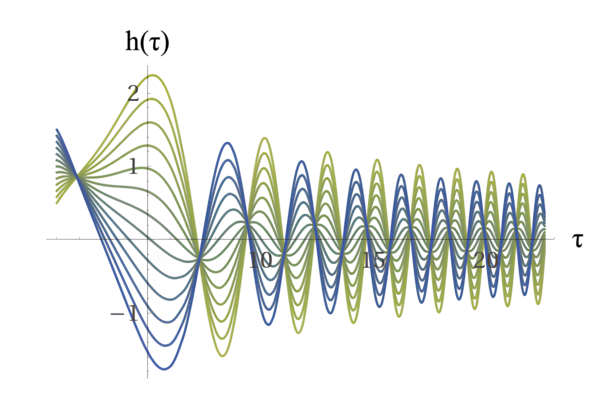} \hspace{-0.35cm}
\includegraphics[scale=0.18]{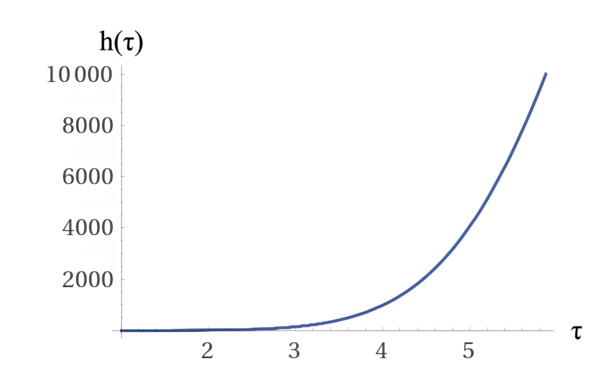} 
\caption{The curves show partial solutions of equation (\ref{Mainmain}) assuming $\eta(\tau) = \eta_n \tau^n$, with $n=1$.
The figures illustrate the inclusion of terms in a stepwise
manner, starting with the quadratic term and systematically adding
each of the following terms separately from the left side of the
equation. Finally, the last figure demonstrates the quadratic term
together with the source term from the right side of the equation.
The parameters are: $g_k = 1$; $g_q = 0.7$; $g_{\Lambda} = 0.333$; $g_r = 0.4$; $g_s =  0.03$; $\alpha = 1/2$; $\gamma = 1$; ${\cal C} = 1$; $n =1$.} \label{sol}
\end{figure*}

\begin{figure*}[htpb]
\centering
\includegraphics[scale=0.18]{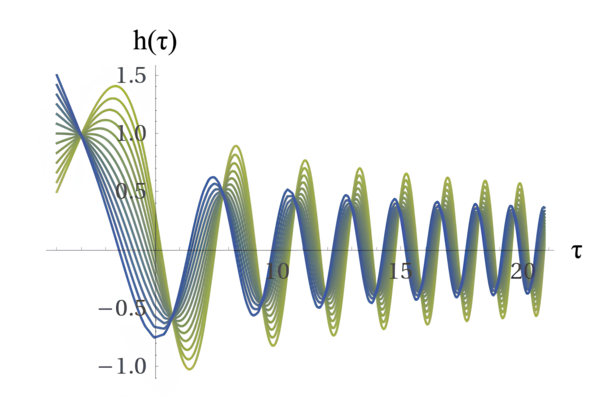} \hspace{-0.35cm}
\includegraphics[scale=0.18]{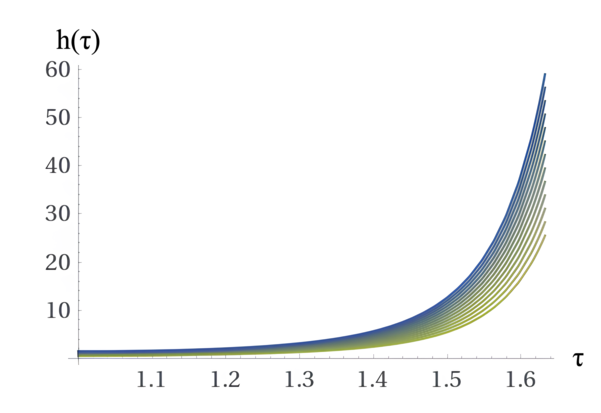} \hspace{-0.25cm}
\includegraphics[scale=0.18]{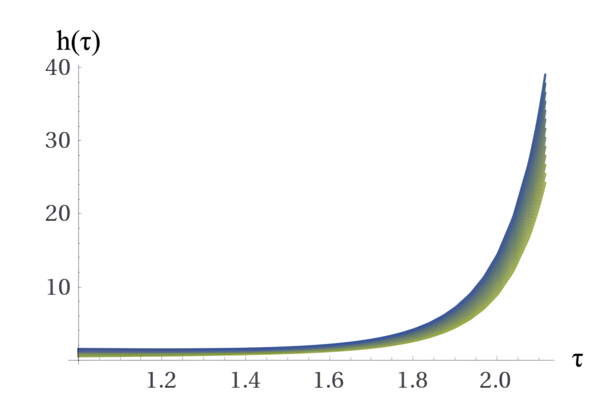} \hspace{-0.35cm}
\includegraphics[scale=0.18]{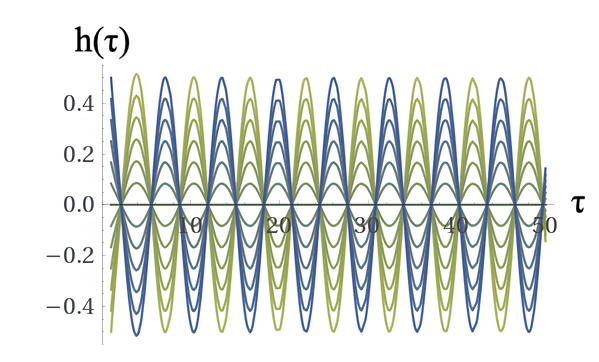}
\caption{The pair of figures on the left displays solutions of  
 equation (\ref{Mainmainmain}) corresponding to $n=2$. The first figure on the right shows  solutions for $n=-1$. 
 The last figure presents a sample solution of equation (\ref{Mainmain}).
The values of the parameters are: $g_k = 1$; $g_q = 0.7$; $g_{\Lambda} = 0.333$; $g_r = 0.4$; $g_s =  0.03$; $\alpha = 1/2$; $\gamma = 1$; ${\cal C} = 1$; $n =-1$.
} \label{sol*}
\end{figure*}

For practical reasons during the calculations performed below, we have organized the following set of expressions based on equations (\ref{etaNC}), (\ref{VetatNew}), (\ref{NC1steq}), and (\ref{vetat}):
\begin{eqnarray}
            \frac{{\eta}''}{\eta} &=& - \frac{1}{2} \Bigl(\frac{\eta'}{\eta}\Bigr)^2 - \frac{3}{2}\frac{\alpha \gamma {\eta'}}{\eta^{3 \alpha +2}} - \frac{1}{2\eta^2}V(\eta,\tau) \, ;
    \label{etaNC2}
\end{eqnarray}
with the potential $V(\eta,\tau)$ defined as 
\begin{equation}
    V(\eta,\tau) \! = \!  g_k + 2 g_q \eta - 3 g_{\Lambda} \eta^2 + \frac{g_r}{\eta^2} + 3 \frac{g_s}{\eta^4} + \frac{\alpha \bigl(3 \alpha - 1 \bigr)}{\eta^{3 \alpha}} - \frac{3 \alpha^2 \tau}{\eta^{6 \alpha -1}}   \, . \label{VetaNew2}\!
\end{equation}
These equations describe the temporal evolution of the scale factor $\eta(t)$ in the non-commutative branch-cut formulation with mixed first and second-order time-derivatives of the scale factor $\eta(t)$.
Additionally, we have
\begin{equation}
    \frac{\eta'}{\eta} = - \frac{1}{\eta}v(\eta,\tau) \, ,\label{NC1steq2}
\end{equation}
with the potential $v(\eta,\tau)$ defined as
\begin{equation}
  \frac{1}{\eta} v(\eta,\tau) =  \frac{1}{3\eta^2} \Biggl\{- \frac{3 \alpha \gamma}{\bigl(3\alpha-1\bigr)\eta^{3\alpha - 1}} + \Biggl( g_k + 2 g_q \eta - 3 g_{\Lambda} \eta^2 + \frac{g_r}{\eta^2} + 3 \frac{g_s}{\eta^4} + \frac{\alpha \bigl(3 \alpha - 1 \bigr)}{\eta^{3 \alpha}} \Biggr) \tau
-  \frac{3}{2}\frac{ \alpha^2 \tau^2}{\eta^{6 \alpha  -1}} \Biggr\} \, . \label{VetatNew3}
\end{equation}
These equations describe the temporal evolution of the scale factor, $\eta(t)$, within the non-commutative branch-cut formulation, accounting for a first-order dependence on the time-derivative of $\eta(t)$.
By combining equations (\ref{etaNC2})
through (\ref{VetatNew3}), we can express equation (\ref{Main}) as
\begin{eqnarray}
         {h}^{''}(\tau) & \!\!+ \!\!& \Biggl\{\!\frac{1}{2} \frac{1}{9\eta^4}\! \Biggl(\!- \frac{3 \alpha \gamma}{\bigl(3\alpha-1\bigr)\eta^{3\alpha - 1}}\! +\!\Biggl[ g_k \!+ 2 g_q \eta \! - 3 g_{\Lambda} \eta^2\! + \frac{g_r}{\eta^2}\! + 3 \frac{g_s}{\eta^4}\! + \frac{\alpha \bigl(3 \alpha - 1 \bigr)}{\eta^{3 \alpha}} \!\Biggr] \!\tau \! - \! \frac{3}{2}\frac{ \alpha^2 \tau^2}{\eta^{6 \alpha  -1}} 
        \! \Biggr)^{\!\!2} \nonumber \\ 
         & \!\!\!\!\!\!\!+\!\!\!\! \!\!& \frac{3}{2}\frac{\alpha \gamma}{\eta^{3 \alpha + 2}} \!\Biggl(\!- \frac{3 \alpha \gamma}{\bigl(3\alpha-1\bigr)\eta^{3\alpha - 1}} \!+ \!\Biggl[ g_k \!+ 2 g_q \eta\! - 3 g_{\Lambda} \eta^2 \!+ \frac{g_r}{\eta^2} \!+ 3 \frac{g_s}{\eta^4}\! + \frac{\alpha \bigl(3 \alpha - 1 \bigr)}{\eta^{3 \alpha}} \!\Biggr] \!\tau \! - \! \frac{3}{2}\frac{ \alpha^2 \tau^2}{\eta^{6 \alpha  -1}} \Biggr) \nonumber \\
         &\!\!\!\!\! + \!\!\!\!& \frac{1}{2\eta^2} \Biggl(g_k + 2 g_q \eta - 3 g_{\Lambda} \eta^2 + \frac{g_r}{\eta^2} + 3 \frac{g_s}{\eta^4} + \frac{\alpha \bigl(3 \alpha - 1 \bigr)}{\eta^{3 \alpha}}  - \frac{3 \alpha^2\tau}{\eta^{6 \alpha -1}} \Biggr)  \Biggr\} {h}(\tau) \nonumber \\
         && \!\!\!\!\!\! = \!  -\frac{16}{3} \pi \eta \Biggl(\! - \frac{3 \alpha \gamma}{\bigl(3\alpha-1\bigr)\eta^{3\alpha - 1}} +\!\Biggl[ g_k \!+ 2 g_q \eta \!- 3 g_{\Lambda} \eta^2 \!+ \frac{g_r}{\eta^2}\! + 3 \frac{g_s}{\eta^4}\! + \frac{\alpha \bigl(3 \alpha - 1 \bigr)}{\eta^{3 \alpha}} \!\Biggr] \!\tau \!
  -\frac{3}{2}\frac{ \alpha^2 \tau^2}{\eta^{6 \alpha  -1}} \!\Biggr).   \nonumber\\
  \label{Mainmain}  
\end{eqnarray}
The curves in figure \ref{sol} display partial solutions of equation (\ref{Mainmain}) with the assumption $\eta(\tau) = \eta_n \tau^n$, with $n=1$.
Starting from the left, we first consider only the presence of the
quadratic term, and then systematically add each of the other terms individually, which are contained on the left side of the equation. In the final figure, we include both the quadratic term and the source term from the right side of the equation. The results suggest that, for $n=1$, it is a reasonable approximation to retain the dominant quadratic term and the source term while disregarding the other.
\begin{eqnarray}
         {h}^{''}(\tau) & \! + \!\! & \frac{1}{2} \frac{1}{9\eta^4} \!\Biggl(\!- \frac{3 \alpha \gamma}{\bigl(3\alpha \!-\!1\bigr)\eta^{3\alpha - 1}}\! +\!\!\Biggl[g_k \!+ \!2 g_q \eta\! - 3 g_{\Lambda} \eta^2\! + \frac{g_r}{\eta^2}\! + 3 \frac{g_s}{\eta^4}\! + \frac{\alpha \bigl(3 \alpha - 1 \bigr)}{\eta^{3 \alpha}}\! \Biggr] \!\tau - \! \frac{3}{2}\frac{ \alpha^2 t^2}{\eta^{6 \alpha  -1}} 
         \! \Biggr)^{\!\!\!2} \!{h}(\tau) \nonumber \\ 
         \!\!\!\!&& \!\! \approx  \!\!  -\frac{16}{3} \pi \eta \!\Biggl(\! - \frac{3 \alpha \gamma}{\bigl(3\alpha \!-\!1\bigr)\eta^{3\alpha - 1}} \!+\!\Biggl[ g_k\! + 2 g_q \eta \!- 3 g_{\Lambda} \eta^2 \! + \frac{g_r}{\eta^2} \!+ 3 \frac{g_s}{\eta^4}\! + \frac{\alpha \bigl(3 \alpha - 1 \bigr)}{\eta^{3 \alpha}} \!\Biggr] \!\tau \!
  -\frac{3}{2}\frac{ \alpha^2 t^2}{\eta^{6 \alpha  -1}} \!\Biggr)  . \nonumber \\
        \label{Mainmainmain} 
\end{eqnarray}
This approach, in view of the formal difficulties in solving  equation (\ref{Mainmain}), allowed us to 
explore the parameter space adopted in the formulation more comprehensively.
The pair of left figures in \ref{sol*}
depicts solutions of equation (\ref{Mainmainmain}) corresponding to $n=2$, while the first figure on the right illustrates solutions of equation (\ref{Mainmainmain}) considering the first two dominant terms corresponding to $n=-1$ The last figure represents a sample solution of equation (\ref{Mainmainmain}). The results in this figure correspond to a homogeneous distribution of primordial gravitational waves.  The sampling solutions consider the following 
boundary conditions: $\Psi(1)=1$, $\Psi'(1)=0$ $\Psi(1)=0$, $\Psi'(1)=1$. 
 
\subsection{Power spectrum and density parameter}
 
The energy density of gravitational waves can be written as
\begin{equation}
    \rho_{GW} = \int_0^{\infty} \frac{d\rho_{GW}}{df} df \, . \label{rhoGWpacheco}
\end{equation}
On the other hand, we can express the energy density of gravitational waves as
\begin{equation}
    \rho_{GW} = \langle T_{00} \rangle = \frac{c^2}{16 \pi G} \Bigl\langle \sum_{\alpha} \Bigl(\frac{\partial h_{\alpha}}{\partial t} \Bigr)^2  \Bigr\rangle \, ,  \label{rhoGWpacheco2}
\end{equation}
where the time average $\langle ... \rangle \equiv \frac{1}{T} \int_{T_i/2}^{T_f/2}...dt$ and the sum is performed over the polarization modes $+$ and $-$.

We can now define the dimensionless amplitude of the wave as
\begin{equation}
    h_{\alpha}(t) = \int_{\infty}^{\infty} \tilde{h}_{\alpha}(t) e^{i2 \pi f t} df \, . 
\end{equation}
Derivation of the equation above and squaring gives
\begin{equation}
  \Bigl(\!\frac{\partial h_{\alpha}(t)}{\partial t} \! \Bigr)^{\!\!2} \!\! = \!\! \int_{-\infty}^{\infty} \! \int_{-\infty}^{\infty} \!\! (-4 \pi^2) f f'  h_{\alpha}(f)
  \tilde{h}_{\alpha}(f') e^{i2 \pi (f + f') t} df df'. 
\end{equation}
Taking the time average from the previous equation
\begin{equation}
     \Bigl\langle \sum_{\alpha} \Bigl(\frac{\partial h_{\alpha}(t)}{\partial t} \Bigr)^2  \Bigr\rangle = \frac{4 \pi^2}{T} \sum_{\alpha} \int_{-\infty}^{\infty} f^2 \Bigl|
     \tilde{h}_{\alpha}(t)
     \Bigr|^2 df \, . \label{average}
\end{equation}
Replacing (\ref{average}) in (\ref{rhoGWpacheco2}) and comparing with (\ref{rhoGWpacheco}), we obtain
\begin{equation}
   f \frac{d\rho_{GW}}{df}  = \frac{\pi c^2}{4GT} f^3 \sum_{\alpha} \Bigl|
     \tilde{h}_{\alpha}(t)
     \Bigr|^2  \, .
\end{equation}
Defining the density parameter for gravitational waves
\begin{equation}
   \Omega_{GW} = \frac{1}{\rho_c} \frac{d \, \rho_{GW}}{d \, \log f}  = \frac{2\pi^2}{3 H_0^2T} f^3 \sum_{\alpha}  \Bigl|
     \tilde{h}_{\alpha}(t)
     \Bigr|^2  \, ,
\end{equation}
with $\rho_{c0}$ defining the closure density of the universe
\begin{equation}
    \rho_{c0} = \frac{3c^2 H_0^2}{8\pi G} \approx 7.6 \times 10^{-9} erg/cm^3 \, , \end{equation}
    with $H_0 = 67.74 km/s/Mpc$ \, . 

Introducing the spectral density by the relation
\begin{equation}
    S_n(f) = \frac{1}{2}f \sum_{\alpha} \Bigl| \tilde{h}_{\alpha}(f) \Bigr|^2  \, , 
\end{equation}
and taking the “averaging” time $T$ as the period of the wave results that $T = 1/ f$. In this conditions
\begin{equation}
    \Omega^{(0)}_{GW}(f) = \frac{4 \pi^2}{3 H_0^2} f^3 S_n(f) \, . \label{Sn}
\end{equation}

For a statistic, homogeneous, isotropic, unpolarized, and Gaussian GW background, the expression of the energy density of gravitational waves can be written  (in a simplified notation), in terms of the wave number $k$, as
\begin{eqnarray}
    \rho_{GW} & \! = \! & \frac{c^4}{32 \pi G \eta^2(\tau)} <\! h_r'(\mathbf{k},\tau)h_p'(\mathbf{q},\tau)\!> \, , 
    \nonumber \\
    & \! = \! &  \frac{c^4}{32 \pi G \eta^2(\tau)}\frac{8\pi^5}{k^3}\!<\! \delta^{(3)}\!(\mathbf{k} - \mathbf{q})\delta_{rp} h'^2\!(k,\tau)\!\!>  \, .
    \nonumber \\
    && \label{rhogw*}
    \end{eqnarray}
   For a general conformal time, taking $h'^2(k,\eta) \simeq  k^2h^2(k,\eta)$, which is approximately valid for $k^2>>H$, we obtain from (\ref{rhogw*})
    \begin{eqnarray}
       \rho_{GW}(k,\tau) & =  & \int_0^{\infty} \frac{dk}{k} \frac{d \,\rho_{GW}}{d\, \log k} \nonumber \\
       & = & \frac{1}{16\pi  \eta^2(\tau)} \int_0^{\infty} k\, dk \, h^2(k,\tau) \, , \label{rhoGW}
    \end{eqnarray}
   where
   \begin{equation}
    \frac{d \rho_{GW}}{d \, \log k} = \frac{k^2 h^2(k,\tau)}{16\pi  \eta^2(\tau)} =  \Omega_{GW}(k,\tau) \,  \rho_{c}\, .
   \end{equation}
From this expression, 
   \begin{equation}
      \Omega_{GW}(k,\tau) =  \frac{1}{\rho_{c}}  \frac{k^2 h^2(k,\tau)}{16\pi  \eta^2(\tau)}  \, ,\label{OmegaGW*} 
   \end{equation}
  where
   \begin{equation}
         \rho_{c} =  \frac{3c^2}{8 \pi G} H^2(\tau) \to    \rho_{c} =  \frac{3}{8 \pi} H^2(\tau)   \quad \mbox{(G = c =1)} \, , \label{rhoc*}
   \end{equation}
   with $\rho_{c}$ defining the density closure of the Universe at time $\tau$ and $H^2 = \Bigl(\frac{\eta'(\tau)}{\eta(\tau)}\Bigr)^2.$  
From expressions (\ref{Sn}) and (\ref{OmegaGW*}) we obtain, for any time $\tau$:
   \begin{eqnarray}
       S_n(k,\tau)  & = &  \frac{3 H^2 (2\pi)}{k^3}  \Omega_{GW}(k,\tau)
        =  \frac{3 H^2 (2\pi)}{k^3}  \frac{1}{\rho_{c}}  \frac{k^2 h^2(k,\tau)}{16\pi  \eta^2(\tau)} \nonumber \\
        & = & \frac{\pi h^2(k,\tau)}{k\eta^2(\tau)} \, , \label{SnOmegaGW}
   \end{eqnarray}
\begin{equation}
 \Omega_{GW}(f,\tau) = \frac{2 \pi^2}{3H^2} f^2 \frac{h^2(f,\tau)}{\eta^2(\tau)} \, , \label{Omegakey}
\end{equation}
and 
\begin{equation}
    S_n(f,\tau) =  \frac{1}{2f} \frac{h^2(f,\tau)}{\eta^2(\tau)} \, . \label{Skey}
\end{equation}

\section{Stochastic gravitational wave background}

Recent observations of gravitational waves by the Advanced LIGO and Virgo detectors indicate the possible existence of a stochastic gravitational wave background, similar to the cosmic microwave background, created through mergers of binary black holes and binary neutron stars throughout cosmic evolution. Doubts persists as to whether this stochastic background can be observed directly, or whether upper limits placed on it in specific frequency bands would imply important astrophysical and cosmological signals to be observed. In what follows we address the previously enumerated mechanisms for generating gravitational relic waves and present some preliminary numerical predictions of spectral intensities.

The first-order electroweak (EW) phase transition plays a pivotal role in electroweak baryogenesis. In the case of a strong first-order EW phase transition, gravitational waves can be generated through three different mechanisms: from the collision of bubbles during their expansion phase, which leads to quadrupole contributions to the stress-energy tensor, thus generating  GWs; through the decay of magneto-hydrodynamic turbulences arising from the movement of bubbles that converts kinetic energy into turbulence,  and, finally, through the propagation of damped sound waves~\cite{Pacheco2023WS}. The latter two mechanisms can persist for multiple Hubble times after the phase transition has occurred, providing additional sources of gravitational waves.
The intensity of gravitational waves is primarily determined by the following parameters: the duration scale of the transition, $\beta^{-1}$, the expansion speed of the bubble wall, $u_{\omega}$, and the ratio between vacuum and thermal energy densities, $\alpha_T$~\citep{Kamionkowski}.



\indent

\subsection{Sub-horizon condition with a matter-energy source}

In what follows, we consider solutions of the field equations that describe the time evolution of the perturbation $h_{\mu\nu}$ of the metric $h_{\mu\nu} = \eta_{\mu\nu} + h_ {\mu\nu}$, within the scope of branch-cut gravity in the sub-horizon mode with a matter-energy source. We refer more precisely to the linearized formulation of BCG, which corresponds to a first-order expansion of perturbations. Diffeomorphism invariance, similarly to General Relativity, due to the ontological character of BCG, is maintained, so that at the linearized level it takes the form of a gauge invariance. The symmetric tensor $h_{\mu\nu}$ comprises in principle ten degrees of freedom that are reduced, due to gauge invariance, to the two polarization orientations of gravitational waves and are most apparent in what is called the transverse traceless gauge (TT gauge), which is only valid in vacuum. In this work we do not delve into details regarding these aspects relating to the polarization of gravitational waves. Our primary intention was to deepen the non-commutative formulation in order to promote studies in the future that are more focused on the observation of signals from primordial gravitational waves. At this level, two regimes stand out, sub-horizon and super-horizon modes. As the spatiotemporal scales grow beyond the horizon, the amplitude of relic gravitational waves freezes and after the end of inflation these scales re-enter the horizon after the end of inflation during radiation and matter dominated era and would lay imprints on the CMB surface. As pointed out recently by~\citep{Odintsov}, sub-horizon and super-horizon modes will be probed distinctly by the future gravitational wave experiments and the current and future CMB-based experiments, and both are relevant for the identification of relic gravitational waves. However, according to~\citep{Profumo}, the most relevant effect of sub-horizon mode is a change in the PBH mass function and formation redshift, which may affect, in turn, relic gravitational wave (GW) observables. The authors found, in particular, that sub-horizon PBH formation enhances the isotropic SGWB energy density and the absolute angular power spectrum. In what follows, we focus our approach to this mode relating to the formation of primordial gravitational waves.

Combining equations (\ref{shc}), (\ref{NC1steq2}), and (\ref{VetatNew3}), the sub-horizon condition leads to the differential equation
\begin{eqnarray}
 {h''}(k,\tau) + k^2 {h}(k,\tau) 
  & = &  -\frac{16}{3} \pi \eta \Biggl( - \frac{3 \alpha \gamma}{\bigl(3\alpha-1\bigr)\eta^{3\alpha - 1}} -\frac{3}{2}\frac{ \alpha^2 \tau^2}{\eta^{6 \alpha  -1}}  \nonumber \\ 
  && +\Biggl[ g_k + 2 g_q \eta - 3 g_{\Lambda} \eta^2 + \frac{g_r}{\eta^2} + 3 \frac{g_s}{\eta^4} + \frac{\alpha \bigl(3 \alpha - 1 \bigr)}{\eta^{3 \alpha}} \Biggr] \tau 
  \Biggr)   \, . \label{equationh}
\end{eqnarray}
The aforementioned field equation describes the evolution over time of the metric disturbance having as its source the potential defined in equation (\ref{NC1steq2}) that configures the dynamic composition of matter in the primordial cosmic period. 
To solve equation (\ref{equationh}), due to the technical difficulties of its resolution, we used powerful computational algorithms, based on the Runge-Kutta-Fehlberg method, that made it possible to obtain algebraic solutions without the need to implement simplifications or any computational approximation (see Appendix C). 

   \subsection{Generation of gravitational waves: bubble collisions}

As an application example of our approach, in what follows we determine the spectral shape of relic gravitational waves produced by bubble collisions~\cite{Caprini2016}:
\begin{equation}
    h_0^2 \Omega_b(\nu/\nu_b) = 1.67 \times 10^{-5}  \Bigl( \frac{H_*}{\beta} \Bigr)^2  \Biggl( \frac{\kappa_{\infty} \alpha_T}{1 + \alpha_T} \Biggr)^2 \Bigl( \frac{100}{g_*} \Bigr)^{1/3}  \Biggl( \frac{0.11 u^3_{\omega}}{0.42+u^2_{\omega}} \Biggr) S_b(\nu/\nu_b) \, , \label{hoOmegacnu}
\end{equation}
In the strong regime, $u_{\omega} \geq 0$ and $\alpha_{\infty}$ represents the critical value for $\alpha_T$~\citep{Pacheco2023WS,Caprini2016}. 
The parameter $\nu_b$ represents the characteristic frequency, which is contingent upon the duration of the transition denoted as $\beta^{-1}$. Corrected for the redshift effect, this quantity is given by~\citep{Pacheco2023WS}
\begin{eqnarray}
  \nu_b  = 1.65 \times  10^{-5} \Bigl( \frac{\beta}{H_*} \Bigr)  \Bigl( \frac{T_*}{100~{\rm GeV}} \Bigr)  \Bigl( \frac{g_*}{100} \Bigr)    
  \Biggl( \frac{0.62}{1.8+u^2_{\omega}-0.1u_{\omega}} \Biggr) Hz \approx 1{\rm mHz} \, .   \label{nuc}
\end{eqnarray}
\begin{figure*}[htpb]
\centering
\includegraphics[scale=0.40]{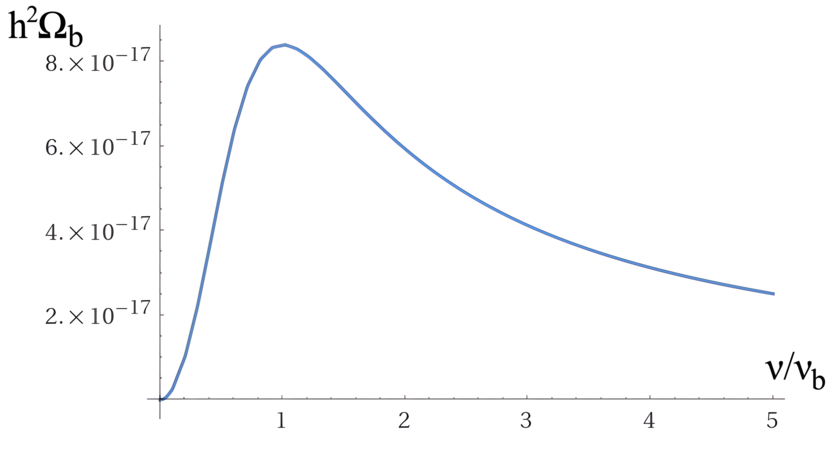} 
\caption{Typical regularized gravitational-waves spectra generated during a strong EW phase transition due to bubble collisions.
} \label{EWMHD}
\end{figure*}
In the following, due to the formal complexity of the solutions of equation (\ref{equationh}), guided by the dominant terms of these solutions in the domain $\nu/\nu_c \to 0$ (see Appendix C), we propose in expressions (\ref{*Sb}) and (\ref{Sb*}) an analytical formulation that maintains the essence of the branch-cut quantum gravity proposal as well as the main requirements of the spectral shape of the signal. Combining equation (\ref{Skey}) and the solution of (\ref{equationh}), by means of a systematic preliminary analytical study summarized in Appendix C, we propose the following expression to represent the spectral shape of the signal:
\begin{equation}
    S_b(\nu/\nu_c) = \sum_{m=0}^{\kappa} (-1)^{m+1} \frac{b}{a^m \bigl(\frac{\nu}{\nu_c}\bigr)^{mb - a}} \, ,  \label{*Sb}
\end{equation}
where $a$ and $b$ represent positive expansion coefficients, $b \equiv a +1$ and $m$ denotes real positive numbers. Under these conditions, the series above may converge to
\begin{equation}
 S_b(\nu/\nu_c) \Rightarrow  \frac{b \times \bigl(\frac{\nu}{\nu_b}\bigr)^a}{a \times \bigl(\frac{\nu}{\nu_b}\bigr)^b + 1} + {\cal O}
 \Bigl( \Bigl(\frac{\nu}{\nu_b} \Bigr)^{\kappa b - a} \Bigr) \, , \label{Sb*}
\end{equation}
in order $m = \kappa$. 
 As for the content of the initial proposal for the branched quantum gravity, by introducing a potential that would describe the main content of the relic Universe in terms of mass and energy contributions, against the backdrop of a non-commutative algebraic structure, sources of the primordial gravitational waves, these aspects are also well covered in the proposal considered. After carrying out the different theoretical steps, we arrive to a formal dependence on the corresponding spectral shape of the signal in terms of a convergent series $(\nu/\nu_c)^{\kappa}$-dependent, in different orders. In short, our theoretical developments indicate that the final expression for the spectral shape of the signal should reduce to a regular function with may be expanded in 
$1/(\nu/\nu_c)^{\kappa}$-dependent series. In addition, the proposal defined in equation (\ref{Sb*}) contemplates the condition that the maximum of the function corresponds to the value of the characteristic frequency, $\nu/\nu_c =1$, as well as additionally makes it possible to fit simulated data with the simple choice of the only formulation parameter, $a$ (see for example~\cite{Pacheco2023WS}, in which the choice of $a$ corresponds to $2.8$ to fit 
simulated data to describe spectrum of the gravitational radiation determined by numerical simulations using the 'envelope approximation'~\cite{Huber}).


\section{Discussion of Results}

There are numerous research works in the literature that address the topic of quantum gravity in a non-commutative environment. However, with regard to the description of the wave function of the Universe based on the formalism of Ho\v{r}ava-Lifshitz quantum gravity and the Wheeler-DeWitt equation, or based on other alternative approaches, the number of articles dealing with this topic are not significant. 
Most of the published works deal with the temporal evolution of the scale factor of the Universe, a topic of studies commonly designated as ``dynamical equations". In some articles the authors deal with both topics. We indicate a few references on these topics~\cite{Bastos,Remo,Mena,Hugo,Neto,Vakili,Obregon,Monerat,Vaz,Lizzi,Haidar,Kristian}). We emphasize once again that concerning the wave function of the Universe in a non-commutative environment, most authors, due to the inherent computational difficulties imposed by the formalism, use approximations that significantly limit their conclusions and direct their studies towards formal aspects without a numerical approach or algebraic results.

 Regarding the dynamical equations, some studies involving non-commutative formulations appear to be more improved but are still limited by significant computational approximations. It is important to mention, although this is a non-central topic in the present work, studies involving strings and non-commutative gauge theories which have contributed significantly to a better understanding of the influence of non-commutative algebra on the deformation of geometric structures and the impacts of these studies in understanding the accelerated evolution of our Universe in string theory. Edward Witten, precursor of this line of investigation~\cite{Witten}, was followed by many scientists who have followed this line of investigation (see for instance ~\cite{Dolan,David} and references therein).
 
In our case, in describing the evolution of the Universe's wave function, as well as in the dynamic equations involving the Universe's scale factor as well as in relic gravitational waves, we focus on consistently numerically solving the resulting equations without using, from the point from a computational point of view, no approximation, through a numerical approach based on the method known as Runge-Kutta-Fehlberg. This method proves to be quite powerful for solving differential equations, without adopting any numerical approximation. This computational approach made possible, despite the extreme formal complexity of the theory and the resulting equations, a broad numerical study of the evolutionary process focused on in the present work, through the exploration of the theory's parameter space. This study thus enabled a broad study of the evolution of the functions $\Psi(\eta)$, $\Psi(\xi)$, $\eta(t)$, and $\xi(t)$ and a wide class of solutions that for reasons of brevity we limited their presentation to just a few figures. The most expressive results involving both the contraction phase and the expansion phase of the branched Universe indicate an accelerating cosmic expansion.


\section{Conclusions}\label{7}


The paradigm supporting the theory of renormalization
groups -organizing physical phenomena based on energy or distance
scales- holds firm in commutative quantum field theories. In their
non-commutative counterparts, however, one encounters an uncertain
territory.
One of the features of non-commutative field theories is the mixing of short and long scales. 
A striking illustration of this phenomenon can be found in UV/IR
mixing~\cite{Ya}. In cases where non-commutativity exists at a small
scale, the UV/IR mixing effect is anticipated to manifest at an
earlier epoch in the Universe's history, thereby contributing to raising new questions about the
hierarchy problem.
Another striking example pertains to the inhomogeneities within the distribution of large-scale structures
and anisotropies observed in the CMB radiation. These anomalies bear traces of the non-commutative nature of the early Universe. Specifically, the power spectrum of these structures becomes direction-dependent in non-commutative spacetime~\cite{Bartlett}.

In 1947,  Hartland Sweet Snyder introduced the concept of quantizing spacetime in a seminal paper~\cite{Snyder}. While this paper has received relatively little citation, it sparked one of the most remarkable inquiries in the realm of physics, namely the possibility of discretizing spacetime.
In alignment  with Snyder's proposal, 
the uncertainty principle of Werner Heisenberg suggests that spacetime possesses a non-commutative structure, which can be represented as
\begin{equation} 
[x^{\mu},x^{\nu }] = i\theta^{\mu \nu} \, . 
\end{equation}
This non-commutative property implies a minimum scale of 
approximately $\sim \sqrt{\theta}$.
From a cosmological perspective, to assess the implications of this concept on the dynamics of the Universe, it is most appealing to investigate how the non-commutativity  of coordinates affects the 
deformation of spacetime algebra. We assume that this equation holds within a comoving frame, a coordinate system in which galaxies are freely falling.

In this work we adopt an alternative approach by considering a non-commutative quantum cosmological scenario based on the deformation of a mini-superspace of variables obeying Poisson algebra. We aim to examine whether such a perspective can help identify the mechanism that drives the acceleration of the Universe~\cite{Riess}. While the results presented here are preliminary, they are promising in suggesting that  non-commutative branch-cut gravity offers an algebraic framework that impacts both the statistical distribution and gravitational dynamics of the matter constituting the primordial Universe. 
As previously emphasized, the implications of non-commutative algebra 
are evident in the solutions presented in this study. These 
implications manifest in various aspects, including the wave function of the early Universe, the dynamical equations involving the cosmic scale factor and its dual counterpart, as well as relic gravitational waves. The results point to a dynamic acceleration driven by a force that in our view, arises from the reconfiguration of matter in the early Universe due to the algebraic structure of non-commutative geometry. This structure captures the intrinsic properties of spacetime at short distances, with significant implications for the dynamical symmetries and conventional duality symmetries of quantum spacetime geometry.

A seminal aspect of our study involves the propagation of relic gravitational waves in the stages before and after the transition involving our Universe and its mirror counterpart. 
In our study, we have identified a similar behavior involving relic gravitational waves and the wave function of the Universe, with respect to a topological quantum leap in the branch-cut transition region, addressed in other studies~\citep{Pacheco,Bodmann2023a,Bodmann2023b}.  Based on 
Bekenstein's universal upper bound of a localized quantum system that establishes that
\begin{equation}
    S \leq \frac{2 \pi k_B R E}{\hbar c} \,,. \label{UpperBound}
\end{equation}
with $E$ representing the total energy of a system enclosed in a circumference with surface area $A$ and radius $R$, we thus developed a
 Generalized Heisenberg Uncertainty Principle (GHUP) version for the Bekenstein's criterion~\citep{Bodmann2023b}, establishing a point of contact with Snider's predictions
\begin{equation}
  \Delta S \Delta E \lesssim \frac{\hbar c}{2}
  \frac{\pi k_B \bigl(1 + \theta \bigr)}{\sqrt{\theta} \ell_P}  
\, , \label{GHUfinal}
\end{equation}
in which $\sqrt{\theta}$ represents the minimum scale in the non-commutative space-time algebra, so the minimum observable GHUP length is assumed as $\Delta x \sim \sqrt{\theta} \ell_P$.

The relic cosmological stochastic background, as expected, is characterized by stability, isotropy, absence of polarization, whose origin are fundamental processes that supposedly occurred in the early Universe~\citep{Pacheco2023WS}. Among these processes, quantum vacuum fluctuations, cosmic phase transitions and cosmic strings stand out~\citep{Caprini}. And insofar, as pointed out earlier, beyond redshift $z\sim  1100$, the early Universe is opaque to electromagnetic radiation, these relic gravitational waves are the only capable messengers of the prevailing physical conditions of the early Universe.

In this contribution, the results obtained for the description of gravitational waves bring a new perspective, in line with the implications of equation (\ref{GHUfinal}) which indicate an uncertainty around the previous boundary conditions, as a result of a minimum scale. And also in line with the implications of the Bekenstein criterion, indicating a quantum leap in the boundary of the branch-cut transition region.  
These results also indicate, in view of the mixture of intensities of the original sources of relic gravitational waves identified by the distribution of colors associated with the quantities that describe the corresponding power spectra and density parameters, two fundamental aspects. The first concerns the primordial vision of a smooth and isotropic Universe, symmetrical to ours. In this description, the stress-energy must take the form of a perfect fluid fully described by an energy density $\rho_{\epsilon}$ and pressure p, in which the variations of the internal equation of state are so small that they make no qualitative difference in the cosmic evolutionary process. This premise seems to collapse when we examine the results obtained in this contribution, which indicate asymmetric Universes, whether in the distribution of intensities of gravitational wave sources, or in the implications of non-symmetric algebra in the short-distance structure of spacetime. The second aspect concerns the transition region between the different phases of the Universe, bringing to light the discussion about the possibility of observing signatures not only corresponding to the era marked by the dissociation of matter and radiation in our Universe, but also the phases associated with the cosmic seeds, sources of gravitational waves, existing in the mirror Universe.

In conclusion, the results suggesting a primordial dynamic acceleration of spacetime demonstrate that non-commutative quantum branch-cut gravity provides a viable theoretical alternative to models such as
inflation~\cite{Guth1981,Guth2004} and bouncing~\cite{Ijjas2014,Ijjas2018,Ijjas2019}. 
This exploration aligns with a fundamental characteristic of non-commutative algebra, namely the interplay between small and large scales.
As a result, if the effects of non-commutative algebra were indeed present in the primordial Universe, it is reasonable to anticipate their persistence in the present day. 

A fundamental question then emerges: Can inhomogeneities in the distribution of large-scale structures and anisotropies in the stochastic gravitational wave background (SGWB), if they indeed exist, carry traces of the non-commutative nature of the early Universe? Our results indicate a scenario during the early stages of the Universe, characterized by an SGWB distribution that deviates significantly from the homogeneity expected to be observed today. 

As a consequence, 
inhomogeneities in the SGWB distribution, if influenced by
traces of non-commutativity, could serve as 
crucial windows into the initial phases of the early Universe,
preceding the recombination era. 
However,  answering this question necessitates further observations,
and this theme will remain the primary focus of our ongoing investigations.

A final word regarding the results obtained to describe relic gravitational waves in a non-commutative formulation: the regular function that describes the spectral shape of the signal is the result, among many other functional possibilities, of the Lagrangian structure of the branched quantum gravity formulation. proposal. There are still aspects to be revealed in the future, in order to introduce a functional formulation, albeit regular, that highlights the behavior of the spectral form of the signal parametrically dependent on the non-commutative structure. This aspect, although in our formulation it is implicitly contained in the functional structure of the expression (\ref{Sb*}), deserves a consistent specific approach.

\section{Acknowledgements}
P.O.H. acknowledges financial support from PAPIIT-DGAPA (IN116824). F.W. is supported by the U.S. National Science Foundation under Grant PHY-2012152.
\appendix
\section{Commuting and non-commuting coordinates} \label{A}
\def\ii{\'\i}
In what follows, we will elaborate on how to find a transformation from non-commuting coordinates to commuting ones. 
We assume that $u$ and $v$ still commute, i.e.,
$\sigma = 0$. 

Let us define the commuting and non-commuting coordinates respectively
as
\begin{equation}
({\tilde x}_i)  =  ({\tilde u}, \tilde{p}_u, {\tilde v}, {\tilde p}_v)  \quad \mbox{and} \quad
(x_i)  =  (u,p_u , v, p_v)
~.
\label{eq-1}
\end{equation}
The commuting and non-commuting coordinates satisfy both the respective 
Poisson-brackets
\begin{equation}
\left\{ {\tilde x}_i , {\tilde x}_j \right\}  =  {\tilde g}_{ij}  \quad \mbox{and}
\quad
\left\{ x_i , x_j \right\}  =  g_{ij} \, 
,
\label{eq-2}
\end{equation}
where on the right side of these equations we have the symplectic metrics, satisfying
respectively
\begin{equation}
{\tilde g}_{ji}  =  -{\tilde g}_{ij}  \quad \mbox{and} \quad
g_{ji}  =  -g_{ij}
~.
\label{eq-3}
\end{equation}
This symmetry property is the reason why we can call it a symplectic space.
The matrix structure of these metrics is
\begin{eqnarray}
\left({\tilde g}\right)  = 
\left(
\begin{array}{cccc}
0 & 1 & 0 & 0 \\
-1 & 0 & 0 & 0 \\
0 & 0 & 0 & 1 \\
0 & 0 & -1 & 0 
\end{array}
\right) \, ; \quad
\left(g\right)  = 
\left(
\begin{array}{cccc}
0 & 1 & 0 & \gamma \\
-1 & 0 & -\chi & \alpha \\
0 & \chi & 0 & 1 \\
-\gamma & -\alpha & -1 
& 0 
\end{array}
\right)
~.
\label{eq-4}
\end{eqnarray}
Using (\ref{eq-2}), this leads to the non-zero Poisson brackets
(only the Poisson brackets are listed which will be non-zero in the
non-commuting case)
\begin{eqnarray}
\left\{ {\tilde u}, {\tilde p}_u \right\} & = & 1 \, ; \quad 
\left\{ {\tilde v}, {\tilde p}_v \right\}  = 1 \, ; \quad  \mbox{and} \quad
\left\{ {\tilde u}, {\tilde p}_v \right\}  =  0 \, ;
\nonumber \\
\left\{ {\tilde v}, {\tilde p}_u \right\} & = & 0 \, ; \quad \mbox{and} \quad 
\left\{ {\tilde p}_u, {\tilde p}_v \right\} = 0 \, ; 
\end{eqnarray}
and for the non-commuting case
\begin{eqnarray}
\left\{ u, p_u \right\} & = & 1 \, ; \quad 
\left\{ v, p_v \right\} =  1 , ; \quad \mbox{and} \quad
\left\{ u, p_v \right\}  =  \gamma \, ; 
\nonumber \\
\left\{ v, p_u \right\} & = & \chi \, ; \quad \mbox{and} \quad 
\left\{ p_u, p_v \right\} =  \alpha
~.
\label{eq-5}
\end{eqnarray}
Now, we look for a transformation 
\begin{eqnarray}
{\tilde x}_i & = & \sum_j M_{ij} x_j \, , 
\label{eq-6}
\end{eqnarray}
such that the above Poisson brackets are satisfied. This gives us conditions
for the matrix elements $M_{ij}$. Equation
(\ref{eq-6}) can be cast into the form
\begin{eqnarray}
{\tilde u} & = & M_{11} u + M_{12} p_u + M_{13} v + M_{14} p_v \, ;
\nonumber \\ 
{\tilde p}_u & = & M_{21} u + M_{22} p_u + M_{23} v + M_{24} p_v \, ; 
\nonumber \\ 
{\tilde v} & = & M_{31} u + M_{32} p_u + M_{33} v + M_{34} p_v \, ; 
\nonumber \\ 
{\tilde p}_v & = & M_{41} u + M_{42} p_u + M_{43} v + M_{44} p_v \, ; 
~.
\label{eq-7}
\end{eqnarray}
Using that $\tilde{u}=u$ and $\tilde{v}=v$, we can reduce the
matrix $M$ to the expression
\begin{eqnarray}
\left( M\right)  & = &
\left(
\begin{array}{cccc}
1 & 0 & 0 & 0 \\
M_{21} & M_{22} & M_{23} & M_{24} \\
0 & 0 & 1 & 0 \\
M_{41} & M_{42} & M_{43} & M_{44}
\end{array}
\right) \, . 
\label{eq-8}
\end{eqnarray}
Using (\ref{eq-5}), and the transformation (\ref{eq-6}) with
(\ref{eq-8}), we obtain
\begin{eqnarray}
\left\{ {\tilde u}, {\tilde p}_u\right\} & = & 1 =
M_{22} + M_{24}\gamma \, ;
\nonumber \\
\left\{ {\tilde v}, {\tilde p}_v\right\} & = & 1 =
M_{42}\chi + M_{44} \, ;
\nonumber \\
\left\{ {\tilde u}, {\tilde p}_v\right\} & = & 0 =
M_{42} + M_{44}\gamma \, ;
\nonumber \\
\left\{ {\tilde v}, {\tilde p}_u\right\} & = & 0 =
M_{22}\chi + M_{24} 
~,
\label{eq-9}
\end{eqnarray}
where the Poisson-bracket $\left\{ \tilde{p}_u, \tilde{p}_v\right\} = 0$
will be calculated later. First of all we resolve the set of equations
(\ref{eq-9}), which leads to
\begin{eqnarray}
M_{22} & = & \frac{1}{1-\gamma \chi} \, ; \quad \quad M_{24} = -\frac{\chi}{1-\gamma\chi} \, ; 
\nonumber \\ 
M_{42} & = & -\frac{\gamma}{1-\gamma \chi} \, ; \quad 
M_{44} = \frac{1}{1-\gamma\chi}
~,
\label{eq-10}
\end{eqnarray}
which leads consequently to the matrix
\begin{eqnarray}
\left( M\right)  & = &
\left(
\begin{array}{cccc}
1 & 0 & 0 & 0 \\
M_{21} & \frac{1}{1-\gamma\chi} & M_{23} & -\frac{\chi}{1-\gamma\chi} \\
0 & 0 & 1 & 0 \\
M_{41} & -\frac{\gamma}{1-\gamma\chi} & M_{43} & \frac{1}{1-\gamma\chi}
\end{array}
\right)
~.
\label{eq-11}
\end{eqnarray}
An additional information we get when calculating the missing commutator:
\begin{equation}
\left\{ {\tilde p}_u, {\tilde p}_v \right\}  =  0
= -M_{41}+M_{23}+\frac{\alpha}{a-\gamma\chi}
~,
\label{eq-12}
\end{equation}
which leads to 
\begin{eqnarray}
M_{41} & = & \frac{\alpha}{1-\gamma\chi} + M_{23}
~~~.
\label{eq-13}
\end{eqnarray}
So, finally we obtain the structure of the transformation matrix, with
still some liberty because $M_{21}$, $M_{23}$ and $M_{43}$ are free to
choose! That the momenta now commute is important, because this
allows us to write these momenta proportional to derivatives of the
conjugate variable.

With the last form of the transformation matrix and an appropriate choice of the remaining matrix elements, we obtain for the
transformation of the momenta:
\begin{eqnarray}
{\tilde p}_u & = & M_{21} u + \frac{1}{1-\gamma\chi} p_u + M_{23}v
-\frac{\chi}{1-\gamma\chi} p_v~;
\nonumber \\
{\tilde p}_v & = & \left(\frac{\alpha}{1-\gamma\chi} + M_{23}\right) u
-\frac{\gamma}{1-\gamma\chi} p_u 
 +  M_{43} v + \frac{1}{1-\gamma\chi} p_v~,
\label{eq-14}
\end{eqnarray}
where we have defined $\Gamma \equiv (\gamma\chi -1)$.

There is still an ambiguity, which we call gauge transformation, to choose the remaining matrix elements. There are two possible paths: The first is to select a particular choice for ${\tilde p}_u$ and ${\tilde p}_v$ in terms of the non-commuting variables. The second one is to invert the matrix and select a particular choice for the non-commuting momenta $p_u$ and $p_v$ in terms of the commuting ones. We chose the second path and the particular choice is listed in the main text. The choice then  gives conditions to the remaining matrix elements, which can be resolved. 

The present deduction also implies that there are several options in 
choosing ${\tilde p}_u$ and ${\tilde p}_v$, depending on the particular choice
of $M_{21}$, $M_{23}$ and $M_{43}$. Still, for each alternative one has to verify 
the relations of the Poisson brackets.

\section{Generation of gravitational waves: bubble collisions} \label{B}

In our calculations for the generation of gravitational waves as a result of bubble collisions, we use $\alpha_{\infty} \simeq 2.71 \times 10^{-2}$, and assume $u_{\omega} \to 1$ (runaway regime)~\citep{Pacheco2023WS}.
The parameter $\kappa_{\infty}$ holds a crucial role in our analysis,
as it quantifies the efficiency of converting latent heat into bulk
motion, a pivotal factor in defining the amplitude of GW signals. In reference to~\cite{Espinosa}, $\kappa_{\infty}$
is approximately given as $\kappa_{\infty}\approx 3.516 \times 10^{-2}$.
When considering the parameter $\alpha_T$, from~\citep{Pacheco2023WS} we use $\alpha_T \approx 3.68 \times 10^{-2} (100/g_*).$
Using the above values for $\kappa_{\infty}$ and $\alpha_T$, we obtain
\begin{equation}
\Biggl(\frac{\alpha_T \, \kappa_{\infty}}{1 + \alpha_T} \Biggr)^2 
 \approx  1.557411 \times 10^{-6} \, , \label{alphaTkinfty}  
\end{equation}
with $g_*$ is given as~\citep{Addazi}
\begin{equation}
g_* = g_*^{\rm{[SM]}} = 100, \quad \mbox{and thus} \quad 
    \Biggl( \frac{100}{g_*} \Biggr)^{1/3} = 1 \,.
\end{equation}
  Assuming the runaway regime ($u_{\omega} \to 1$), we get  
  \begin{equation}
     \Bigl( \frac{0.11 u^3_{\omega}}{0.42+u^2_{\omega}} \Bigr) \approx 7.746479 \times 10^{-2} \,, \label{uomega}
\end{equation}
and 
\begin{equation}
     \Bigl( \frac{0.62}{1.8+u^2_{\omega} -0.1n_{\omega}} \Bigr) \approx 2.29629 \times 10^{-1} \, . 
\end{equation}
    In a first order phase transition, the bubble nucleation process is fixed by the tunnelling probability between the two vacua states of the effective potential~(for the details see ~\citep{Pacheco2023WS})
    \begin{equation}
        V(T,\phi) = \frac{\gamma}{2} \Bigl(T^2 - T_0^2 \Bigr) \phi^2 - \frac{\sigma}{3} T \phi^3 + \frac{\lambda}{4} \phi^4 \,. 
    \end{equation}
   The usual standard model potential does not generates a strong transition required to produce a significant background. In general, the potential must be modified and a minimum change implies in additional gauge bosons (at least two new ones), what in practice means to modify the $\phi T^3$ term.
   The solution for the vacuum states permits the evaluation of the nucleation temperature, that is of the order of 
$T_* = 166$ GeV~\citep{Pacheco2023WS}. 
During the EW phase transition, a fraction of the latent heat is used to excite sound waves, turbulence, and the bulk motion of bubbles, which are able to generate gravitational waves. Thus, the physical conditions of transition must be used for all mechanisms, which are not independent.
Once the nucleation temperature $T_*$ is computed (or fixed), the duration of the transition can be estimated from~\citep{Pacheco2023WS}
\begin{equation}
    \Biggl( \frac{\beta}{H_*} \Biggr) \simeq 4 \, \ln\Biggl( \frac{M_P}{T_*} \Biggr) \, . 
\end{equation}
However, fixing the nucleation temperature to be $T_* = 166 \, GeV$,
the relation above implies $(\beta/H_*) \approx 155 .$

\section{Gravitational waves spectral shape of the signal}

In this appendix we briefly describe the steps to obtain the gravitational waves spectra described by the spectral shape of the signal, considering the sub-horizon condition with a matter-energy source given by equation (\ref{equationh}).
To solve this equation, due to the technical difficulties of its resolution, as we stressed before, we used powerful computational algorithms, based on the Runge-Kutta-Fehlberg method, that made it possible to obtain algebraic solutions without the need to implement simplifications or any computational approximation. The corresponding solution of the field equation that describes the evolution over time of the metric disturbance having as its source the potential defined in equation (\ref{NC1steq2}), that configures the dynamic composition of matter in the primordial cosmic period, is
\begin{eqnarray}
    {h}(k,\tau) &\! = \!& c_2(\eta) \sin(k \tau) + c_1(\eta) \cos(k \tau) 
    + \frac{1}{k^4\eta^3}  \Bigl[k^2 \Bigl(6.28319 \tau^2 \eta^2 
    + \tau \bigl(-4.18879 \eta^{5/2} \! + 16.7384 \eta^6  \nonumber \\
    &&- 23.4572 \eta^5  -  16.7552 \eta^4
     -  6.70206  \eta^2 - 0.150796\bigr)+ 50.2655 \eta^{7/2} \Bigr) - 12.5664 \eta^2 \Bigr]. \nonumber \\ && \label{equationh*}
\end{eqnarray}
$S_b(\nu/\nu_c)$, the spectral shape of the signal, is determined in terms of the solution (\ref{equationh*}) of equation (\ref{equationh}), after integrating out the time dependence of $S_b(\nu/\nu_c, \tau)$, taking the limits $0$ to $\beta/H_*$,
with $k \to \nu/\nu_c$:
\begin{eqnarray}
 \!\!\!\!\! \!\!\!\!\!    S_b(\nu/\nu_c) & = & {\cal A}(\nu/\nu_c)\int_0^{\beta/H_*} \!\!\!\!\!  S_b(\nu/\nu_c,\tau) d\tau \nonumber \\
    & = & {\cal A}(\nu/\nu_c)\int_0^{\beta/H_*} \!\!\!\!\!\!\!\!\!\!  \frac{1}{4\pi (\nu/\nu_c)} \frac{h^2(\nu/\nu_c,\tau)}{\eta^2(\tau)} d\tau
    \, , \label{Skey*}
\end{eqnarray}
where the regulatory function ${\cal A}(\nu/\nu_c)$ guarantees the normalization of the spectral shape, the boundary condition $S_n(\nu/ \nu_c) \to 0$, as well as the position of the maximum of the function at the point $\nu/\nu_c = 1$. 
The final expression for the 
spectral shape of the signal is quite intricate and difficult to manipulate. 
As a preliminary study, in the following we insert a generic scale power law behavior for the branch-cut scale factor,  $\eta(\tau) = a_n \tau^n$, which covers the cases of radiation ($n=1$) and matter ($n=2$) domination, as well as the de Sitter inflation ($n=-1$) to determine the main ingredients of the time-independent spectral shape of the signal, expressed as $S_b(\nu/\nu_c)$. Taking for simplicity the case $n=1$, we show below a series expansion of the resulting integral, limiting for convergence reasons the final expression to the $(\nu/\nu_c)^6$- and $(\beta/H_*)^6$-dependent terms. The mean ingredients of this calculation, except for determining the regulatory function ${\cal A}(\nu/\nu_c)$,  
are synthesized in the following expression:
\begin{eqnarray}
 S_n(\nu/\nu_c) & \approx & \Bigl( \sin\Bigl(k\frac{\beta}{H_*}\Bigr) + \cos\Bigl(k\frac{\beta}{H_*}\Bigr) \Bigr) \Biggl\{ 0.0795775 
 \Bigl(\frac{\nu_c}{\nu}\frac{H_*}{\beta} \Bigr) - 0.159155 \Bigl(\frac{\nu_c}{\nu}\frac{H_*}{\beta} \Bigr)^{\!\!3} \nonumber \\ && +  1.90986 \Bigl(\frac{\nu_c}{\nu}\frac{H_*}{\beta} \Bigr)^{\!\!5}  \! \Biggr) + {\cal O} \Bigl( \Bigl( \frac{H_*}{\beta} \Bigr)^6 \Bigr) 
  +  \Bigl(\frac{\nu_c}{\nu}\Bigr)^5 \Biggl[ 3.15843 \Bigl(\frac{\beta}{H_*} \Bigr) + 92.4467 \Bigl(\frac{\beta}{H_*} \Bigr)^{1/2} \nonumber \\ &&
  + 169.119 log\Bigl(\frac{\beta}{H_*} \Bigr) + 0.150401 \Bigl(\frac{H_*}{\beta} \Bigr) + 0.733704 \Bigl(\frac{H_*}{\beta} \Bigr)^{3/2} 
 \! -  0.000596831 \Bigl(\frac{H_*}{\beta}  \Bigr)^{3} \! \Biggr] \nonumber \\ && + 0.0795775 log\Bigl(\frac{\nu_c}{\nu}\Bigr) -
 0.0397887 log\Bigl(\Bigl(\frac{\nu_c}{\nu}\Bigr)^2\Bigr) + 0.125 + {\cal O} \Bigl( \Bigl( \frac{H_*}{\beta} \Bigr)^{11/2} \Bigr)
\nonumber \\
&& + \Biggl[\Bigl( \sin\Bigl(k\frac{\beta}{H_*}\Bigr) - \cos\Bigl(k\frac{\beta}{H_*}\Bigr) \Bigr) \Biggl( 0.0795775  \Bigl(\frac{\nu_c}{\nu}\frac{H_*}{\beta} \Bigr)^2 -  0.477465 \Bigl(\frac{\nu_c}{\nu}\frac{H_*}{\beta} \Bigr)^4 \Biggr) \nonumber \\ && 
 +  \Bigl( \sin\Bigl(k\frac{\beta}{H_*}\Bigr) + \cos\Bigl(k\frac{\beta}{H_*}\Bigr) \Bigr) \Biggl(-
0.0795775 
 \Bigl(\frac{\nu_c}{\nu}\frac{H_*}{\beta} \Bigr) \Biggr)+  {\cal O} \Bigl( \Bigl( \frac{H_*}{\beta} \Bigr)^6 \Biggr]
\Biggr\} . \nonumber \\ &&
\end{eqnarray}

\section*{Author Contributions}
Conceptualization, C.A.Z.V.; methodology, C.A.Z.V. and B.A.L.B. and P.O.H and J.A.deF.P. and D.H. and F.W. and M.M.; software, C.A.Z.V. and B.A.L.B. and M.R. and M.M.; validation, C.A.Z.V. and B.A.L.B. and D.H. and P.O.H. and J.A.deF.P. and F.W.; formal analysis, C.A.Z.V. and B.A.L.B. and P.O.H. and J.A.deF.P. and D.H. and F.W.; investigation, C.A.Z.V. and B.A.L.B. and P.O.H. and J.A.deF.P. and M.R. and M.M. and F.W.; resources, C.A.Z.V.; data curation, C.A.Z.V. and B.A.L.B.; writing—original draft preparation, C.A.Z.V.; writing—review and editing, C.A.Z.V. and B.A.L.B. and P.O.H. and J.A.deF.P. and D.H. and M.R. and M.M. and F.W.; visualization, C.A.Z.V. and B.A.L.B.; supervision, C.A.Z.V.; project administration, C.A.Z.V.; funding acquisition (no funding acquisition). All authors have read and agreed to the published version of the manuscript.

\end{document}